\documentclass[floats,floatfix,showpacs,amssymb,physrev,twocolumn,superscriptaddress,reprint,
nofootinbib, longbibliography]{revtex4-2}

\usepackage{amssymb,amsmath,verbatim,mathtools,needspace,enumitem,etoolbox,graphicx,physics,microtype,afterpage,xspace}

\usepackage[dvipsnames, usenames]{xcolor}
\usepackage{standalone}
\definecolor{linkcolor}{rgb}{0.0,0.3,0.5}
\usepackage[unicode, colorlinks=true, linkcolor=linkcolor, citecolor=linkcolor, filecolor=linkcolor,urlcolor=linkcolor, pdfusetitle]{hyperref}
\usepackage[all]{hypcap}
\usepackage[T1]{fontenc}
\usepackage{tikz}
\usepackage{tabularx}
\usepackage{xcolor}
\usepackage{natbib}
\usepackage[caption=false]{subfig}
\definecolor{britishracinggreen}{rgb}{0.0, 0.26, 0.15}
\definecolor{cadmiumgreen}{rgb}{0.0, 0.42, 0.24}
\definecolor{forestgreen}{rgb}{0.13, 0.55, 0.13}
\allowdisplaybreaks
\newcommand{\rad}{\mathrm{rad}}
\newcommand{\tl}{\tilde{\ell}}
\newcommand{\tm}{\tilde{m}}
\newcommand{\rmd}{\mathrm{d}}
\defcitealias{Sperhake:2019wwo}{Phys.~Rev.~D \textbf{101} (2020) 024044}

\begin{document}

\title{Anomalies in the gravitational recoil of eccentric black-hole mergers with unequal mass ratios}

\author{Miren Radia}
\email{m.r.radia@damtp.cam.ac.uk}
\affiliation{Department of Applied Mathematics and Theoretical Physics, 
Centre for Mathematical Sciences, University of Cambridge, Wilberforce Road,
Cambridge CB3 0WA, United Kingdom}

\author{Ulrich Sperhake}
\email{u.sperhake@damtp.cam.ac.uk}
\affiliation{Department of Applied Mathematics and Theoretical Physics, 
Centre for Mathematical Sciences, University of Cambridge, Wilberforce Road,
Cambridge CB3 0WA, United Kingdom}
\affiliation{California Institute of Technology, Pasadena, California 91125, 
USA}

\author{Emanuele Berti}
\email{berti@jhu.edu}
\affiliation{Department of Physics and Astronomy, Johns Hopkins University, 
3400 N. Charles Street, Baltimore, Maryland, 21218, USA}

\author{Robin Croft}
\affiliation{Department of Applied Mathematics and Theoretical Physics, 
Centre for Mathematical Sciences, University of Cambridge, Wilberforce Road,
Cambridge CB3 0WA, United Kingdom}

\date{\today}

\begin{abstract}
  The radiation of linear momentum imparts a recoil (or ``kick'') to
  the center of mass of a merging black-hole binary system. Recent
  numerical relativity calculations have shown that
  eccentricity can lead to an approximate 25\% increase in recoil
  velocities for equal-mass, spinning binaries with spins lying in the
  orbital plane (``superkick'' configurations)~[U.~Sperhake et al.~\citetalias{Sperhake:2019wwo}].
  Here we investigate the impact of nonzero eccentricity on the kick
  magnitude and gravitational-wave emission of nonspinning,
  unequal-mass black hole binaries. We confirm that nonzero
  eccentricities at merger can lead to kicks which are larger by
  up to $\sim 25\,\%$ relative to the quasicircular case.
  We also find that the kick velocity $v$ has an oscillatory
  dependence on eccentricity, which we interpret as a consequence of
  changes in the angle between the infall direction at merger and the
  apoapsis (or periapsis) direction.
\end{abstract}

\maketitle

\section{Introduction}

Gravitational waves (GWs) carry energy, angular momentum and linear
momentum away from the source with potentially observable
consequences. The radiated energy corresponds to an often enormous
mass deficit in the source; for example the first ever detected
black-hole (BH) binary merger, GW150914~\cite{Abbott:2016blz}, radiated
$\Delta M\approx 3\,M_{\odot}$, or about $4.6\,\%$ of the total mass
of the source. A tiny fraction of this energy is deposited into GW
interferometers, thus enabling us to detect and characterize the
signal~\cite{Saulson:2010zz}. The angular momentum radiated in GWs
reduces the rotation rate of possible merger remnants and---at least
in four spacetime dimensions---plays a critical role in avoiding the
formation of naked singularities in the form of BHs spinning above the
Kerr limit; see
e.g.~Refs.~\cite{Campanelli:2006uy,Sperhake:2009jz}. Therefore, GW emission
is a necessary ingredient of the theory of general relativity, in the
sense that it avoids the formation of spacetime singularities and
preserves its predictive power.

In this paper, we focus on the radiated linear momentum, which
imparts a recoil (commonly referred to as a \emph{kick}) on the center
of mass of the emitting
system~\cite{Bonnor1961-wy,Peres:1962zz,Bekenstein:1973zz}.

Whereas GWs inevitably carry energy and angular momentum---provided
their sources do---the radiation of linear momentum requires some
degree of asymmetry, as realized in nonspherical supernova explosions
or unequal-mass ratios and/or spin misalignments in binary BH
mergers. The inspiral of two equal-mass, nonspinning BHs, for example,
radiates energy and angular momentum, whereas the emitted linear
momentum is zero by symmetry. By turning these considerations around,
we may also regard the study of recoiling GW emitters as a guided
search for characteristic (in some loose sense ``asymmetric'')
features in their orbital dynamics which, in turn, might help us to
better understand astrophysical sources through GW observations.
A recoiling postmerger BH, for example, can induce a blue (or
red) shift in parts of its GW signal that may be exploited in future
GW observations to directly measure BH kicks
\cite{Gerosa:2016vip,CalderonBustillo:2018zuq,Lousto:2019lyf}, and
the effect of kicks should be taken into account in future ringdown
tests of general relativity with third-generation GW detectors to
avoid systematic biases~\cite{Varma:2020nbm}.
The asymmetric emission of GWs is not the only mechanism that 
can contribute to recoils; if there is an accretion disk or some other
astrophysical background, this can also impart a kick on the remnant
BH that can be $\mathcal{O}(100)\;\mathrm{km/s}$ \cite{Cardoso:2020lxx}.

For binary BH mergers, early estimates of the recoil speeds of the
remnant BH relied on a variety of approximations, including
post-Newtonian (PN) theory~\cite{Fitchett1983-xq,Blanchet:2005rj}, BH
perturbation theory~\cite{Hughes:2004ck}, the effective-one-body
formalism~\cite{Damour:2006tr}, the close-limit
approximation~\cite{Sopuerta:2006wj,Sopuerta:2006et}, and combinations
thereof~\cite{LeTiec:2009yg}. Not long afterwards, during the
numerical relativity (NR) gold rush, several groups obtained more
accurate results for the kick velocity from the merger of nonspinning
BHs along
quasicircular
orbits~\cite{Baker:2006vn,Gonzalez:2006md,Herrmann:2007cwl}.
These calculations were followed by the discovery that the merger of
spinning BHs can lead to kick velocities of $\sim 3000$~km/s when the
spins lie in the orbital plane and point in opposite directions
(``superkick''
configurations~\cite{Gonzalez:2007hi,Campanelli:2007cga,Campanelli:2007ew}),
and to even larger kicks of order $\sim 5000$~km/s when the spins are
partially aligned with the orbital angular momentum (``hang-up kick''
configurations~\cite{Lousto:2011kp}). 
The probability of such large
recoils occurring in nature depends therefore on spin alignment, and
this has been studied by several authors (see, e.g.,
Refs.~\cite{Schnittman:2007sn,Dotti:2009vz,Kesden:2010ji,Lousto:2012su,Berti:2012zp,Lousto:2012su}).

The possible occurrence of superkicks has important consequences for
astrophysical BHs and their
environments~\cite{Komossa:2012cy,Colpi:2014poa,Blecha:2015baa,Barack:2018yly}.
It is pertinent to compare the recoil velocities obtained from NR
simulations with the escape velocities of various astrophysical
environments \cite{Merritt:2004xa}. For example, stellar-mass BH
binaries are believed to form dynamically in globular clusters
\cite{Benacquista:2011kv}. In this case the escape velocities are
generally $\mathcal{O}(10)\;\mathrm{km/s}$, smaller than the
$\mathcal{O}(100)\;\mathrm{km/s}$ kicks predicted for quasicircular,
nonspinning binaries~\cite{Gonzalez:2006md}. Then relativistic recoils
can affect the proportion of BH merger remnants that are retained by
globular clusters even if the BHs are
nonspinning~\cite{Morawski:2018kfs}. At the other end of the scale,
the recoil velocities of supermassive BHs can be used to constrain
theories of their growth at the center of dark matter halos
\cite{Haiman:2004ve}. Kicked remnants in the accretion disk of an
active galactic nucleus may also lead to detectable electromagnetic
counterparts for stellar-origin BH
mergers~\cite{Graham:2020gwr,Chen:2020gek}.

As mentioned above, a net gravitational recoil requires some asymmetry in the
system, so that the GW emission is anisotropic.
A natural way to
accentuate the asymmetry is through the addition of orbital
eccentricity. Early calculations in the close-limit
approximation~\cite{Sopuerta:2006et} predicted a kick proportional to
$1+e$ for small eccentricities, $e\lesssim 0.1$.  More recently,
numerical relativity calculations led to the conclusion that
eccentricity can lead to an approximate 25\% increase in recoil
velocities for superkick configurations with moderate
eccentricities~\cite{Sperhake:2019wwo}.

The main goal of this study is to investigate the impact of nonzero
eccentricity on the kick magnitude and the corresponding GW emission
of nonspinning, unequal-mass BH binaries.  As we shall see, the
eccentricity has a subtle but significant effect on the kick
magnitude, which manifests itself in corresponding patterns in the GW
signal, especially in subdominant multipoles.

For isolated binary systems with large initial separations, the
emission of GWs acts to circularize the orbit by the time the signal
enters the frequency band of ground-based detectors.  However, viable
dynamical formation channels of stellar-origin BH binaries could
result in a non-negligible population of merging BHs that still retain
moderate eccentricities at frequencies relevant for ground-based GW
detection (see,
e.g.,Refs.~\cite{Samsing:2017rat,Samsing:2017xmd,Samsing:2017oij,Rodriguez:2018pss,Samsing:2020tda,Tagawa:2020jnc}).
Furthermore, the presence of astrophysical media such as
accretion disks may increase the eccentricity during the inspiral
\cite{Cardoso:2020iji}.
Most of the events observed by the LIGO/Virgo Collaboration show no
evidence of significant eccentricities~\cite{Salemi:2019owp} but the
extraordinary GW190521 event~\cite{Abbott:2020tfl} is potentially
consistent with an eccentricity as high as
$e\approx0.7$~\cite{Romero-Shaw:2020thy,Gayathri:2020coq}.

Orbital eccentricity is expected to be a distinguishing feature of
stellar-origin BH binaries that form dynamically, but a nonzero
eccentricity is more likely at the low frequencies accessible by LISA,
where gravitational radiation reaction has less time to circularize
the binary~\cite{Nishizawa:2016jji,Breivik:2016ddj,Nishizawa:2016eza}.
If confirmed, a nonzero eccentricity would hint at a possible
dynamical origin for this event~\cite{Romero-Shaw:2020thy}.

Eccentricity is expected to play an even more prominent role for
massive BH binaries: the dynamics of these binaries in stellar and
gaseous environments is expected to lead to distinct (but generically
nonzero) orbital eccentricities by the time the binaries enter the
LISA sensitivity window (see Ref.~\cite{Roedig:2011rn} and references
therein). Even larger eccentricities are possible if BH binary
coalescence occurs through the interaction with a third
BH~\cite{Bonetti:2018tpf}.

Our work is an exploration of the effect of large eccentricities near
merger, and it differs in several ways from the catalog of eccentric,
unequal-mass simulations presented in Ref.~\cite{Huerta:2019oxn}.
While their study considered a larger range of mass ratios (in our
notation, $1/10\leq q\leq 1$), they carried out fewer simulations for
each value of $q$. The binaries in their simulations have initial
eccentricities smaller than $e_0 = 0.18$ 15 cycles before merger,
and since they start at larger orbital separations,
their eccentricity will have further decreased by the time of merger.
As we will see below, the larger initial eccentricities in
our simulations allow us to highlight interesting periodicities in the
emission of gravitational radiation and the behavior of the recoil
velocity.

The remainder of this paper is organized as follows.
In Sec.~\ref{sec:NR} we discuss our two numerical codes (\textsc{Lean}
and \textsc{GRChombo}), the computational framework, and the catalog
of simulations we produced for this study. In Sec.~\ref{sec:results}
we present the main results of our simulations. In
Sec.~\ref{sec:concl} we summarize these results and point out possible
directions for future work. In Appendix \ref{sec:accuracy} we
detail our tests for numerical accuracy and verify that
our two codes give comparable results.
Finally, in Appendix \ref{sec:tagging} we 
discuss the tagging of cells for adaptive mesh refinement used in one
of our numerical codes (\textsc{GRChombo}).
Throughout this work we use geometrical units ($G=c=1$).

\section{Computational framework and set of simulations}
\label{sec:NR}
\subsection{Numerical methods}

The simulations reported in this work have been performed with the
\textsc{GRChombo}~\cite{Clough:2015sqa, GRChomboWebsite} and 
\textsc{Lean}~\cite{Sperhake:2006cy} codes.
We estimate the error budget of our simulations from both codes to be up to 3.5\,\%. Details of our convergence analyses are provided in 
Appendix~\ref{sec:accuracy}. Though different codes were
used for each sequence of configurations, we undertook comparison tests 
in order to ensure consistent results, and these can also be found in
Appendix~\ref{sec:accuracy}.

\subsubsection{\textsc{GRChombo} setup}
\label{sec:grchombo}
\textsc{GRChombo}~\cite{Clough:2015sqa} is a finite difference
numerical relativity code which uses the method of lines with
fourth-order Runge-Kutta time stepping. In contrast to previous studies
with \textsc{GRChombo} we have implemented sixth-order spatial
stencils in order to improve phase accuracy~\cite{Husa:2007hp}. The
Einstein equations are solved by evolving the covariant and conformal
Z4 (CCZ4) formulation~\cite{Alic:2011gg} with the prescription
described in Sec.~F of~\cite{Alic:2013xsa}, namely the replacement
$\kappa_1 \to \kappa_1/\alpha$, in order to stably evolve BHs
and maintain spatial covariance. After this replacement and in the
notation of Ref.~\cite{Alic:2011gg}, we use the constraint damping
parameters $\kappa_1=0.1$, $\kappa_2=0$ and $\kappa_3=1$ in all
simulations.  However, unlike Refs.~\cite{Clough:2015sqa,Alic:2011gg},
we use the conformal factor defined by
\begin{equation}
    \chi = \det(\gamma_{ij})^{-1/3},\label{eq:con-fac}
\end{equation}
where $\gamma_{ij}$ is the physical spatial metric.
\textsc{GRChombo} is built on the \textsc{Chombo}~\cite{ChomboReport} 
library for solving partial differential equations with 
block-structured adaptive mesh refinement (AMR) which supports
nontrivial mesh hierarchies using Berger-Rigoutsos grid generation
\cite{Berger1991}. The grid comprises a hierarchy of cell-centered 
Cartesian meshes consisting of $L+1$ refinement levels labeled from 
$l=0,\ldots,L$, each with grid spacing $h_l=h_0/2^l$. Given the 
AMR, the grid configuration changes dynamically during the simulation.
The regridding is controlled by the tagging of cells for refinement 
in the Berger-Rigoutsos algorithm~\cite{Berger1991}, with cells being 
tagged if the tagging criterion $C$ exceeds a specified threshold 
value $t_R$. Details of the tagging criterion used in this work are 
provided in Appendix \ref{sec:tagging}. The Berger-Oliger scheme~\cite{Berger1991} 
is used for time stepping on the mesh hierarchy, and we take a 
Courant-Friedrichs-Lewy (CFL) factor of $1/4$ in all simulations. 
Due to the inherent symmetry of the configurations considered, we 
employ bitant symmetry in order to reduce the computational expense.

\subsubsection{\textsc{Lean} setup}
\label{sec:lean}
The \textsc{Lean} code~\cite{Sperhake:2006cy} is based on the 
\textsc{Cactus} computational toolkit~\cite{Goodale2002a} and uses 
the method of lines with fourth-order Runge-Kutta time stepping and 
sixth-order spatial stencils for improved phase accuracy 
\cite{Husa:2007hp}. The Einstein equations are implemented in the form 
of the Baumgarte-Shapiro-Shibata-Nakamura-Oohara-Kojima (BSSNOK) 
formulation~\cite{Nakamura:1987zz,Shibata:1995we,Baumgarte:1998te} 
with the moving-puncture gauge~\cite{Campanelli:2005dd,Baker:2005vv}. 
The \textsc{Carpet} driver~\cite{Schnetter:2003rb} provides AMR using 
the technique of ``moving boxes.'' We use bitant symmetry to exploit 
the symmetry of the simulations and reduce computational expense. 
The computational domain 
comprises a hierarchy of $L+1$ refinement levels labeled from 
$l=0,\ldots l_F,\ldots,L$, each with grid spacing $h_l=h_0/2^l$. 
Before applying the symmetry, for $l\leq l_F$ each level consists of 
a single fixed cubic grid of half-length\footnote{In one departure from
this rule, we enhance $R_2$ by a factor of 4/3 for the simulations
of sequence \texttt{lq1:2} of Table \ref{tab:sequences}.}
$R_l=R_0/2^l$, and for 
$l_F<l\leq L$, each level consists of two cubic components of 
half-length $R_l=2^{L-l}R_L$ centered around each BH. We adopt this 
notation for consistency with that used to describe \textsc{GRChombo}. 
This translates into the more conventional \textsc{Lean} grid setup 
notation (cf. Ref.~\cite{Sperhake:2006cy}) as
\begin{equation}
    \left\{(R_0,\ldots,2^{-l_F}R_0) \times 
    (2^{L-l_F-1}R_L,\ldots,R_L),h_L\right\}.
\end{equation}
A CFL factor of $1/2$ is used in all simulations, and apparent horizons 
are computed with \textsc{AHFinderDirect}
\cite{Thornburg:1995cp,Thornburg:2003sf}.

\subsubsection{Initial data}
For both codes, we use puncture data~\cite{Brandt:1997tf} of
Bowen-York \cite{Bowen:1980yu} type provided by the spectral solver of
Ref.~\cite{Ansorg:2004ds} in the form of the \textsc{Cactus} thorn
\textsc{TwoPunctures} for \textsc{Lean}, and a standalone version
integrated into \textsc{GRChombo}.  In the latter case, we take
advantage of the improvements made in Ref.~\cite{Paschalidis:2013oya}
to use spectral interpolation.

\subsection{Black-hole binary configurations}
\label{sec:configs}
We follow the construction of sequences of BH binary configurations
and the notation of Ref.~\cite{Sperhake:2007gu}. In particular, we 
denote by $M_1$ 
and $M_2$ the initial BH masses. Without loss of generality, since 
we are only considering unequal masses ($M_1 \neq M_2$), we take $M_2 > M_1$
and denote their sum by $M = M_1 + M_2$. The reduced mass is 
$\mu = M_1 M_2 /M$ and to quantify the mass ratio, we use either
\begin{equation}
    q = \frac{M_1}{M_2}
    \label{eq:mass-ratio}
\end{equation}
or the symmetric mass ratio $\eta = \mu/M$. Finally, the 
total Arnowitt-Deser-Misner (ADM) mass~\cite{Arnowitt:1962hi} is denoted by 
$M_{\mathrm{ADM}}$.

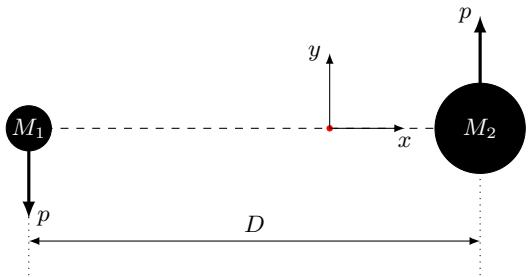
\begin{figure}[t]
    \centering
    \begin{tikzpicture}
        \draw[dashed] (-4,0) -- (2,0);
        \filldraw[color=white, fill=red] (0,0) circle (0.05);
    
        \filldraw[fill = black] (-4,0) circle (0.3);
        \draw[very thick,-latex] (-4,0) -- (-4,-1.2);
        \node[anchor=west] at (-4,-1.2) {$p$};
        \node[color=white] at (-4,0) {$M_1$};
        
        \filldraw[fill = black] (2,0) circle (0.6);
        \draw[very thick,-latex] (2,0) -- (2,1.5);
        \node[anchor=east] at (2,1.5) {$p$};
        \node[color=white] at (2,0) {$M_2$};
    
        \draw[dotted] (2,0) -- (2,-2);
        \draw[dotted] (-4,0) -- (-4,-2);
        \draw[latex-latex] (-4,-1.5) -- (2,-1.5);
        \node[anchor=south] at (-1,-1.5) {$D$};
        
        \draw[-latex] (0,0) -- (0,1);
        \node [anchor=east] at (0,1) {$y$};
        \draw[-latex] (0,0) -- (1,0);
        \node[anchor=north] at (1,0) {$x$};
    \end{tikzpicture}
    \caption{Schematic diagram of the initial BH binary setup for 
    an arbitrary configuration in one of the sequences.}
    \label{fig:bh-setup}
\end{figure}

In order to construct a sequence for a fixed mass ratio, we first 
determine an initial quasicircular configuration. We specify the initial
coordinate separation $D/M$ along the $x$ axis, and the scale in the codes 
is fixed by choosing $M_1=0.5$. Next, Eq.~(65) in 
Ref.~\cite{Brugmann:2008zz} is used to calculate the initial tangential 
momentum of each BH, $\mathbf{p}=(0,\pm p,0)$ (as shown in 
Fig.~\ref{fig:bh-setup}). We use a 
Newton-Raphson method to iteratively solve for the Bowen-York bare mass 
parameters that give the desired BH masses. The binding energy 
of this quasicircular configuration is then computed using
\begin{equation}
    E_{\mathrm{b}}=M_{\mathrm{ADM}}-M.\label{eq:binding-energy}
\end{equation}
The rest of the sequence with increasing orbital eccentricity is 
constructed by fixing the binding energy and gradually reducing the initial
linear momentum parameter $p$.
We decide to reduce the linear momentum rather than, for example, 
altering its direction,
so that the $x$ axis is fixed as the initial apoapsis for all configurations.
For a given configuration with fixed $p$, we 
iteratively solve for the separation $D$ and bare masses that give the 
required binding energy and BH masses. The choice to keep the 
binding energy constant as the momentum parameter (and thus the initial 
kinetic energy) is reduced means that the initial separation increases along 
the sequence. This ensures an inspiral phase of comparable duration as the 
eccentricity increases.
The initial orbital angular momentum of the system 
is given by $L=Dp$~\cite{York:1989jn}. Even though $D$ increases as $p$ 
decreases, the initial angular momentum of the system monotonically decreases 
as $p$ decreases for all but the least one or two eccentric configurations 
in a sequence.

We have parametrized the configurations within a sequence by their
initial tangential momentum $p$, but we would like to measure the
eccentricity of these configurations. Unfortunately, there is no
gauge-invariant measure of eccentricity~\cite{Loutrel:2018ydu} and the
ambiguity in any definition is particularly pronounced in the late
stages of inspiral from which our simulations start. Following
Ref.~\cite{Sperhake:2007gu}, we use the formalism in
Ref.~\cite{Memmesheimer:2004cv} to obtain a PN estimate for the
eccentricity. Note that this formalism has three eccentricity
parameters---$e_t$, $e_r$ and $e_\phi$---and employs two
different types of coordinates: ADM-like and harmonic. The choice of
which parameter and coordinate type to use is somewhat arbitrary.
We mostly focus on the eccentricity parameter $e_t$ in harmonic coordinates\footnote{
The ADM-like estimate of Ref.~\cite{Memmesheimer:2004cv} differs
by only a few percent for $e_t \lesssim 0.8$, and
would not significantly alter our results.} 
as in Ref.~\cite{Sperhake:2019wwo}. This estimate should be 
taken with a pinch of salt due to the relatively small initial binary 
separations $D$ in our simulations. Furthermore, $e_t$ has an infinite 
gradient as a function of the initial orbital angular momentum
in the quasicircular limit (see Fig.~1 in Ref.~\cite{Sperhake:2007gu}),
such that values of $e_t\lesssim0.1$ are difficult to realize in practice,
unless the BHs start from large initial distance.
In the head-on limit $e_t$ diverges, and a Keplerian/Newtonian
interpretation ceases to be valid. Despite these shortcomings, this 
estimate provides us with a helpful approximation of the eccentricity
and a criterion to quantify deviations away from quasicircularity.

The sequences considered in this work are given in
Table~\ref{tab:sequences}.  Note that there are two sequences
corresponding to the mass ratio $q=1/2$.  The sequence \texttt{lq1:2}
has a longer inspiral phase compared to the other sequences. For the
nearly quasicircular configurations, the binary completes about six
orbits before merger in the \texttt{lq1:2} sequence, and about three
orbits in all other sequences. The longer sequence of simulations was
conducted in order to identify any possible artifacts in the shorter
sequences due to the exclusion of the earlier inspiral phase. In
addition to the labeling of sequences in Table~\ref{tab:sequences},
we refer to individual simulations within a sequence by appending
``\texttt{-p}'' to the sequence label followed by a four digit integer
which is given by $10^3p/M$ truncated appropriately; for example,
\texttt{sq1:2-p0100} denotes the simulation in sequence
\texttt{sq1:2} with initial tangential momentum $p = 0.1M$.

\begin{table}[t]
    {
    \caption{Sequences of binary BH configurations studied in this 
    work with their mass ratio, binding energy $E_{\mathrm{b}}/M$,
    and the GW extraction radius $r_{\rm ex}$.
    For reference, we also
    list for each sequence the kick velocities $v_c$ in the
    quasicircular limit. These values agree, within the numerical
    uncertainties, with the results of
    Ref.~\cite{Gonzalez:2006md}.}
    \centering
    \begin{ruledtabular}
    \begin{tabular}{cccccc}
        Sequence & Code & $q$ & $E_{\mathrm{b}}/M$ & $r_{\mathrm{ex}}/M$ 
        & $v_c$ (km/s)\\
        \hline
        \texttt{sq2:3} & \textsc{GRChombo} & $2/3$ & $-0.0113386$ & $88$ & 102 \\
        \texttt{sq1:2} & \textsc{Lean} & $1/2$  & $-0.0106964$ & $80$ & 149\\
        \texttt{lq1:2} & \textsc{Lean} & $1/2$ & $-0.0090858$ & 80 & 150 \\
        \texttt{sq1:3} & \textsc{GRChombo} & $1/3$ & $-0.0093684$ & $65$ & 178 \\
    \end{tabular}
    \end{ruledtabular}
    }
    \captionsetup{justification=justified}
    \label{tab:sequences}
\end{table}

\subsection{Diagnostics}
For all simulations, we have extracted values of the Weyl scalar $\Psi_4$ 
on spheres of finite coordinate radius given in Table \ref{tab:sequences} 
for each sequence. We also computed the dominant terms in the multipolar
decomposition,
\begin{equation}\label{eq:psi4-multipoles}
    \Psi_4(t,r,\theta,\phi) = \sum_{\ell=2}^\infty\sum_{m=-\ell}^{\ell} 
    \psi_{\ell, m}(t,r)\left[{}_{-2}Y^{\ell,m}(\theta,\phi)\right],
\end{equation}
where ${}_{-2}Y^{\ell,m}$ are the usual spin-weight $-2$ spherical 
harmonics.

Our main diagnostics are the energy, linear momentum and angular momentum 
radiated in GWs, which are computed directly from the 
extracted $\Psi_4$ values on the spheres using standard methods. For
completeness, we reproduce the formulae here.

The radiated energy $E^{\rad}$ is given by
\cite{Campanelli:1998jv,Lousto:2007mh}
\begin{equation}
    E^{\rad}(t) = \lim_{r\to\infty}\frac{r^2}{16\pi}\int_{t_0}^t\rmd t^\prime
    \oint_{S^2_r}\rmd\Omega\,\left|\int_{-\infty}^{t^\prime}\rmd t^{\prime\prime}\,
    \Psi_4\right|^2.\label{eq:Erad}
\end{equation}
The radiated linear momentum $\mathbf{P}^{\rad}$ is given by
\begin{equation}
    \mathbf{P}^{\rad}(t) = \lim_{r\to\infty}\frac{r^2}{16\pi}
    \int_{t_0}^t\rmd t^\prime\oint_{S^2_r}\rmd\Omega\,
    \hat{\mathbf{e}}_r\left|\int_{-\infty}^{t^\prime}\rmd t^{\prime\prime}\,
    \Psi_4\right|^2,\label{eq:Prad}
\end{equation}
where $\hat{\mathbf{e}}_r$ is the flat-space unit radial vector
\begin{equation}
    \hat{\mathbf{e}}_r = (\sin\theta\cos\phi,\sin\theta\sin\phi,\cos\theta).
\end{equation}
Finally, the radiated angular momentum $\mathbf{J}^{\rad}$ is given by
\begin{multline}
    \mathbf{J}^{\rad}(t) = - 
    \lim_{r\to\infty}\frac{r^2}{16\pi}\mathrm{Re}\int_{t_0}^{t}\rmd t^\prime
    \left\{\oint_{S^2_r}\left(\int_{-\infty}^{t^\prime}\rmd t^{\prime\prime}\,
    \bar{\Psi}_4\right)\right.\\
    \left. \times\hat{\mathbf{J}}\left(\int_{-\infty}^{t^\prime}
    \rmd t^{\prime\prime}\int_{-\infty}^{t^{\prime\prime}}
    \rmd t^{\prime\prime\prime}\,\Psi_4\right)\,\rmd\Omega\right\},
    \label{eq:Jrad}
\end{multline}
where the angular momentum operator $\hat{\mathbf{J}}$ for spin weight 
$s=-2$ is given by
\begin{equation}
    \hat{\mathbf{J}}=\left(\mathrm{Re}\,\hat{\mathbf{J}}_+,\mathrm{Im}\,
    \hat{\mathbf{J}}_+,\frac{\partial}{\partial\phi}\right)\,,
\end{equation}
and
\begin{equation}
    \hat{\mathbf{J}}_+=\mathrm{e}^{\mathrm{i}\phi}\left(\mathrm{i}
    \frac{\partial}{\partial\theta} - \cot\theta\frac{\partial}{\partial\phi} 
    + 2\mathrm{i}\csc\theta\right).
\end{equation}

Additionally, we compute the radiated linear momentum from the multipolar
amplitudes $\psi_{\ell, m}$ in Eq.~\eqref{eq:psi4-multipoles} using the 
formulae of Ref.~\cite{Ruiz:2007yx}. From the symmetry of our 
configurations, the $z$ component vanishes identically: $P_z^{\rad}=0$.
For the components in the orbital plane, we write
$P_+^{\rad}=P_x^{\rad}+\mathrm{i}P_y^{\rad}$. Then,
\begin{equation}
     P_+^{\rad}(t) = \sum_{\tl=2}^{\infty}
     \sum_{\tm=-\tl}^{\tl}P_+^{\tl,\tm},
     \label{eq:P+rad}
\end{equation}
where
\begin{multline}
    P_+^{\tl,\tm}(t) = \lim_{r\rightarrow \infty}
    \frac{r^2}{8\pi}\int_{t_0}^t\rmd t^{\prime}
    \left\{\left(\int^{t^\prime}_{-\infty}\rmd t^{\prime\prime}\, 
    \psi_{\tl,\tm}\right)\right. \\
    \times\left(\int_{-\infty}^{t^\prime} 
    \left[a_{\tl,\tm}\bar{\psi}_{\tl,\tm+1}+ 
    b_{\tl,-\tm}\bar{\psi}_{\tl-1,\tm+1}\right. \right.\\
    \left.\left.\vphantom{\int_{-\infty}^t}\left.-
    b_{\tl+1,\tm+1}\bar{\psi}_{\tl+1,\tm+1}\right]
    \rmd t^{\prime\prime}\right)\right\},
    \label{eq:Plmrad}
\end{multline}
and the coefficients $a_{\ell,m}$ and $b_{\ell,m}$ are given by
\begin{align}
    a_{\ell, m} &= \frac{\sqrt{\left(\ell - m\right)\left(\ell +
    m+1\right)}}{\ell\left(\ell + 1\right)},\\
    b_{\ell, m} &= \frac{1}{2\ell}\sqrt{\frac{\left(\ell - 2\right) 
    \left(\ell + 2\right) \left(\ell + m\right) \left(\ell + m - 
    1\right)}{\left(2\ell - 1\right)\left(\ell + 1\right)}}.
\end{align}
We will find it helpful to define the partial sums,
\begin{align}
    P_+^{\tl} &= \sum_{\tm = -\tl}^{\tl} P_+^{\tl,\tm},\\
    P_+^{\leq\tl} &= \sum_{\tl^\prime=2}^{\tl} P_+^{\tl^\prime}. 
    \label{eq:partialkick}
\end{align}
In practice, we do not evaluate the limit in Eqs.~\eqref{eq:Erad}, 
\eqref{eq:Prad}, \eqref{eq:Jrad} and \eqref{eq:Plmrad}, but rather just 
evaluate them at the finite extraction radius $r=r_{\mathrm{ex}}$, as given 
in Table \ref{tab:sequences}. A discussion of the error this introduces is 
given in the following section.

In order to exclude the spurious radiation inherent in Bowen-York initial 
data, we start the integration in Eqs.~\eqref{eq:Erad}, \eqref{eq:Prad}, 
\eqref{eq:Jrad}, and \eqref{eq:Plmrad} at $t_0 = 50M + r_{\mathrm{ex}}$. The 
recoil velocity is computed from the radiated momentum according to 
\begin{equation}
    \mathbf{v}=-\frac{\mathbf{P}^{\rad}}{M_{\mathrm{fin}}},
\end{equation}
where $M_{\mathrm{fin}}$ is the mass of the BH merger remnant. The 
quantity $M_{\mathrm{fin}}$ can be computed using energy balance:
\begin{equation}
    M_{\mathrm{fin}} = M_{\mathrm{ADM}} - \tilde{E}^{\rad}\,,
\end{equation}
where $\tilde{E}^{\rm rad}$ denotes the radiated energy
{\it including} the spurious radiation.
We similarly compute the spin of the final BH $\chi_{\mathrm{fin}}$ (which,
by symmetry, must be in the $z$ direction) using the radiated angular 
momentum:
\begin{equation}
    \chi_{\mathrm{fin}} = \frac{L-J^{\rad}_z}{M_{\mathrm{fin}}^2},
\end{equation}
where the initial angular momentum is $L=pD$. For \textsc{Lean} simulations, 
we have compared $M_{\rm fin}$ and $\chi_{\rm fin}$ with the
corresponding values derived from the apparent horizon properties,
and find agreement to within $\leq 0.1\,\%$.

\section{Results}
\label{sec:results}
\begin{figure*}[t]
    \subfloat
    {
        \includegraphics[width=0.48\linewidth]{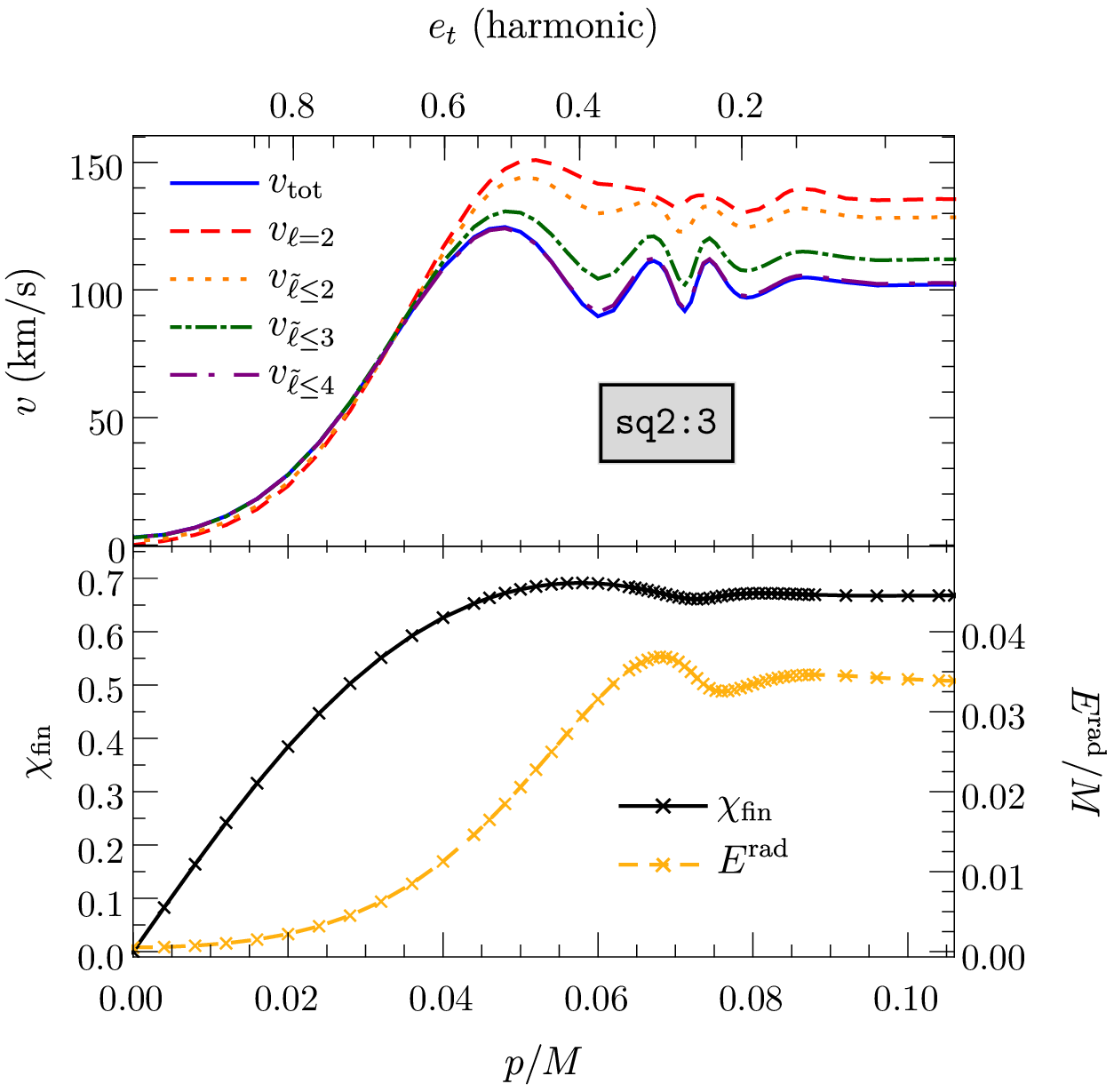}
    }
    \hfill
    \subfloat
    {
        \includegraphics[width=0.48\linewidth]{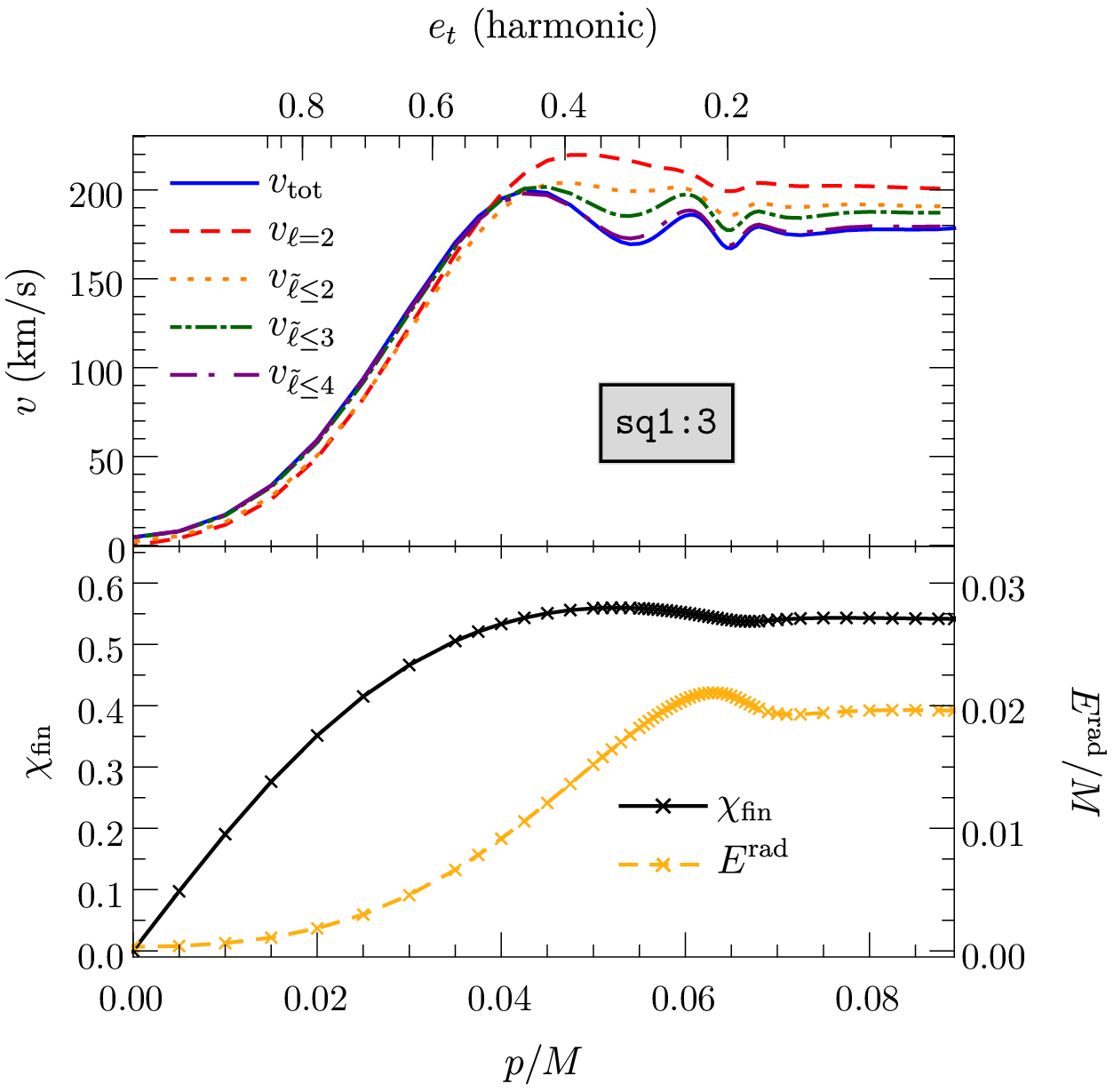}
    }
    \hspace{0 mm}
    \subfloat
    {
        \includegraphics[width=0.48\linewidth]{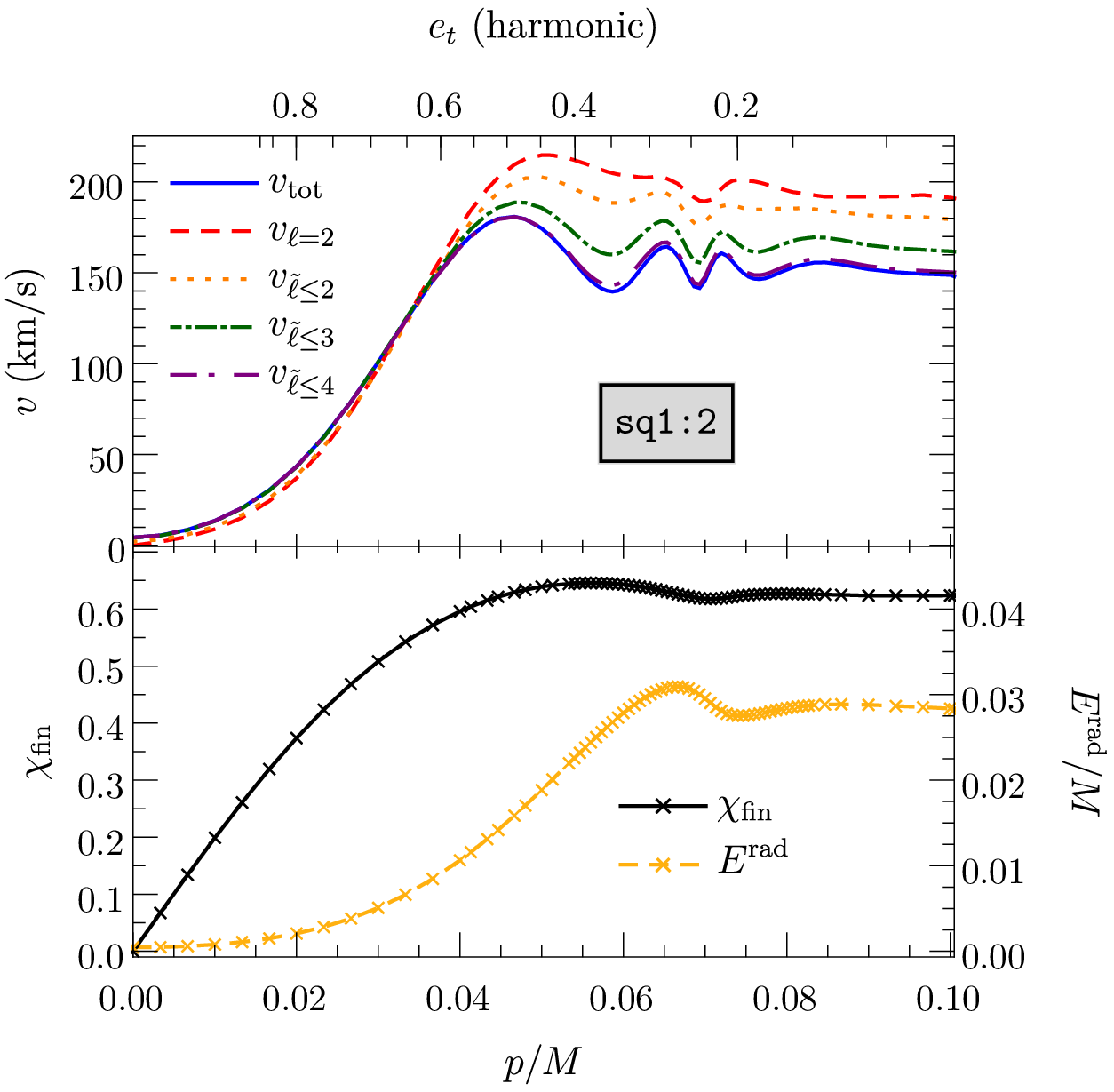}
    }
    \hfill
    \subfloat
    {
        \includegraphics[width=0.48\linewidth]{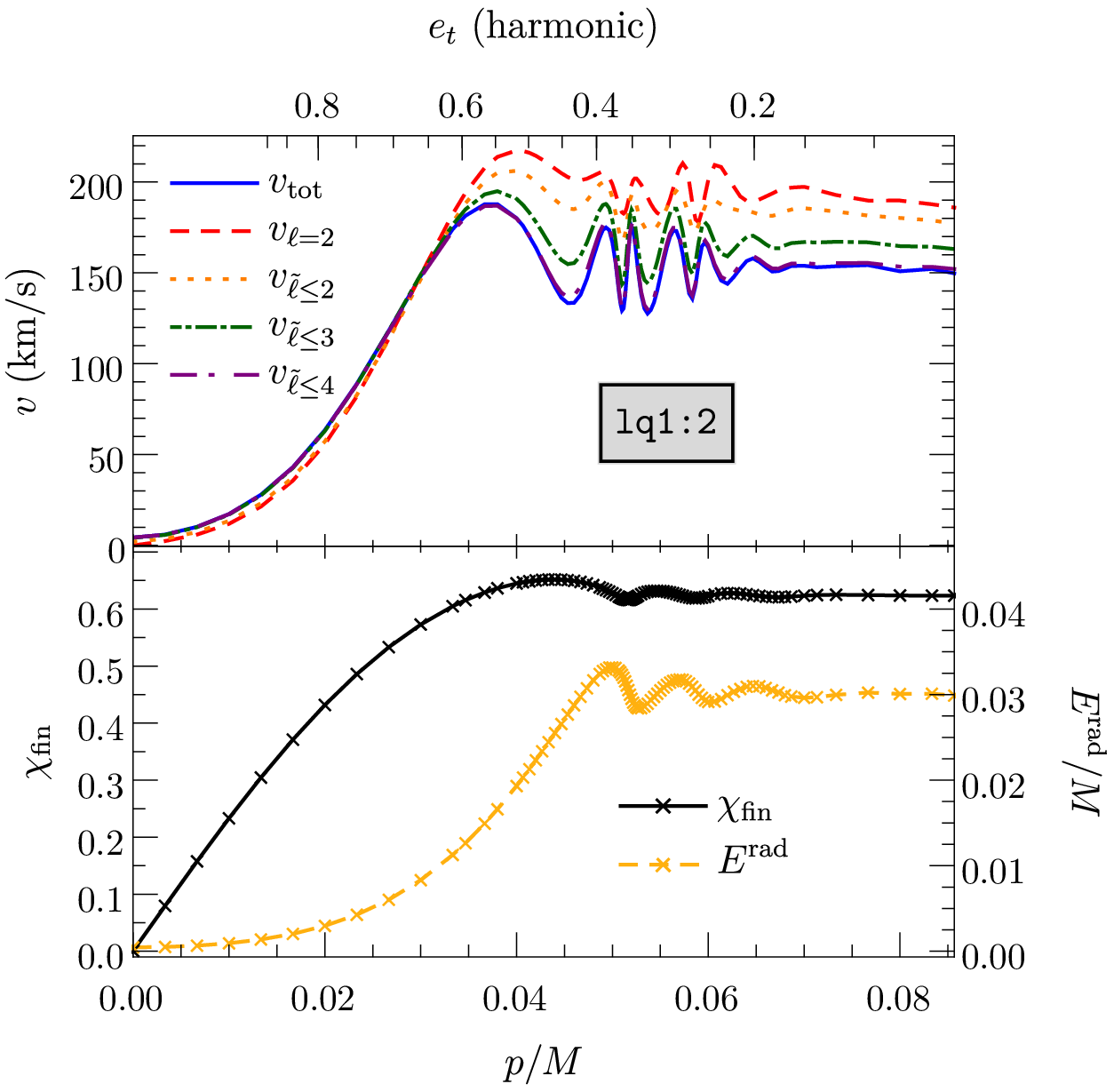}
    }
    \caption{For each sequence of simulations in Table \ref{tab:sequences}:
    Top panel: the recoil velocity $v$ is plotted as a function 
    of the initial tangential momentum $p/M$. The individual curves 
    represent the total kick $v_\mathrm{tot}$ (blue, solid), the 
    contribution to the kick from $\ell=2$ modes of $\Psi_4$, $\psi_{2,m}$, 
    only in Eqs.~(\ref{eq:P+rad})--(\ref{eq:Plmrad}) $v_{\ell=2}$ 
    (red, dashed), and the contributions to the kick from 
    $P_+^{\leq \tilde{\ell}^\prime}$ defined in Eq.~\eqref{eq:partialkick}
    $v_{\tilde{\ell}\leq\tilde{\ell}^\prime} $ for $\tilde{\ell}^\prime=2$ 
    (orange, dotted), $\tilde{\ell}^\prime=3$ (green, dot-dashed) 
    and $\tilde{\ell}^\prime=4$ (purple, long dot-dashed). 
    Our estimate of the eccentricity (see Sec.~\ref{sec:configs}) is 
    provided on the upper horizontal axis.
    Bottom panel: The final BH spin $\chi_{\mathrm{fin}}$ 
    (black, solid) and the energy radiated in GWs $E^\rad$ (gold, dashed) 
    are also plotted as functions of $p/M$. For both curves, the 
    individual simulations performed for this analysis are shown by 
    $\times$ symbols.}
    \label{fig:kicks2}
\end{figure*}
\begin{figure}[t]
    \centering
    \includegraphics[width=\columnwidth]{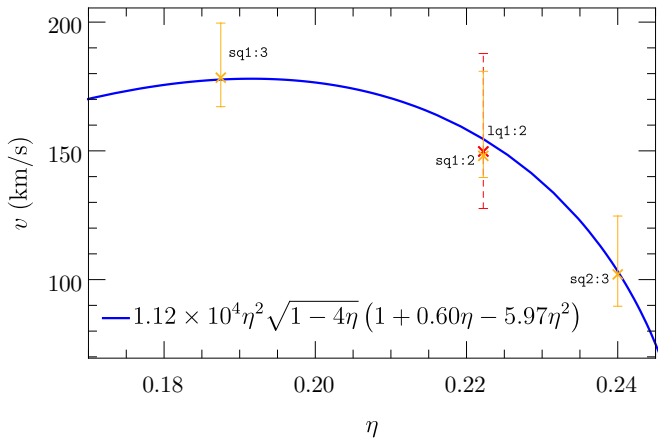}
    \caption{The range of recoil velocities obtained for each sequence
    is plotted against the symmetric mass ratio $\eta$. Note that for
    each sequence we exclude the configurations with 
    $p<p_{\text{max}}$ (i.e. the head-on limit), where 
    $p=p_{\text{max}}$ is the tangential momentum that maximizes the kick.
    The three short sequences are marked in gold and the long sequence is 
    marked in red (dashed). A fitted formula for the quasicircular
    kick as a function of $\eta$ from Ref.~\cite{Healy:2017mvh} is
    also shown in blue for comparison. 
    }
    \label{fig:quasicircular-fit-comparison}
\end{figure}

Using the framework summarized in the previous section,
we have simulated four sequences of nonspinning BH binaries,
characterized by their mass ratio (\ref{eq:mass-ratio})
and binding energy (\ref{eq:binding-energy}).
The parameters of these sequences are listed in Table~\ref{tab:sequences}.
We have selected our mass ratios such that they cover the
regime of maximum recoil, realized for $\eta=0.195$ or
$q=1/2.77$ (cf. Fig.~\ref{fig:quasicircular-fit-comparison}).
Recall that sequences \texttt{sq2:3}, \texttt{sq1:2} and \texttt{sq1:3}
complete about three orbits and sequence \texttt{lq1:2} completes
about six orbits, respectively, in the quasicircular limit.

Our main results are displayed in Fig.~\ref{fig:kicks2}, where we plot
for all sequences the total recoil speed $v_{\rm tot}$, various
truncations of the multipolar contributions to the total recoil
according to Eqs.~(\ref{eq:P+rad})--(\ref{eq:partialkick}), the
total radiated GW energy $E^{\rm rad}$ and the dimensionless spin
$\chi_{\rm fin}$ of the BH resulting from the merger.

Let us first focus on the total recoil $v_{\rm tot}$, displayed in each
of the figure's top panels as the blue solid line.  For each mass
ratio, the global maximum of the kick velocity is realized for
moderate eccentricities
$e_t\approx 0.5$.
We also illustrate this kick variation in
Fig.~\ref{fig:quasicircular-fit-comparison}, where
the solid blue curve shows the
quasicircular kick as a function
of the symmetric mass ratio $\eta$ according to
Fit 3 in Table V of Ref.~\cite{Healy:2017mvh}. The velocity ranges
obtained for our eccentric binaries are
overlayed as the vertical bars for each of our sequences.
The bar for each constant-$\eta$ sequence
is obtained by starting at the quasicircular limit on the 
right of each panel in Fig.~\ref{fig:kicks2} and identifying
the minimum and maximum of $v(p)$, excluding the 
plunge regime to the left of the global maximum.

For our sequences \texttt{sq2:3}, \texttt{sq1:2} and
\texttt{lq1:2}, the 
magnification of the kick through moderate values
of the orbital eccentricity is similar to the
enhancement by up to 25\,\% reported in
Ref.~\cite{Sperhake:2019wwo}
for the so-called superkick configurations
\cite{Gonzalez:2007hi,Campanelli:2007cga}.
For \texttt{sq1:3} the effect is milder, with a
$\sim 12\,\%$ amplification, but still well above
the uncertainty estimates of our simulations.
On the other hand, as evidenced by the oscillatory pattern
of the function $v(p)$ in Fig.~\ref{fig:kicks2}, appropriate
nonzero values of the eccentricity can also
lead to a {\it reduction} of the maximum kick at a given mass ratio
by $\sim 10\,\%$.
This overall modification of the gravitational recoil
in the merger of eccentric, nonspinning BH binaries
is the first main result of our study.

Besides the global maximum, we also note a number of local minima and
maxima in the kick velocity as we vary the eccentricity in
Fig.~\ref{fig:kicks2}. For all mass ratios $(q=2/3,\,1/2,\,1/3)$ we
see about five 
local extrema in $v(p)$ in our three short sequences,
corresponding to the two upper panels and the bottom-left panel.
We notice a similar, albeit less pronounced,
oscillatory pattern in the functions
$E_{\rm rad}(p)$ and $\chi_{\rm fin}(p)$ for the radiated
energy and final spin in the lower subpanels in
Fig.~\ref{fig:kicks2}. Our results display no systematic
correlation, however, between the extrema of the respective
quantities; neither global nor local extrema in $v$,
$E_{\rm rad}$ or $\chi_{\rm fin}$ coincide in
magnitude or their eccentricity values.
We believe this diversity is due to the qualitatively
different dependence of the radiated quantities on the
GW multipoles: overlaps of {\it different} multipoles for the
kick, a sum of terms $\propto \psi_{lm}^2$ for the energy,
and the interaction of first and second time integrals
for the angular momentum in Eq.~(\ref{eq:Jrad}).

We added
to our study the $q = 1/2$ sequence of longer BH binary inspirals to
investigate whether these anomalies in $v=v(p)$ might merely result
from ignoring in our simulations the earlier inspiral phase. The
remarkable outcome of this test, however, is that the oscillatory
behavior in the kick as a function of eccentricity is \emph{more}
pronounced in the long sequence.  The solid blue curve in the
bottom-right panel of Fig.~\ref{fig:kicks2} displays significantly more
rapid oscillations in the eccentricity regime
$0.2\lesssim e_t \lesssim 0.4$ as compared to the shorter inspiral
sequences.  This oscillatory behavior, and the apparent increase in
the number of oscillations as we increase the initial separation of
the BHs, is the second of our results.

We next attempt to gain insight into the origin of this
behavior. For this purpose, we have computed the
multipolar contributions to the total kick according to
Eqs.~(\ref{eq:Plmrad})--(\ref{eq:partialkick}). The resulting
velocities are displayed in Fig.~\ref{fig:kicks2} by the additional
dashed, dotted and dash-dotted curves. 
Here, the curves labeled
$v_{\ell=2}$ have been computed from the $\ell=2$ modes of $\Psi_4$
($\psi_{2,m}$ only) in Eqs.~(\ref{eq:P+rad})-(\ref{eq:Plmrad}).
We computed this additional contribution (red dashed curves in the
figure) to determine whether the oscillatory behavior is also present
in the pure quadrupole signal. The answer is yes: the oscillations are
clearly perceptible in $v_{\ell=2}$, even though they are a bit milder
than in the total kick $v_{\mathrm{tot}}$. Considering all
(cumulative) multipolar contributions shown in Fig.~\ref{fig:kicks2},
we notice the following behavior:
\begin{enumerate}[label=(\arabic*)]
 \item The oscillatory dependence of the kick on eccentricity
       is present at any level of truncating the multipolar
       contributions in the cumulative sum (\ref{eq:partialkick}).
 \item The partial sum of the kick up to $\tilde{\ell}=4$
       barely differs from the total kick, indicating that
       higher-order overlap terms do not significantly
       contribute to the kick.
 \item The higher-order contributions $\tilde{\ell}>2$
       to the cumulative kick (\ref{eq:partialkick})
       systematically decrease the kick, counteracting
       the pure quadrupole contribution $v_{\ell=2}$.
\end{enumerate}
In short, we have not identified any specific multipoles
dominating the variation in the kick function $v=v_{\mathrm{tot}}(p)$.

In our search for an explanation, we turn next to the infall direction
of the BH binary just before merger.
\begin{figure*}[t]
    \subfloat
    {
    \includegraphics[width=\columnwidth]{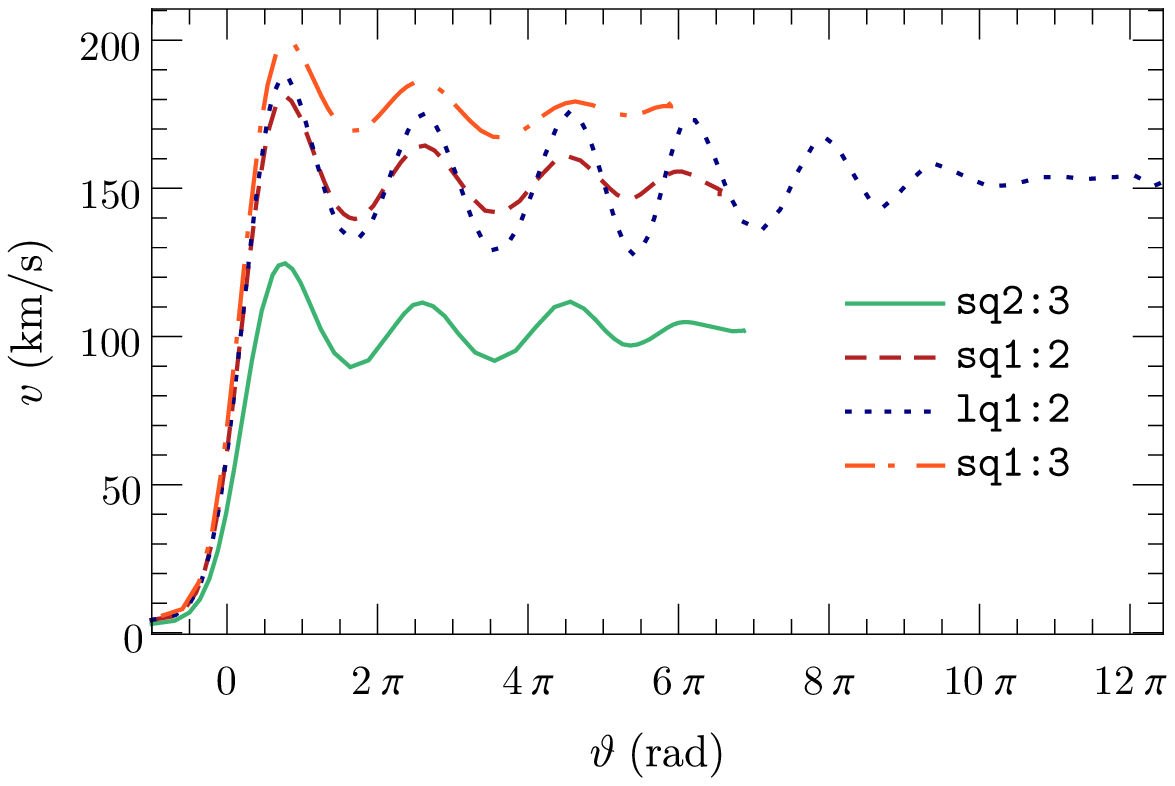}
    \label{fig:kick-theta}
    }
    \hfill
    \subfloat
    {
    \includegraphics[width=\columnwidth]{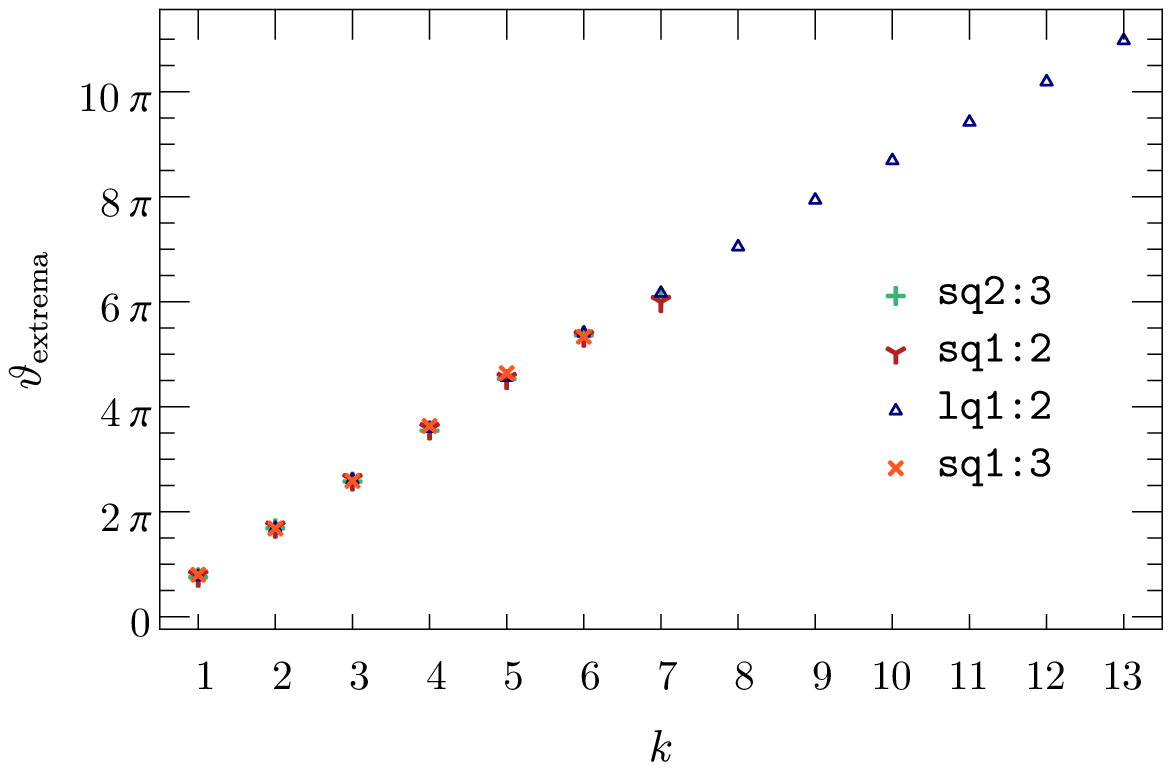}
    \label{fig:theta-extrema}
    }
    \caption{
    Plots involving the angle of the kick $\vartheta$ for 
    all sequences. In the left panel we plot the BH recoil 
    velocity $v$ against $\vartheta$. In the right panel we plot the location of the local extrema $\vartheta_\text{extrema}$ 
    of the left panel against the index of the extrema $k$ counting rightwards from
    the global maximum on the left.
    }
    \label{fig:theta-plots}
\end{figure*}
A well-known feature of the superkicks generated in the inspiral
of BHs with opposite spins $\boldsymbol{S}_1=-\boldsymbol{S}_2$
pointing in the orbital
plane is the sinusoidal variation with the initial azimuthal angle
of the spin vectors; cf.~Fig.~4 in Ref.~\cite{Bruegmann:2007zj}.
The initial orientation of the spins can, alternatively, be
interpreted as a measure for the angle between the in-plane
spin components and the BH binary's infall direction at merger
\cite{Lousto:2009ka}. The superkick is therefore commonly
determined by simulating otherwise identical BH binary configurations
for different values of this angle and fitting the resulting data
with a cosine function; see, e.g.,~Sec.~III A in
Ref.~\cite{Sperhake:2019wwo}. For the eccentric, nonspinning BH
binaries considered in this work, it is the
initial apsis (either a periapsis or an apoapsis)
that defines a reference direction. Unfortunately,
neither the apsis nor a ``binary infall direction''
are rigorously defined
quantities in the strong-field regime of general relativity,
and we consider instead the orientation of the final
kick relative to the $x$ axis, defined by
\begin{equation}
  \tilde{\vartheta} = \mathrm{arg}(v_x + \mathrm{i}v_y)\,.
\end{equation}
For convenience, we define
\begin{equation}
    \vartheta = \tilde{\vartheta} + 2n\pi,
\end{equation}
where $n\geq 0$ is chosen minimally for each configuration in order to obtain
$\vartheta$ as a monotonic function of the initial tangential momentum $p$
for each sequence. We will interchangeably refer to $\vartheta$ 
and $\tilde{\vartheta}$ as the angle of the kick.
Since all of our simulations start with the BHs located on the $x$
axis with purely tangential initial momentum $\mathbf{p}=(0,\pm p,0)$
(Fig.~\ref{fig:bh-setup}), the $x$ direction can be regarded as the
initial direction of the apoapsis. If we furthermore interpret the
gravitational recoil to be predominantly generated by the excess
beaming of the GWs in the direction of the smaller and faster BH (see
Fig.~3 in Ref.~\cite{Wiseman:1992dv}) during the short merger phase,
the kick direction can serve as an approximate measure for the infall
direction of the binary. 

We can test this prediction by computing the kick magnitude
as a function of the angle $\vartheta$; if correct, we would
expect a periodic variation with a period close to $2\pi$.
We do not expect an exact $2\pi$ periodicity because the
relevant periapsis (or apoapsis) direction should be the last one before merger,
and will shift away from the $x$ axis during the inspiral
due to 
apsidal precession---the BH analog
of Mercury's perihelion precession around the Sun. More
specifically, we would expect deviations from a $2\pi$ periodicity
to be more pronounced for longer inspirals, i.e.,~lower eccentricity
and/or larger initial separations, but only mildly
dependent on the mass ratio $q$.
Quite remarkably, all of these
features are borne out by the functions $v = v(\vartheta)$
displayed for our four sequences in the left panel of 
Fig.~\ref{fig:theta-plots}
and the location of the extrema in this plot shown in 
the right panel of Fig.~\ref{fig:theta-plots}.
For all sequences we observe the same approximate $2\pi$ periodicity,
with deviations from this value increasing at larger
$\vartheta$, i.e.~for longer inspirals.
Note also that $\vartheta=-\pi$ in the head-on limit, as expected
for our initial configurations, that start with the heavier BH
located on the positive $x$ axis.

While short of a rigorous proof, this result provides considerable
evidence in favor of interpreting the oscillatory dependence of the
kick on the eccentricity as a consequence of the corresponding
variation in the infall direction as measured relative to the last
apoapsis (or periapsis) of the eccentric binary. This interpretation
also explains why the longer sequence \texttt{lq1:2} exhibits more
oscillations than the shorter sequences \texttt{sq1:3}, \texttt{sq1:2}
and \texttt{sq2:3}. Let us consider for this purpose two binary
configurations that only differ by a tiny amount of eccentricity
$\delta e$. The longer the inspiral phase, the more time these two
binaries have to build up a considerable phase difference and, hence,
a different kick and merger GW signal. Note the potentially dramatic
consequences of this behavior for the GW emission from eccentric
binaries over astrophysical time scales. For long astrophysical
inspirals retaining some eccentricity near merger,
the kick and GW merger signal should exhibit critical dependence on
the eccentricity. In terms of our Fig.~\ref{fig:kicks2}, the function
$v=v(e_t)$ would display a huge number of oscillations rather than the
handful observed in our case, and the resulting curve would look like
a ``band'' rather than a single line. Within the band, a very small
change $\delta e_t$ in eccentricity can produce a finite change in the
kick and merger waveform.

\begin{figure}[t]
    \centering
    \includegraphics[width=\columnwidth,clip=true]{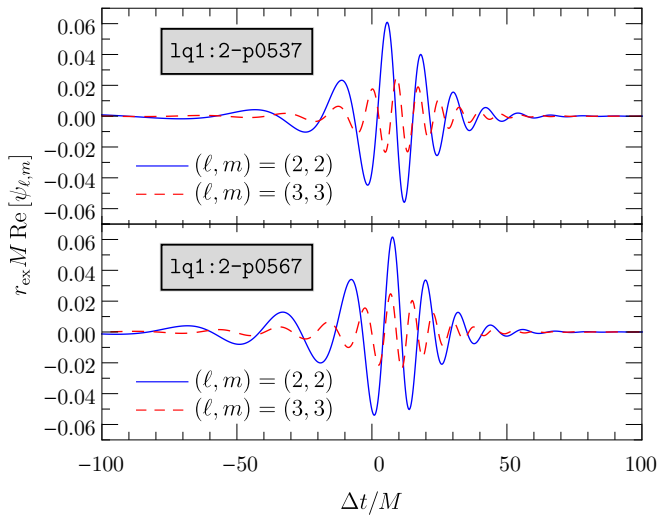}    
    \caption{The real parts of the $(\ell,m)=(2,2)$ and $(3,3)$ modes of $\Psi_4$
             are shown as functions of time for the two binaries
             of sequence \texttt{lq1:2} with $p/M=0.537$ and $p/M=0.567$,
             resulting in kick velocities of $v = 128$
             and $173\,{\rm km/s}$, respectively.
            }
             \label{fig:psi4modes}
\end{figure}

As indicated by our analysis of the multipolar contributions
to the total recoil, the variations in the GW signal are of a
complex nature. We defer a more comprehensive analysis of
the GW pattern to future work, but merely illustrate with an 
example the type of variations that are encountered.
For this purpose, we show in Fig.~\ref{fig:psi4modes}
the $(\ell,m)=(2,2)$ and $(3,3)$ multipoles of the
GW signal around merger for the configurations \texttt{lq1:2-p0537} 
and \texttt{lq1:2-p0567}, corresponding to a local minimum and 
maximum in the
kick, respectively; cf.~the bottom-right panel of Fig.~\ref{fig:kicks2}. In Fig.~\ref{fig:psi4modes},
the time has been shifted such that $\Delta t =0$ corresponds
to the first occurrence of a common apparent horizon. The main
difference perceptible in the figure is the relative phase shift of
the (3,3) mode relative to the dominant quadrupole (2,2). For the
case $p=0.567M$ with maximal kick, the global peaks of both multipoles
are aligned, whereas for $p=0.537M$ with minimal kick, the global peak
of the $(2,2)$ mode coincides with a minimum in $(\ell,m)=(3,3)$. We
have made similar observations for other pairs of modes such as
$(2,2)$ and $(2,1)$, and find these pairs to dominate the oscillatory
variation in the multipolar series expansion (\ref{eq:partialkick}).
%

\section{Conclusions}
\label{sec:concl}
In this paper we have studied the gravitational recoil and GW emission
of sequences of nonspinning BH binaries with mass ratios $q=2/3$,
$1/2$ and $1/3$,
and eccentricity varying from the quasicircular to the head-on limit.
For this purpose we have evolved $274$ configurations
with the {\sc GRChombo} and {\sc Lean} codes. Both codes yield
convergent results for the recoil with a total error budget of
$3$-$4\,\%$ and exhibit excellent agreement, well within this uncertainty
estimate, for a verification configuration simulated with both codes.
In order to estimate the impact of variations in the overall length
of the inspirals, we have evolved two sequences for the case $q=1/2$ which
complete about three and six orbits, respectively, in the quasicircular limit.

The findings of our study are summarized as follows.
\begin{enumerate}[label=(\roman*)]
 \item For all sequences, the total recoil reaches
 a global maximum for moderate eccentricities $e \sim 0.5$.
 As in the case of the enhancement of superkicks studied
 in Ref.~\cite{Sperhake:2019wwo}, the maximum kick is
 enhanced by up to about $25\,\%$ relative to the value obtained
 for quasicircular configurations.
\item Besides this global maximum, we observe an oscillatory
  dependence of the kick $v$ as a function of eccentricity, with
  several local minima and maxima in the function $v = v(e)$. Appropriate
  nonzero values of the eccentricity can lead to a
  {\it reduction} of the kick by $\sim 10\,\%$ relative to the quasicircular
  value instead of an increase. By
  splitting the kick into separate multipolar contributions, we notice
  that this oscillatory dependence is already present, albeit in a
  slightly weaker form, when we consider only quadrupole terms in the
  series expansion \eqref{eq:P+rad}.  Further contributions involving
  $\ell \ge 2$ multipoles tend to decrease the overall kick and mildly
  enhance the oscillatory variation; see Fig.~\ref{fig:kicks2}.
 \item We interpret this oscillatory variation in the kick
 as a consequence of changes in the angle between the infall
 direction at merger and the apoapsis (or periapsis)
 direction. In the absence of rigorous definitions for either
 of these directions, we approximate this angular variation
 by considering the direction of the final kick and the
 $x$ axis, assuming that the former is related via relativistic
 GW beaming to the infall direction and by taking into account
 that our BHs start on the $x$ axis with zero radial momentum.
 Displayed as a function of this angle, the kick displays
 the expected periodic behavior with a period close to but
 mildly deviating from $2\pi$, presumably due to periapsis
 precession.
 \item We have explored the dependence of this oscillatory behavior
 of the recoil by simulating an additional sequence of eccentric
 binaries with mass ratio $q=1/2$, but less negative binding energy,
 corresponding to about six orbits in the quasicircular limit.
 We find the oscillations in $v=v(e)$ to be more pronounced
 and numerous than in the shorter sequence. We attribute this
 feature to the longer available time window during which otherwise
 identical binaries with tiny differences in the initial
 eccentricity build up a phase difference prior to merger.
 This observation raises the intriguing possibility that the
 total recoil depends highly sensitively on the initial eccentricity.
 \item The variations in the kick velocity are accompanied by
 relative time shifts in the peak amplitudes of subdominant
 multipoles relative to the peaks of the (2,2) mode;
 cf.~Fig.~\ref{fig:psi4modes}. For configurations with a large (small)
 kick, the peak amplitude of subdominant multipoles tends to
 be aligned (misaligned) with the quadrupole peak.
\end{enumerate}
Our findings point to a variety of future investigations. While our
simulations indicate an increased sensitivity of the GW merger signal
to the initial eccentricity for larger initial separations
(i.e.~longer inspirals), it is not clear how this will be affected by
the circularizing nature of GW emission. In this context, it will also
be important to analyze in more quantitative terms the differences in
the GW signals and possible implications for parameter inference in GW
observations. A thorough investigation of long eccentric inspirals on
astrophysical time scales will likely require PN methods and may
benefit greatly from a multi-time-scale analysis in phase space, as
applied to spin-precessing BH binaries in
Refs.~\cite{Kesden:2014sla,Gerosa:2015tea} or to the dynamics of
binary systems in external gravitational background potentials in
Refs.~\cite{Hamilton:2019a,Hamilton:2019b}. If there is a single
conclusion to draw from the results of this work, it is the
surprisingly rich phenomenology of the GW signals of eccentric compact
binaries---even in the absence of spins---which merits as much as it
requires further investigation.

\begin{acknowledgments}
We thank Michalis Agathos, Vishal Baibhav, Vitor Cardoso, Thomas
Helfer, and Nicholas Speeney for useful discussions. 
We also thank Chris Moore, Carlos Lousto and Juan Calder\'on 
Bustillo for helpful comments on this manuscript.
M.R. thanks the
GRChombo collaboration \cite{GRChomboWebsite} for their code
development, and particularly Katy Clough and Tiago Fran\c{c}a. M.R. is
supported by a Science and Technology Facilities Council (STFC)
studentship. U.S. is supported by the European Union’s H2020 ERC
Consolidator Grant ``Matter and strong-field gravity: new frontiers in
Einstein's theory'' Grant No.~MaGRaTh--646597, and the STFC
Consolidator Grant No. ST/P000673/1. E.B. is supported by NSF Grant
No. PHY-1912550, NSF Grant No. AST-2006538, NASA ATP Grant No.
17-ATP17-0225 and NASA ATP Grant No. 19-ATP19-0051. This work has
received funding from the European Union's Horizon 2020 research and
innovation programme under the Marie Sk\l odowska-Curie Grant
No.~690904. This work was supported by the GWverse COST Action
CA16104, ``Black holes, gravitational waves and fundamental physics''.
Computational work was performed on
the San Diego Supercomputer Center Comet and Texas Advanced Computing 
Center (TACC) Stampede2 clusters at the University of California San Diego 
and the University of Texas at Austin (UT Austin), respectively, through
NSF-XSEDE Grant No.~PHY-090003; the Cambridge Service for Data 
Driven Discovery (CSD3) system at the University of Cambridge 
through STFC capital Grants No.~ST/P002307/1 and No.~ST/R002452/1,
and STFC operations Grant No.~ST/R00689X/1;
the TACC Frontera cluster at UT Austin; and
the JUWELS cluster at GCS@FZJ, Germany through PRACE Grant No.~2020225359.

\end{acknowledgments}
\appendix

\section{Numerical accuracy}
\label{sec:accuracy}
\begin{figure*}[t]
    \subfloat
    {
        \includegraphics[width=0.48\linewidth]{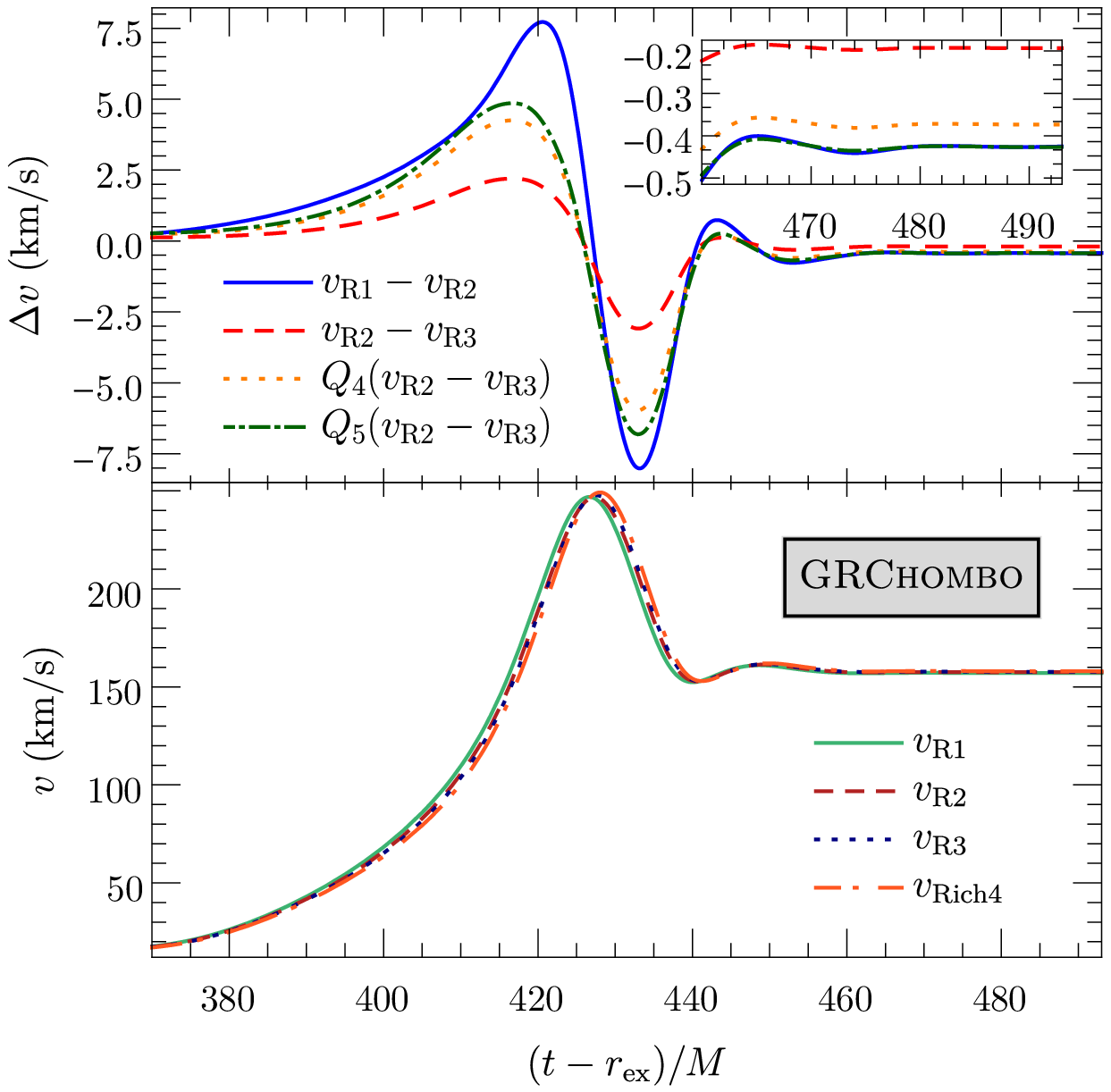}
    }
    \hfill
    \subfloat
    {
        \includegraphics[width=0.48\linewidth]{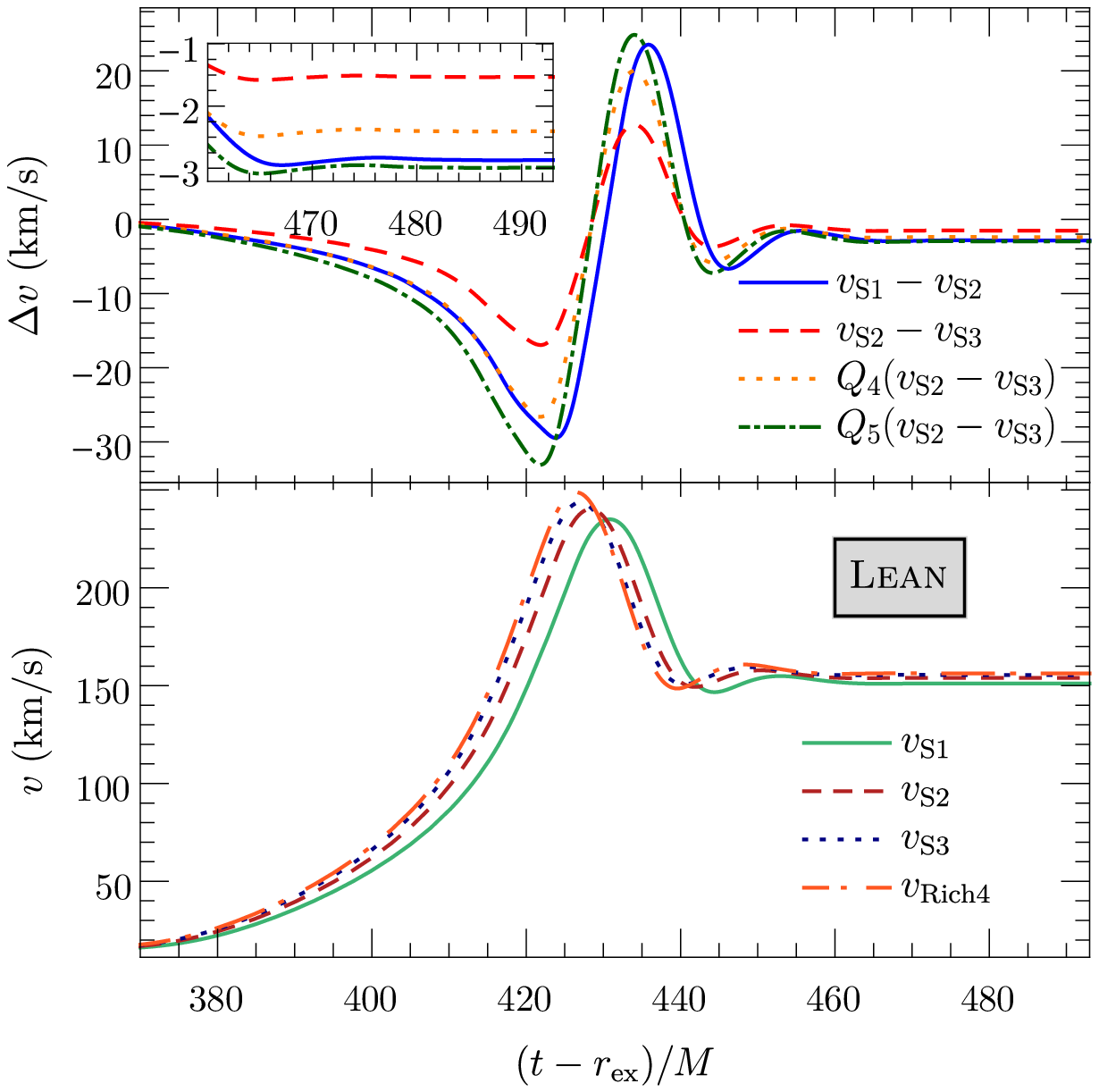}
    }
    \caption{For each code, we show convergence plots for the
      accumulated linear momentum radiated from \textsf{sq1:2-p0100}
      by plotting the BH recoil velocity in the bottom panels. The
      Richardson extrapolated curve, $v_{\mathrm{Rich4}}$, assuming
      fourth-order convergence, is also shown in the bottom panel. The
      grid configurations are given in Table~\ref{tab:grchombo-grids}
      for \textsc{GRChombo} and in Table~\ref{tab:lean-grids} for
      \textsc{Lean}. The top panel shows the difference between the
      configurations along with rescalings corresponding to fourth- and
      fifth-order convergence. The inset shows a magnification of the
      right side of the plot: the final value of the recoil
      velocity is what we show in Fig.~\ref{fig:kicks2}.} 
    \label{fig:convergence}
\end{figure*}
As in Ref.~\cite{Sperhake:2019wwo}, the uncertainty in our numerical results 
for the recoil velocities has two predominant contributions: the 
discretization error and the finite extraction radii for the Weyl scalar
$\Psi_4$.

To estimate the uncertainty arising from the latter, we have selected a
representative sample of the simulations from each sequence and extrapolated 
the cumulative radiated momentum to infinity from about six extraction
radii in the range $r_{\rm ex}/2 \le r_{\rm ex}$ 
using a Taylor series in $1/r$ as in 
Ref.~\cite{Sperhake:2011zz}. We report the results from the finite 
extraction radii given in Table~\ref{tab:sequences} and estimate the 
error by comparing with the linear-order extrapolation.
For both codes, we estimate that the contribution from this error is about
2\% 
for all sequences.

In order to estimate the error contribution from finite differencing and
verify that our codes give consistent results, we have performed simulations
of \texttt{sq1:2-p0100} (the binary in sequence \texttt{sq1:2} with 
$p/M = 0.1$) with both codes. We discuss the analyses of the convergence
of each code separately before comparing.

\subsection{\textsc{GRChombo} convergence}
\begin{table}[b]
{   
    \caption{Grid configurations used for \textsc{GRChombo} simulations.
    As explained in Sec.~\ref{sec:grchombo} and Appendix \ref{sec:tagging},
    the total 
    number of refinement levels is $L+1$, the number of cells along each
    dimension on the coarsest level is $N$, $t_R$ is the regridding 
    threshold value, $b$ is the BH tagging buffer parameter
    that we
    set proportional to the mass $M_i$ ($i=1,\,2$)
    of the nearest BH for all configurations except R4, 
    and $h_L$ denotes the grid spacing.
    }
    \centering
    \begin{ruledtabular}
    \begin{tabular}{lcccccc}
        Label & $L$ & $N$ & $t_R$ & $b$ & $h_L/M_1$ & tagging\\
        \hline
        R1 & 7 & 320 & 0.012 & $0.5M_i$ & 3/80 & Spherical \\
        R2 & 7 & 368 & 0.01043 & $0.5M_i$ & 3/92 & Spherical\\
        R3 & 7 & 416 & 0.00923 & $0.5M_i$ & 3/104 & Spherical\\
        R4 & 7 & 352 & 0.01091 & 0.7 & 3/88 & Box\\
    \end{tabular}
    \end{ruledtabular}
}
    \label{tab:grchombo-grids}
\end{table}
For \textsc{GRChombo}, we have performed the simulations of 
\texttt{sq1:2-p0100} with resolutions $h_L=3M_1/80$, $3M_1/92$ and 
$3M_1/104$, and we refer to the configurations corresponding to these 
resolutions as R1, R2 and R3, respectively.
The full grid configurations 
are given in Table~\ref{tab:grchombo-grids} and the results of this 
analysis are shown in the left panel of Fig.~\ref{fig:convergence}. Around
merger, at $(t-r_{\rm ex})/M\sim 420$, our results exhibit mild
overconvergence
in the top-left panel of Fig.~\ref{fig:convergence}.
The important results for our analysis in Fig.~\ref{fig:kicks2},
however, are the final kick values after the merged BH has settled down.
As can be seen from the inset, the convergence here is close to 
fifth order. 
From our convergence 
analysis, the difference between the result obtained from the R1 simulation 
and the more conservative fourth-order Richardson-extrapolated result leads 
to an estimate of the discretization error of about $1\%$. A similar error 
estimate is also obtained for the radiated energy, $E^{\rad}$.
From experience, we have found smaller values for the mass ratio $q<1$ 
more challenging to accurately simulate than larger values, and we 
therefore feel justified in using this error estimate (for a $q=1/2$ 
configuration) as a conservative estimate for the error in the 
\texttt{sq2:3} sequence simulations ($q=2/3$). We therefore used the R1 grid
configuration for this sequence with  $l_1^{\max}=l_2^{\max}=L=7$ 
(both BHs are covered by 
the finest level; see Appendix \ref{sec:tagging} for details).

For the \texttt{sq1:3} simulations, we used the R4 grid configuration
(see Table \ref{tab:grchombo-grids}) with $l_1^{\max}=L=7$ and
$l_2^{\max}=L-1=6$ (the larger BH is not covered by the finest level:
see Appendix \ref{sec:tagging} for details). This corresponds to a
resolution of $h_L=3M_1/88$. We performed a separate convergence
analysis of \texttt{sq1:3-p0089}, which led to an estimated 1\%
discretization error.

Combining both the finite extraction radius and discretization errors, 
our estimate for the total error budget of the \textsc{GRChombo} 
simulations is about $3\%$.

\subsection{\textsc{Lean} convergence}
\begin{table}[b]
    {
    \caption{Grid configurations used for \textsc{Lean} simulations. As 
    explained in Sec.~\ref{sec:lean}, the total number of refinement 
    levels is $L+1$, the number of fixed refinement levels is $l_F+1$, 
    $R_0$ is the half-length of the outer grid, $R_L$ is the half-length 
    of one cubic component of the innermost grid, and $h_L$ is the grid 
    spacing on the finest level.}
    \centering
    \begin{ruledtabular}
    \begin{tabular}{lccccc}
        Label & $L$ & $l_F$ & $R_0$ & $R_L$ & $h_L/M_1$\\
        \hline
        S1 & 7 & 4 & 384 & 1 & 1/20\\
        S2 & 7 & 4 & 384 & 1 & 1/24\\
        S3 & 7 & 4 & 384 & 1 & 1/32\\
        S4 & 7 & 4 & 384 & 1 & 1/28\\
    \end{tabular}
    \end{ruledtabular}
    }
    \label{tab:lean-grids}
\end{table}

With \textsc{Lean}, we have simulated \texttt{sq1:2-p0100} with
resolutions $h_L = M_1/20$, $M_1/24$ and $M_1/32$. We refer to these
grid configurations as S1, S2 and S3, respectively
(cf. Table~\ref{tab:lean-grids}). The right panel of Fig.~\ref{fig:convergence}
shows convergence between fourth and fifth order. For simulations in
\texttt{sq1:2}, we used the S2 grid configuration. From the
convergence analysis, the difference between the result obtained from
the S2 simulation and the fourth-order Richardson extrapolation leads
to an estimate of the discretization error of about $1.5\%$.

For the \texttt{lq1:2} simulations, we have undertaken a separate
convergence analysis of \texttt{lq1:2-p0086} using the same grid
setup as in
Table~\ref{tab:lean-grids}, but using higher resolutions
$h_L/M_1=1/24$, $1/28$ and $1/32$. We observe convergence close to
fourth order and obtain an error estimate of $1\,\%$ from
the Richardson-extrapolated kick for the medium resolution
$h_L/M_1=1/28$.

In summary, the {\sc Lean} simulations of sequence \texttt{sq1:2} are
performed with resolution grid S2 of Table~\ref{tab:lean-grids} and an
error budget of $3.5\,\%$, and those of sequence \texttt{lq1:2} with
grid S4 of Table~\ref{tab:lean-grids} and an error budget of $3\,\%$.

\subsection{Comparison between \textsc{GRChombo} and \textsc{Lean}}
\label{sec:code-comparison}
\begin{figure}[t]
    {
    \centering
    \includegraphics[width=\columnwidth]{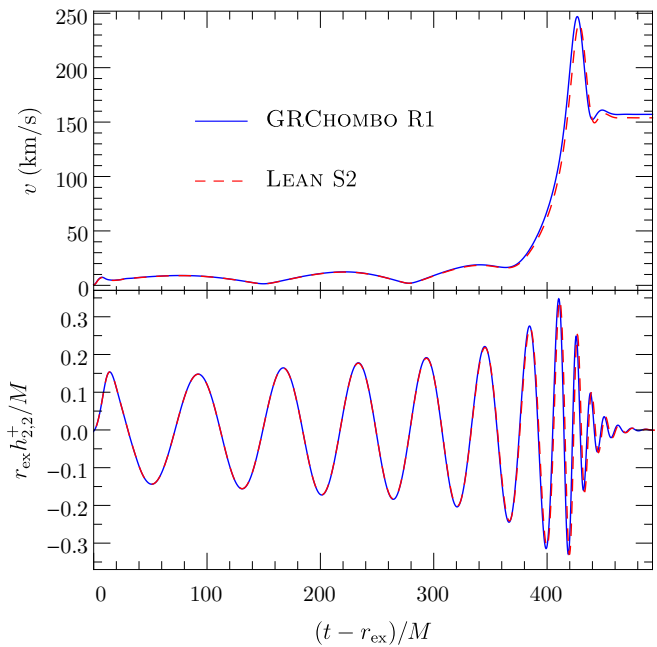}
    }
    \caption{Comparison between \textsc{GRChombo} and \textsc{Lean} 
    for the accumulated linear momentum radiated in GWs in simulations 
    of \texttt{sq1:2-p0100} with $e_t=0.10$.
    We compare the BH recoil velocity (top panel)
    and the corresponding plus-polarized $\ell=m=2$ strain
    amplitude (bottom panel).
    }
    \label{fig:grchombo-lean-comparison}
\end{figure}
A comparison of the recoil velocity computed from \textsc{GRChombo} 
and \textsc{Lean} simulations of \texttt{sq1:2-p0100} with the grid 
configurations R1 and S2 (used for the \texttt{sq2:3} and \texttt{sq1:2} runs) 
respectively, is shown in the top panel of Fig.~\ref{fig:grchombo-lean-comparison}.
The eccentricity estimate for this system is $e_t=0.10$. We have chosen
this configuration for two reasons. First, to determine appropriate
resolutions, we had to calibrate our codes' accuracy at the start of
our exploration, which we began in the regime of mild eccentricities to
acquire an intuitive understanding of their behavior. Second, 
configurations with mild eccentricity have a longer inspiral phase
than highly eccentric ones, and therefore impose a stronger requirement
on phase accuracy. A mildly eccentric binary is therefore ideally suited
to obtain a conservative estimate of the numerical accuracy, which is representative across the targeted parameter space.

The final recoil velocities obtained for this configuration with
our two codes differ by about 2\%, which is well within the error budget 
of each code.
We also show the quadrupole contribution
$h^+_{2,2}$ to the `+' polarization strain defined by \cite{Bishop:2016lgv}
\begin{multline}
    h^+_{\ell,m}(t,r)-\mathrm{i}h^\times_{\ell,m}(t,r) \\
    = a_{\ell,m} + b_{\ell,m}t + \int_0^t\rmd t^\prime\int_0^{t^\prime}
    \rmd t^{\prime\prime}\psi_{\ell,m}(t^{\prime\prime},r),
    \label{eq:strain}
\end{multline}
where the constants $a_{\ell,m}$ and $b_{\ell,m}$ are chosen to minimize 
linear drift,
in the bottom panel of the figure, to better illustrate the agreement
between the codes for these grid configurations.

In Fig.~\ref{fig:convergence} the differences between the results of 
different resolutions with \textsc{Lean} are greater than that of
\textsc{GRChombo}. However, we found that \textsc{Lean} entered the 
convergent regime at lower resolutions than GRChombo. This is compatible 
with the observations of Ref.~\cite{Alic:2011gg} that higher resolutions 
were required for convergence with CCZ4 compared to BSSNOK.

\section{\textsc{GRChombo} tagging criterion}
\label{sec:tagging}
As explained in Sec.~\ref{sec:grchombo}, the regridding is controlled by 
the tagging of cells for refinement in the Berger-Rigoutsos algorithm 
\cite{Berger1991}, with cells being tagged if the tagging criterion $C$ 
exceeds the specified threshold value $t_R$ as given in Table 
\ref{tab:grchombo-grids}. For this work, we use the 
tagging criterion
\begin{widetext}
\begin{equation}
    C=
    \begin{cases}
        0, &\text{if }l\geq l_{\mathrm{BH}}^{\max}
        \text{ and } r_{\mathrm{BH}} < (M_{\mathrm{BH}}+b),\\
        \max(C_{\chi},C_{\mathrm{punc}},C_{\mathrm{ex}}),&
        \text{otherwise},
    \end{cases}
\end{equation}
\end{widetext}
where $l_{\mathrm{BH}}^{\max}$ is a specifiable maximum level parameter 
for each BH (so that it is not unnecessarily over resolved), 
$r_{\mathrm{BH}}$ is the coordinate distance to the puncture, 
$M_{\mathrm{BH}}$ is the mass of the corresponding BH, $b$ is a buffer 
parameter, and $C_\chi$, $C_{\mathrm{punc}}$, and $C_{\mathrm{ex}}$ are 
given as follows:
\begin{enumerate}[label=(\roman*)]
    \item 
        $C_{\chi}$ tags regions in which the gradients of the conformal 
        factor $\chi$ become steep. It is given by
        \begin{equation}
            C_{\chi} = h_l\sqrt{\sum_{i,j}\left(
            \partial_i\partial_j\chi\right)^2}\,,
        \end{equation}
        where $h_l$ is the grid spacing on refinement level $l$.
    \item 
        $C_{\mathrm{punc}}$ tags within spheres around each puncture in 
        order to ensure the horizon is suitably well resolved. It is given 
        by
        \begin{equation}
            C_{\mathrm{punc}} = 
            \begin{cases}
                100, &\text{if } r_{\mathrm{BH}} < 
                (M_{\mathrm{BH}}+b) 2^{\max(l_\mathrm{BH}^{\max}-l-1, 2)}, \\
                0, & \text{otherwise}.
            \end{cases}
        \end{equation}
    \item
        $C_{\mathrm{ex}}$ ensures each sphere on which we extract the 
        Weyl scalar $\Psi_4$ is suitably well resolved. It is given by
        \begin{equation}
            C_{\mathrm{ex}} = 
            \begin{cases}
                100, &\text{if } r < 1.2r_{\mathrm{ex}} \text{ and } 
                l < l_{\mathrm{ex}}, \\
                0, & \text{otherwise},
            \end{cases}
        \end{equation}
         where $r$ is the coordinate distance to the center of mass,
         $r=r_{\mathrm{ex}}$ gives the location of the extraction sphere, 
         and $l_{\mathrm{ex}}$ is a specifiable extraction level parameter 
         for each sphere.
\end{enumerate}
We also used this tagging criterion with the replacement
$r_{\mathrm{BH}}\to
\max(x_{\mathrm{BH}},y_{\mathrm{BH}},z_{\mathrm{BH}})$, where,
e.g. $x_{\mathrm{BH}}$ is the distance to the puncture in the $x$
direction. We refer to this as ``box'' tagging and the original as
``spherical'' tagging. Naively, one might hope that $C_{\chi}$ is
sufficient to ensure suitable refinement around the BHs, since the
gradients of $\chi$ become increasingly steep close to the
punctures. However we found empirically that, without
$C_{\text{punc}}$, the horizons are perturbed significantly by the
refinement boundaries, leading to lower accuracy.


\bibliography{bibliography}

\begin{thebibliography}{99}%
\makeatletter
\providecommand \@ifxundefined [1]{%
 \@ifx{#1\undefined}
}%
\providecommand \@ifnum [1]{%
 \ifnum #1\expandafter \@firstoftwo
 \else \expandafter \@secondoftwo
 \fi
}%
\providecommand \@ifx [1]{%
 \ifx #1\expandafter \@firstoftwo
 \else \expandafter \@secondoftwo
 \fi
}%
\providecommand \natexlab [1]{#1}%
\providecommand \enquote  [1]{``#1''}%
\providecommand \bibnamefont  [1]{#1}%
\providecommand \bibfnamefont [1]{#1}%
\providecommand \citenamefont [1]{#1}%
\providecommand \href@noop [0]{\@secondoftwo}%
\providecommand \href [0]{\begingroup \@sanitize@url \@href}%
\providecommand \@href[1]{\@@startlink{#1}\@@href}%
\providecommand \@@href[1]{\endgroup#1\@@endlink}%
\providecommand \@sanitize@url [0]{\catcode `\\12\catcode `\$12\catcode
  `\&12\catcode `\#12\catcode `\^12\catcode `\_12\catcode `\%12\relax}%
\providecommand \@@startlink[1]{}%
\providecommand \@@endlink[0]{}%
\providecommand \url  [0]{\begingroup\@sanitize@url \@url }%
\providecommand \@url [1]{\endgroup\@href {#1}{\urlprefix }}%
\providecommand \urlprefix  [0]{URL }%
\providecommand \Eprint [0]{\href }%
\providecommand \doibase [0]{https://doi.org/}%
\providecommand \selectlanguage [0]{\@gobble}%
\providecommand \bibinfo  [0]{\@secondoftwo}%
\providecommand \bibfield  [0]{\@secondoftwo}%
\providecommand \translation [1]{[#1]}%
\providecommand \BibitemOpen [0]{}%
\providecommand \bibitemStop [0]{}%
\providecommand \bibitemNoStop [0]{.\EOS\space}%
\providecommand \EOS [0]{\spacefactor3000\relax}%
\providecommand \BibitemShut  [1]{\csname bibitem#1\endcsname}%
\let\auto@bib@innerbib\@empty
\bibitem [{\citenamefont {Sperhake}\ \emph {et~al.}(2020)\citenamefont
  {Sperhake}, \citenamefont {Rosca-Mead}, \citenamefont {Gerosa},\ and\
  \citenamefont {Berti}}]{Sperhake:2019wwo}%
  \BibitemOpen
  \bibfield  {author} {\bibinfo {author} {\bibfnamefont {U.}~\bibnamefont
  {Sperhake}}, \bibinfo {author} {\bibfnamefont {R.}~\bibnamefont
  {Rosca-Mead}}, \bibinfo {author} {\bibfnamefont {D.}~\bibnamefont {Gerosa}},\
  and\ \bibinfo {author} {\bibfnamefont {E.}~\bibnamefont {Berti}},\ }\href
  {https://doi.org/10.1103/PhysRevD.101.024044} {\bibfield  {journal} {\bibinfo
   {journal} {Phys. Rev. D}\ }\textbf {\bibinfo {volume} {101}},\ \bibinfo
  {pages} {024044} (\bibinfo {year} {2020})},\ \Eprint
  {https://arxiv.org/abs/1910.01598} {arXiv:1910.01598 [gr-qc]} \BibitemShut
  {NoStop}%
\bibitem [{\citenamefont {Abbott}\ \emph {et~al.}(2016)\citenamefont {Abbott}
  \emph {et~al.}}]{Abbott:2016blz}%
  \BibitemOpen
  \bibfield  {author} {\bibinfo {author} {\bibfnamefont {B.~P.}\ \bibnamefont
  {Abbott}} \emph {et~al.},\ }\href
  {https://doi.org/10.1103/PhysRevLett.116.061102} {\bibfield  {journal}
  {\bibinfo  {journal} {Phys. Rev. Lett.}\ }\textbf {\bibinfo {volume} {116}},\
  \bibinfo {pages} {061102} (\bibinfo {year} {2016})},\ \Eprint
  {https://arxiv.org/abs/1602.03837} {arXiv:1602.03837 [gr-qc]} \BibitemShut
  {NoStop}%
\bibitem [{\citenamefont {Saulson}(2011)}]{Saulson:2010zz}%
  \BibitemOpen
  \bibfield  {author} {\bibinfo {author} {\bibfnamefont {P.~R.}\ \bibnamefont
  {Saulson}},\ }\href {https://doi.org/10.1007/s10714-011-1237-z} {\bibfield
  {journal} {\bibinfo  {journal} {Gen. Rel. Grav.}\ }\textbf {\bibinfo {volume}
  {43}},\ \bibinfo {pages} {3289} (\bibinfo {year} {2011})}\BibitemShut
  {NoStop}%
\bibitem [{\citenamefont {Campanelli}\ \emph
  {et~al.}(2006{\natexlab{a}})\citenamefont {Campanelli}, \citenamefont
  {Lousto},\ and\ \citenamefont {Zlochower}}]{Campanelli:2006uy}%
  \BibitemOpen
  \bibfield  {author} {\bibinfo {author} {\bibfnamefont {M.}~\bibnamefont
  {Campanelli}}, \bibinfo {author} {\bibfnamefont {C.~O.}\ \bibnamefont
  {Lousto}},\ and\ \bibinfo {author} {\bibfnamefont {Y.}~\bibnamefont
  {Zlochower}},\ }\href {https://doi.org/10.1103/PhysRevD.74.041501} {\bibfield
   {journal} {\bibinfo  {journal} {Phys. Rev. D}\ }\textbf {\bibinfo {volume}
  {74}},\ \bibinfo {pages} {041501(R)} (\bibinfo {year}
  {2006}{\natexlab{a}})},\ \Eprint {https://arxiv.org/abs/gr-qc/0604012}
  {arXiv:gr-qc/0604012} \BibitemShut {NoStop}%
\bibitem [{\citenamefont {Sperhake}\ \emph {et~al.}(2009)\citenamefont
  {Sperhake}, \citenamefont {Cardoso}, \citenamefont {Pretorius}, \citenamefont
  {Berti}, \citenamefont {Hinderer},\ and\ \citenamefont
  {Yunes}}]{Sperhake:2009jz}%
  \BibitemOpen
  \bibfield  {author} {\bibinfo {author} {\bibfnamefont {U.}~\bibnamefont
  {Sperhake}}, \bibinfo {author} {\bibfnamefont {V.}~\bibnamefont {Cardoso}},
  \bibinfo {author} {\bibfnamefont {F.}~\bibnamefont {Pretorius}}, \bibinfo
  {author} {\bibfnamefont {E.}~\bibnamefont {Berti}}, \bibinfo {author}
  {\bibfnamefont {T.}~\bibnamefont {Hinderer}},\ and\ \bibinfo {author}
  {\bibfnamefont {N.}~\bibnamefont {Yunes}},\ }\href
  {https://doi.org/10.1103/PhysRevLett.103.131102} {\bibfield  {journal}
  {\bibinfo  {journal} {Phys. Rev. Lett.}\ }\textbf {\bibinfo {volume} {103}},\
  \bibinfo {pages} {131102} (\bibinfo {year} {2009})},\ \Eprint
  {https://arxiv.org/abs/0907.1252} {arXiv:0907.1252 [gr-qc]} \BibitemShut
  {NoStop}%
\bibitem [{\citenamefont {Bonnor}\ and\ \citenamefont
  {Rotenberg}(1961)}]{Bonnor1961-wy}%
  \BibitemOpen
  \bibfield  {author} {\bibinfo {author} {\bibfnamefont {W.~B.}\ \bibnamefont
  {Bonnor}}\ and\ \bibinfo {author} {\bibfnamefont {M.~A.}\ \bibnamefont
  {Rotenberg}},\ }\href {https://doi.org/10.1098/rspa.1961.0226} {\bibfield
  {journal} {\bibinfo  {journal} {Proc. R. Soc. A}\ }\textbf {\bibinfo {volume}
  {265}},\ \bibinfo {pages} {109} (\bibinfo {year} {1961})}\BibitemShut
  {NoStop}%
\bibitem [{\citenamefont {Peres}(1962)}]{Peres:1962zz}%
  \BibitemOpen
  \bibfield  {author} {\bibinfo {author} {\bibfnamefont {A.}~\bibnamefont
  {Peres}},\ }\href {https://doi.org/10.1103/PhysRev.128.2471} {\bibfield
  {journal} {\bibinfo  {journal} {Phys. Rev.}\ }\textbf {\bibinfo {volume}
  {128}},\ \bibinfo {pages} {2471} (\bibinfo {year} {1962})}\BibitemShut
  {NoStop}%
\bibitem [{\citenamefont {Bekenstein}(1973)}]{Bekenstein:1973zz}%
  \BibitemOpen
  \bibfield  {author} {\bibinfo {author} {\bibfnamefont {J.~D.}\ \bibnamefont
  {Bekenstein}},\ }\href {https://doi.org/10.1086/152255} {\bibfield  {journal}
  {\bibinfo  {journal} {Astrophys. J.}\ }\textbf {\bibinfo {volume} {183}},\
  \bibinfo {pages} {657} (\bibinfo {year} {1973})}\BibitemShut {NoStop}%
\bibitem [{\citenamefont {Gerosa}\ and\ \citenamefont
  {Moore}(2016)}]{Gerosa:2016vip}%
  \BibitemOpen
  \bibfield  {author} {\bibinfo {author} {\bibfnamefont {D.}~\bibnamefont
  {Gerosa}}\ and\ \bibinfo {author} {\bibfnamefont {C.~J.}\ \bibnamefont
  {Moore}},\ }\href {https://doi.org/10.1103/PhysRevLett.117.011101} {\bibfield
   {journal} {\bibinfo  {journal} {Phys. Rev. Lett.}\ }\textbf {\bibinfo
  {volume} {117}},\ \bibinfo {pages} {011101} (\bibinfo {year} {2016})},\
  \Eprint {https://arxiv.org/abs/1606.04226} {arXiv:1606.04226 [gr-qc]}
  \BibitemShut {NoStop}%
\bibitem [{\citenamefont {Calder\'on~Bustillo}\ \emph
  {et~al.}(2018)\citenamefont {Calder\'on~Bustillo}, \citenamefont {Clark},
  \citenamefont {Laguna},\ and\ \citenamefont
  {Shoemaker}}]{CalderonBustillo:2018zuq}%
  \BibitemOpen
  \bibfield  {author} {\bibinfo {author} {\bibfnamefont {J.}~\bibnamefont
  {Calder\'on~Bustillo}}, \bibinfo {author} {\bibfnamefont {J.~A.}\
  \bibnamefont {Clark}}, \bibinfo {author} {\bibfnamefont {P.}~\bibnamefont
  {Laguna}},\ and\ \bibinfo {author} {\bibfnamefont {D.}~\bibnamefont
  {Shoemaker}},\ }\href {https://doi.org/10.1103/PhysRevLett.121.191102}
  {\bibfield  {journal} {\bibinfo  {journal} {Phys. Rev. Lett.}\ }\textbf
  {\bibinfo {volume} {121}},\ \bibinfo {pages} {191102} (\bibinfo {year}
  {2018})},\ \Eprint {https://arxiv.org/abs/1806.11160} {arXiv:1806.11160
  [gr-qc]} \BibitemShut {NoStop}%
\bibitem [{\citenamefont {Lousto}\ and\ \citenamefont
  {Healy}(2019)}]{Lousto:2019lyf}%
  \BibitemOpen
  \bibfield  {author} {\bibinfo {author} {\bibfnamefont {C.~O.}\ \bibnamefont
  {Lousto}}\ and\ \bibinfo {author} {\bibfnamefont {J.}~\bibnamefont {Healy}},\
  }\href {https://doi.org/10.1103/PhysRevD.100.104039} {\bibfield  {journal}
  {\bibinfo  {journal} {Phys. Rev. D}\ }\textbf {\bibinfo {volume} {100}},\
  \bibinfo {pages} {104039} (\bibinfo {year} {2019})},\ \Eprint
  {https://arxiv.org/abs/arXiv:1908.04382 [gr-qc]} {arXiv:1908.04382 [gr-qc]}
  \BibitemShut {NoStop}%
\bibitem [{\citenamefont {Varma}\ \emph {et~al.}(2020)\citenamefont {Varma},
  \citenamefont {Isi},\ and\ \citenamefont {Biscoveanu}}]{Varma:2020nbm}%
  \BibitemOpen
  \bibfield  {author} {\bibinfo {author} {\bibfnamefont {V.}~\bibnamefont
  {Varma}}, \bibinfo {author} {\bibfnamefont {M.}~\bibnamefont {Isi}},\ and\
  \bibinfo {author} {\bibfnamefont {S.}~\bibnamefont {Biscoveanu}},\ }\href
  {https://doi.org/10.1103/PhysRevLett.124.101104} {\bibfield  {journal}
  {\bibinfo  {journal} {Phys. Rev. Lett.}\ }\textbf {\bibinfo {volume} {124}},\
  \bibinfo {pages} {101104} (\bibinfo {year} {2020})},\ \Eprint
  {https://arxiv.org/abs/2002.00296} {arXiv:2002.00296 [gr-qc]} \BibitemShut
  {NoStop}%
\bibitem [{\citenamefont {Cardoso}\ and\ \citenamefont
  {Macedo}(2020)}]{Cardoso:2020lxx}%
  \BibitemOpen
  \bibfield  {author} {\bibinfo {author} {\bibfnamefont {V.}~\bibnamefont
  {Cardoso}}\ and\ \bibinfo {author} {\bibfnamefont {C.~F.~B.}\ \bibnamefont
  {Macedo}},\ }\href {https://doi.org/10.1093/mnras/staa2396} {\bibfield
  {journal} {\bibinfo  {journal} {Mon. Not. R. Astron. Soc.}\ }\textbf
  {\bibinfo {volume} {498}},\ \bibinfo {pages} {1963} (\bibinfo {year}
  {2020})},\ \Eprint {https://arxiv.org/abs/2008.01091} {arXiv:2008.01091}
  \BibitemShut {NoStop}%
\bibitem [{\citenamefont {Fitchett}(1983)}]{Fitchett1983-xq}%
  \BibitemOpen
  \bibfield  {author} {\bibinfo {author} {\bibfnamefont {M.~J.}\ \bibnamefont
  {Fitchett}},\ }\href {https://doi.org/10.1093/mnras/203.4.1049} {\bibfield
  {journal} {\bibinfo  {journal} {Mon. Not. R. Astron. Soc.}\ }\textbf
  {\bibinfo {volume} {203}},\ \bibinfo {pages} {1049} (\bibinfo {year}
  {1983})}\BibitemShut {NoStop}%
\bibitem [{\citenamefont {Blanchet}\ \emph {et~al.}(2005)\citenamefont
  {Blanchet}, \citenamefont {Qusailah},\ and\ \citenamefont
  {Will}}]{Blanchet:2005rj}%
  \BibitemOpen
  \bibfield  {author} {\bibinfo {author} {\bibfnamefont {L.}~\bibnamefont
  {Blanchet}}, \bibinfo {author} {\bibfnamefont {M.~S.}\ \bibnamefont
  {Qusailah}},\ and\ \bibinfo {author} {\bibfnamefont {C.~M.}\ \bibnamefont
  {Will}},\ }\href {https://doi.org/10.1086/497332} {\bibfield  {journal}
  {\bibinfo  {journal} {Astrophys. J.}\ }\textbf {\bibinfo {volume} {635}},\
  \bibinfo {pages} {508} (\bibinfo {year} {2005})},\ \Eprint
  {https://arxiv.org/abs/astro-ph/0507692} {arXiv:astro-ph/0507692}
  \BibitemShut {NoStop}%
\bibitem [{\citenamefont {Hughes}\ \emph {et~al.}(2004)\citenamefont {Hughes},
  \citenamefont {Favata},\ and\ \citenamefont {Holz}}]{Hughes:2004ck}%
  \BibitemOpen
  \bibfield  {author} {\bibinfo {author} {\bibfnamefont {S.~A.}\ \bibnamefont
  {Hughes}}, \bibinfo {author} {\bibfnamefont {M.}~\bibnamefont {Favata}},\
  and\ \bibinfo {author} {\bibfnamefont {D.~E.}\ \bibnamefont {Holz}},\ }in\
  \href {https://doi.org/10.1007/11403913_64} {\emph {\bibinfo {booktitle}
  {{Conference on Growing Black Holes: Accretion in a Cosmological Context}}}}\
  (\bibinfo  {publisher} {Springer},\ \bibinfo {address} {Berlin, Heidelberg,
  Germany},\ \bibinfo {year} {2004})\ \Eprint
  {https://arxiv.org/abs/astro-ph/0408492} {arXiv:astro-ph/0408492}
  \BibitemShut {NoStop}%
\bibitem [{\citenamefont {Damour}\ and\ \citenamefont
  {Gopakumar}(2006)}]{Damour:2006tr}%
  \BibitemOpen
  \bibfield  {author} {\bibinfo {author} {\bibfnamefont {T.}~\bibnamefont
  {Damour}}\ and\ \bibinfo {author} {\bibfnamefont {A.}~\bibnamefont
  {Gopakumar}},\ }\href {https://doi.org/10.1103/PhysRevD.73.124006} {\bibfield
   {journal} {\bibinfo  {journal} {Phys. Rev. D}\ }\textbf {\bibinfo {volume}
  {73}},\ \bibinfo {pages} {124006} (\bibinfo {year} {2006})},\ \Eprint
  {https://arxiv.org/abs/gr-qc/0602117} {arXiv:gr-qc/0602117} \BibitemShut
  {NoStop}%
\bibitem [{\citenamefont {Sopuerta}\ \emph {et~al.}(2006)\citenamefont
  {Sopuerta}, \citenamefont {Yunes},\ and\ \citenamefont
  {Laguna}}]{Sopuerta:2006wj}%
  \BibitemOpen
  \bibfield  {author} {\bibinfo {author} {\bibfnamefont {C.~F.}\ \bibnamefont
  {Sopuerta}}, \bibinfo {author} {\bibfnamefont {N.}~\bibnamefont {Yunes}},\
  and\ \bibinfo {author} {\bibfnamefont {P.}~\bibnamefont {Laguna}},\ }\href
  {https://doi.org/10.1103/PhysRevD.78.049901} {\bibfield  {journal} {\bibinfo
  {journal} {Phys. Rev. D}\ }\textbf {\bibinfo {volume} {74}},\ \bibinfo
  {pages} {124010} (\bibinfo {year} {2006})},\ \bibinfo {note} {[Erratum:
  Phys.Rev.D 75, 069903(E) (2007), Erratum: Phys.Rev.D 78, 049901 (2008)]},\
  \Eprint {https://arxiv.org/abs/astro-ph/0608600} {arXiv:astro-ph/0608600}
  \BibitemShut {NoStop}%
\bibitem [{\citenamefont {Sopuerta}\ \emph {et~al.}(2007)\citenamefont
  {Sopuerta}, \citenamefont {Yunes},\ and\ \citenamefont
  {Laguna}}]{Sopuerta:2006et}%
  \BibitemOpen
  \bibfield  {author} {\bibinfo {author} {\bibfnamefont {C.~F.}\ \bibnamefont
  {Sopuerta}}, \bibinfo {author} {\bibfnamefont {N.}~\bibnamefont {Yunes}},\
  and\ \bibinfo {author} {\bibfnamefont {P.}~\bibnamefont {Laguna}},\ }\href
  {https://doi.org/10.1086/512067} {\bibfield  {journal} {\bibinfo  {journal}
  {Astrophys. J. Lett.}\ }\textbf {\bibinfo {volume} {656}},\ \bibinfo {pages}
  {L9} (\bibinfo {year} {2007})},\ \Eprint
  {https://arxiv.org/abs/astro-ph/0611110} {arXiv:astro-ph/0611110}
  \BibitemShut {NoStop}%
\bibitem [{\citenamefont {Le~Tiec}\ \emph {et~al.}(2010)\citenamefont
  {Le~Tiec}, \citenamefont {Blanchet},\ and\ \citenamefont
  {Will}}]{LeTiec:2009yg}%
  \BibitemOpen
  \bibfield  {author} {\bibinfo {author} {\bibfnamefont {A.}~\bibnamefont
  {Le~Tiec}}, \bibinfo {author} {\bibfnamefont {L.}~\bibnamefont {Blanchet}},\
  and\ \bibinfo {author} {\bibfnamefont {C.~M.}\ \bibnamefont {Will}},\ }\href
  {https://doi.org/10.1088/0264-9381/27/1/012001} {\bibfield  {journal}
  {\bibinfo  {journal} {Classical Quantum Gravity.}\ }\textbf {\bibinfo
  {volume} {27}},\ \bibinfo {pages} {012001} (\bibinfo {year} {2010})},\
  \Eprint {https://arxiv.org/abs/0910.4594} {arXiv:0910.4594 [gr-qc]}
  \BibitemShut {NoStop}%
\bibitem [{\citenamefont {Baker}\ \emph
  {et~al.}(2006{\natexlab{a}})\citenamefont {Baker}, \citenamefont {Centrella},
  \citenamefont {Choi}, \citenamefont {Koppitz}, \citenamefont {van Meter},\
  and\ \citenamefont {Miller}}]{Baker:2006vn}%
  \BibitemOpen
  \bibfield  {author} {\bibinfo {author} {\bibfnamefont {J.~G.}\ \bibnamefont
  {Baker}}, \bibinfo {author} {\bibfnamefont {J.}~\bibnamefont {Centrella}},
  \bibinfo {author} {\bibfnamefont {D.-I.}\ \bibnamefont {Choi}}, \bibinfo
  {author} {\bibfnamefont {M.}~\bibnamefont {Koppitz}}, \bibinfo {author}
  {\bibfnamefont {J.~R.}\ \bibnamefont {van Meter}},\ and\ \bibinfo {author}
  {\bibfnamefont {M.}~\bibnamefont {Miller}},\ }\href
  {https://doi.org/10.1086/510448} {\bibfield  {journal} {\bibinfo  {journal}
  {Astrophys. J. Lett.}\ }\textbf {\bibinfo {volume} {653}},\ \bibinfo {pages}
  {L93} (\bibinfo {year} {2006}{\natexlab{a}})},\ \Eprint
  {https://arxiv.org/abs/astro-ph/0603204} {arXiv:astro-ph/0603204}
  \BibitemShut {NoStop}%
\bibitem [{\citenamefont {Gonzalez}\ \emph {et~al.}(2007)\citenamefont
  {Gonzalez}, \citenamefont {Sperhake}, \citenamefont {Bruegmann},
  \citenamefont {Hannam},\ and\ \citenamefont {Husa}}]{Gonzalez:2006md}%
  \BibitemOpen
  \bibfield  {author} {\bibinfo {author} {\bibfnamefont {J.~A.}\ \bibnamefont
  {Gonzalez}}, \bibinfo {author} {\bibfnamefont {U.}~\bibnamefont {Sperhake}},
  \bibinfo {author} {\bibfnamefont {B.}~\bibnamefont {Bruegmann}}, \bibinfo
  {author} {\bibfnamefont {M.}~\bibnamefont {Hannam}},\ and\ \bibinfo {author}
  {\bibfnamefont {S.}~\bibnamefont {Husa}},\ }\href
  {https://doi.org/10.1103/PhysRevLett.98.091101} {\bibfield  {journal}
  {\bibinfo  {journal} {Phys. Rev. Lett.}\ }\textbf {\bibinfo {volume} {98}},\
  \bibinfo {pages} {091101} (\bibinfo {year} {2007})},\ \Eprint
  {https://arxiv.org/abs/gr-qc/0610154} {arXiv:gr-qc/0610154} \BibitemShut
  {NoStop}%
\bibitem [{\citenamefont {Herrmann}\ \emph {et~al.}(2007)\citenamefont
  {Herrmann}, \citenamefont {Hinder}, \citenamefont {Shoemaker},\ and\
  \citenamefont {Laguna}}]{Herrmann:2007cwl}%
  \BibitemOpen
  \bibfield  {author} {\bibinfo {author} {\bibfnamefont {F.}~\bibnamefont
  {Herrmann}}, \bibinfo {author} {\bibfnamefont {I.}~\bibnamefont {Hinder}},
  \bibinfo {author} {\bibfnamefont {D.}~\bibnamefont {Shoemaker}},\ and\
  \bibinfo {author} {\bibfnamefont {P.}~\bibnamefont {Laguna}},\ }\href
  {https://doi.org/10.1088/0264-9381/24/12/S04} {\bibfield  {journal} {\bibinfo
   {journal} {Classical Quantum Gravity.}\ }\textbf {\bibinfo {volume} {24}},\
  \bibinfo {pages} {S33} (\bibinfo {year} {2007})}\BibitemShut {NoStop}%
\bibitem [{\citenamefont {Gonz{\'a}lez}\ \emph {et~al.}(2007)\citenamefont
  {Gonz{\'a}lez}, \citenamefont {Hannam}, \citenamefont {Sperhake},
  \citenamefont {Br{\"u}gmann},\ and\ \citenamefont {Husa}}]{Gonzalez:2007hi}%
  \BibitemOpen
  \bibfield  {author} {\bibinfo {author} {\bibfnamefont {J.~A.}\ \bibnamefont
  {Gonz{\'a}lez}}, \bibinfo {author} {\bibfnamefont {M.~D.}\ \bibnamefont
  {Hannam}}, \bibinfo {author} {\bibfnamefont {U.}~\bibnamefont {Sperhake}},
  \bibinfo {author} {\bibfnamefont {B.}~\bibnamefont {Br{\"u}gmann}},\ and\
  \bibinfo {author} {\bibfnamefont {S.}~\bibnamefont {Husa}},\ }\href
  {https://doi.org/10.1103/PhysRevLett.98.231101} {\bibfield  {journal}
  {\bibinfo  {journal} {Phys. Rev. Lett.}\ }\textbf {\bibinfo {volume} {98}},\
  \bibinfo {pages} {231101} (\bibinfo {year} {2007})},\ \Eprint
  {https://arxiv.org/abs/gr-qc/0702052} {arXiv:gr-qc/0702052} \BibitemShut
  {NoStop}%
\bibitem [{\citenamefont {Campanelli}\ \emph
  {et~al.}(2007{\natexlab{a}})\citenamefont {Campanelli}, \citenamefont
  {Lousto}, \citenamefont {Zlochower},\ and\ \citenamefont
  {Merritt}}]{Campanelli:2007cga}%
  \BibitemOpen
  \bibfield  {author} {\bibinfo {author} {\bibfnamefont {M.}~\bibnamefont
  {Campanelli}}, \bibinfo {author} {\bibfnamefont {C.~O.}\ \bibnamefont
  {Lousto}}, \bibinfo {author} {\bibfnamefont {Y.}~\bibnamefont {Zlochower}},\
  and\ \bibinfo {author} {\bibfnamefont {D.}~\bibnamefont {Merritt}},\ }\href
  {https://doi.org/10.1103/PhysRevLett.98.231102} {\bibfield  {journal}
  {\bibinfo  {journal} {Phys. Rev. Lett.}\ }\textbf {\bibinfo {volume} {98}},\
  \bibinfo {pages} {231102} (\bibinfo {year} {2007}{\natexlab{a}})},\ \Eprint
  {https://arxiv.org/abs/gr-qc/0702133} {arXiv:gr-qc/0702133} \BibitemShut
  {NoStop}%
\bibitem [{\citenamefont {Campanelli}\ \emph
  {et~al.}(2007{\natexlab{b}})\citenamefont {Campanelli}, \citenamefont
  {Lousto}, \citenamefont {Zlochower},\ and\ \citenamefont
  {Merritt}}]{Campanelli:2007ew}%
  \BibitemOpen
  \bibfield  {author} {\bibinfo {author} {\bibfnamefont {M.}~\bibnamefont
  {Campanelli}}, \bibinfo {author} {\bibfnamefont {C.~O.}\ \bibnamefont
  {Lousto}}, \bibinfo {author} {\bibfnamefont {Y.}~\bibnamefont {Zlochower}},\
  and\ \bibinfo {author} {\bibfnamefont {D.}~\bibnamefont {Merritt}},\ }\href
  {https://doi.org/10.1086/516712} {\bibfield  {journal} {\bibinfo  {journal}
  {Astrophys. J. Lett.}\ }\textbf {\bibinfo {volume} {659}},\ \bibinfo {pages}
  {L5} (\bibinfo {year} {2007}{\natexlab{b}})},\ \Eprint
  {https://arxiv.org/abs/gr-qc/0701164} {arXiv:gr-qc/0701164 [gr-qc]}
  \BibitemShut {NoStop}%
\bibitem [{\citenamefont {Lousto}\ and\ \citenamefont
  {Zlochower}(2011)}]{Lousto:2011kp}%
  \BibitemOpen
  \bibfield  {author} {\bibinfo {author} {\bibfnamefont {C.~O.}\ \bibnamefont
  {Lousto}}\ and\ \bibinfo {author} {\bibfnamefont {Y.}~\bibnamefont
  {Zlochower}},\ }\href {https://doi.org/10.1103/PhysRevLett.107.231102}
  {\bibfield  {journal} {\bibinfo  {journal} {Phys. Rev. Lett.}\ }\textbf
  {\bibinfo {volume} {107}},\ \bibinfo {pages} {231102} (\bibinfo {year}
  {2011})},\ \Eprint {https://arxiv.org/abs/1108.2009} {arXiv:1108.2009
  [gr-qc]} \BibitemShut {NoStop}%
\bibitem [{\citenamefont {Schnittman}\ and\ \citenamefont
  {Buonanno}(2007)}]{Schnittman:2007sn}%
  \BibitemOpen
  \bibfield  {author} {\bibinfo {author} {\bibfnamefont {J.~D.}\ \bibnamefont
  {Schnittman}}\ and\ \bibinfo {author} {\bibfnamefont {A.}~\bibnamefont
  {Buonanno}},\ }\href {https://doi.org/10.1086/519309} {\bibfield  {journal}
  {\bibinfo  {journal} {Astrophys. J. Lett.}\ }\textbf {\bibinfo {volume}
  {662}},\ \bibinfo {pages} {L63} (\bibinfo {year} {2007})},\ \Eprint
  {https://arxiv.org/abs/astro-ph/0702641} {arXiv:astro-ph/0702641 [ASTRO-PH]}
  \BibitemShut {NoStop}%
\bibitem [{\citenamefont {Dotti}\ \emph {et~al.}(2010)\citenamefont {Dotti},
  \citenamefont {Volonteri}, \citenamefont {Perego}, \citenamefont {Colpi},
  \citenamefont {Ruszkowski},\ and\ \citenamefont {Haardt}}]{Dotti:2009vz}%
  \BibitemOpen
  \bibfield  {author} {\bibinfo {author} {\bibfnamefont {M.}~\bibnamefont
  {Dotti}}, \bibinfo {author} {\bibfnamefont {M.}~\bibnamefont {Volonteri}},
  \bibinfo {author} {\bibfnamefont {A.}~\bibnamefont {Perego}}, \bibinfo
  {author} {\bibfnamefont {M.}~\bibnamefont {Colpi}}, \bibinfo {author}
  {\bibfnamefont {M.}~\bibnamefont {Ruszkowski}},\ and\ \bibinfo {author}
  {\bibfnamefont {F.}~\bibnamefont {Haardt}},\ }\href
  {https://doi.org/10.1111/j.1365-2966.2009.15922.x} {\bibfield  {journal}
  {\bibinfo  {journal} {Mon. Not. R. Astron. Soc.}\ }\textbf {\bibinfo {volume}
  {402}},\ \bibinfo {pages} {682} (\bibinfo {year} {2010})},\ \Eprint
  {https://arxiv.org/abs/0910.5729} {arXiv:0910.5729 [astro-ph.HE]}
  \BibitemShut {NoStop}%
\bibitem [{\citenamefont {Kesden}\ \emph {et~al.}(2010)\citenamefont {Kesden},
  \citenamefont {Sperhake},\ and\ \citenamefont {Berti}}]{Kesden:2010ji}%
  \BibitemOpen
  \bibfield  {author} {\bibinfo {author} {\bibfnamefont {M.}~\bibnamefont
  {Kesden}}, \bibinfo {author} {\bibfnamefont {U.}~\bibnamefont {Sperhake}},\
  and\ \bibinfo {author} {\bibfnamefont {E.}~\bibnamefont {Berti}},\ }\href
  {https://doi.org/10.1088/0004-637X/715/2/1006} {\bibfield  {journal}
  {\bibinfo  {journal} {Astrophys. J.}\ }\textbf {\bibinfo {volume} {715}},\
  \bibinfo {pages} {1006} (\bibinfo {year} {2010})},\ \Eprint
  {https://arxiv.org/abs/1003.4993} {arXiv:1003.4993 [astro-ph.CO]}
  \BibitemShut {NoStop}%
\bibitem [{\citenamefont {Lousto}\ \emph {et~al.}(2012)\citenamefont {Lousto},
  \citenamefont {Zlochower}, \citenamefont {Dotti},\ and\ \citenamefont
  {Volonteri}}]{Lousto:2012su}%
  \BibitemOpen
  \bibfield  {author} {\bibinfo {author} {\bibfnamefont {C.~O.}\ \bibnamefont
  {Lousto}}, \bibinfo {author} {\bibfnamefont {Y.}~\bibnamefont {Zlochower}},
  \bibinfo {author} {\bibfnamefont {M.}~\bibnamefont {Dotti}},\ and\ \bibinfo
  {author} {\bibfnamefont {M.}~\bibnamefont {Volonteri}},\ }\href
  {https://doi.org/10.1103/PhysRevD.85.084015} {\bibfield  {journal} {\bibinfo
  {journal} {Phys. Rev. D}\ }\textbf {\bibinfo {volume} {85}},\ \bibinfo
  {pages} {084015} (\bibinfo {year} {2012})},\ \Eprint
  {https://arxiv.org/abs/1201.1923} {arXiv:1201.1923 [gr-qc]} \BibitemShut
  {NoStop}%
\bibitem [{\citenamefont {Berti}\ \emph {et~al.}(2012)\citenamefont {Berti},
  \citenamefont {Kesden},\ and\ \citenamefont {Sperhake}}]{Berti:2012zp}%
  \BibitemOpen
  \bibfield  {author} {\bibinfo {author} {\bibfnamefont {E.}~\bibnamefont
  {Berti}}, \bibinfo {author} {\bibfnamefont {M.}~\bibnamefont {Kesden}},\ and\
  \bibinfo {author} {\bibfnamefont {U.}~\bibnamefont {Sperhake}},\ }\href
  {https://doi.org/10.1103/PhysRevD.85.124049} {\bibfield  {journal} {\bibinfo
  {journal} {Phys. Rev. D}\ }\textbf {\bibinfo {volume} {85}},\ \bibinfo
  {pages} {124049} (\bibinfo {year} {2012})},\ \Eprint
  {https://arxiv.org/abs/1203.2920} {arXiv:1203.2920 [astro-ph.HE]}
  \BibitemShut {NoStop}%
\bibitem [{\citenamefont {Komossa}(2012)}]{Komossa:2012cy}%
  \BibitemOpen
  \bibfield  {author} {\bibinfo {author} {\bibfnamefont {S.}~\bibnamefont
  {Komossa}},\ }\href {https://doi.org/10.1155/2012/364973} {\bibfield
  {journal} {\bibinfo  {journal} {Adv. Astron.}\ }\textbf {\bibinfo {volume}
  {2012}},\ \bibinfo {pages} {364973} (\bibinfo {year} {2012})},\ \Eprint
  {https://arxiv.org/abs/1202.1977} {arXiv:1202.1977 [astro-ph]} \BibitemShut
  {NoStop}%
\bibitem [{\citenamefont {Colpi}(2014)}]{Colpi:2014poa}%
  \BibitemOpen
  \bibfield  {author} {\bibinfo {author} {\bibfnamefont {M.}~\bibnamefont
  {Colpi}},\ }\href {https://doi.org/10.1007/s11214-014-0067-1} {\bibfield
  {journal} {\bibinfo  {journal} {Space Sci. Rev.}\ }\textbf {\bibinfo {volume}
  {183}},\ \bibinfo {pages} {189} (\bibinfo {year} {2014})},\ \Eprint
  {https://arxiv.org/abs/1407.3102} {arXiv:1407.3102 [astro-ph.GA]}
  \BibitemShut {NoStop}%
\bibitem [{\citenamefont {Blecha}\ \emph {et~al.}(2016)\citenamefont {Blecha},
  \citenamefont {Sijacki}, \citenamefont {Kelley}, \citenamefont {Torrey},
  \citenamefont {Vogelsberger}, \citenamefont {Nelson}, \citenamefont
  {Springel}, \citenamefont {Snyder},\ and\ \citenamefont
  {Hernquist}}]{Blecha:2015baa}%
  \BibitemOpen
  \bibfield  {author} {\bibinfo {author} {\bibfnamefont {L.}~\bibnamefont
  {Blecha}}, \bibinfo {author} {\bibfnamefont {D.}~\bibnamefont {Sijacki}},
  \bibinfo {author} {\bibfnamefont {L.~Z.}\ \bibnamefont {Kelley}}, \bibinfo
  {author} {\bibfnamefont {P.}~\bibnamefont {Torrey}}, \bibinfo {author}
  {\bibfnamefont {M.}~\bibnamefont {Vogelsberger}}, \bibinfo {author}
  {\bibfnamefont {D.}~\bibnamefont {Nelson}}, \bibinfo {author} {\bibfnamefont
  {V.}~\bibnamefont {Springel}}, \bibinfo {author} {\bibfnamefont
  {G.}~\bibnamefont {Snyder}},\ and\ \bibinfo {author} {\bibfnamefont
  {L.}~\bibnamefont {Hernquist}},\ }\href
  {https://doi.org/10.1093/mnras/stv2646} {\bibfield  {journal} {\bibinfo
  {journal} {Mon. Not. R. Astron. Soc.}\ }\textbf {\bibinfo {volume} {456}},\
  \bibinfo {pages} {961} (\bibinfo {year} {2016})},\ \Eprint
  {https://arxiv.org/abs/1508.01524} {arXiv:1508.01524 [astro-ph.GA]}
  \BibitemShut {NoStop}%
\bibitem [{\citenamefont {Barack}\ \emph {et~al.}(2019)\citenamefont {Barack}
  \emph {et~al.}}]{Barack:2018yly}%
  \BibitemOpen
  \bibfield  {author} {\bibinfo {author} {\bibfnamefont {L.}~\bibnamefont
  {Barack}} \emph {et~al.},\ }\href {https://doi.org/10.1088/1361-6382/ab0587}
  {\bibfield  {journal} {\bibinfo  {journal} {Classical Quantum Gravity.}\
  }\textbf {\bibinfo {volume} {36}},\ \bibinfo {pages} {143001} (\bibinfo
  {year} {2019})},\ \Eprint {https://arxiv.org/abs/1806.05195}
  {arXiv:1806.05195 [gr-qc]} \BibitemShut {NoStop}%
\bibitem [{\citenamefont {Merritt}\ \emph {et~al.}(2004)\citenamefont
  {Merritt}, \citenamefont {Milosavljevic}, \citenamefont {Favata},
  \citenamefont {Hughes},\ and\ \citenamefont {Holz}}]{Merritt:2004xa}%
  \BibitemOpen
  \bibfield  {author} {\bibinfo {author} {\bibfnamefont {D.}~\bibnamefont
  {Merritt}}, \bibinfo {author} {\bibfnamefont {M.}~\bibnamefont
  {Milosavljevic}}, \bibinfo {author} {\bibfnamefont {M.}~\bibnamefont
  {Favata}}, \bibinfo {author} {\bibfnamefont {S.~A.}\ \bibnamefont {Hughes}},\
  and\ \bibinfo {author} {\bibfnamefont {D.~E.}\ \bibnamefont {Holz}},\ }\href
  {https://doi.org/10.1086/421551} {\bibfield  {journal} {\bibinfo  {journal}
  {Astrophys. J. Lett.}\ }\textbf {\bibinfo {volume} {607}},\ \bibinfo {pages}
  {L9} (\bibinfo {year} {2004})},\ \Eprint
  {https://arxiv.org/abs/astro-ph/0402057} {arXiv:astro-ph/0402057}
  \BibitemShut {NoStop}%
\bibitem [{\citenamefont {Benacquista}\ and\ \citenamefont
  {Downing}(2013)}]{Benacquista:2011kv}%
  \BibitemOpen
  \bibfield  {author} {\bibinfo {author} {\bibfnamefont {M.~J.}\ \bibnamefont
  {Benacquista}}\ and\ \bibinfo {author} {\bibfnamefont {J.~M.}\ \bibnamefont
  {Downing}},\ }\href {https://doi.org/10.12942/lrr-2013-4} {\bibfield
  {journal} {\bibinfo  {journal} {Living Rev. Relaltivity}\ }\textbf {\bibinfo
  {volume} {16}},\ \bibinfo {pages} {4} (\bibinfo {year} {2013})},\ \Eprint
  {https://arxiv.org/abs/1110.4423} {arXiv:1110.4423 [astro-ph.SR]}
  \BibitemShut {NoStop}%
\bibitem [{\citenamefont {Morawski}\ \emph {et~al.}(2018)\citenamefont
  {Morawski}, \citenamefont {Giersz}, \citenamefont {Askar},\ and\
  \citenamefont {Belczynski}}]{Morawski:2018kfs}%
  \BibitemOpen
  \bibfield  {author} {\bibinfo {author} {\bibfnamefont {J.}~\bibnamefont
  {Morawski}}, \bibinfo {author} {\bibfnamefont {M.}~\bibnamefont {Giersz}},
  \bibinfo {author} {\bibfnamefont {A.}~\bibnamefont {Askar}},\ and\ \bibinfo
  {author} {\bibfnamefont {K.}~\bibnamefont {Belczynski}},\ }\href
  {https://doi.org/10.1093/mnras/sty2401} {\bibfield  {journal} {\bibinfo
  {journal} {Mon. Not. R. Astron. Soc.}\ }\textbf {\bibinfo {volume} {481}},\
  \bibinfo {pages} {2168} (\bibinfo {year} {2018})},\ \Eprint
  {https://arxiv.org/abs/1802.01192} {arXiv:1802.01192 [astro-ph.GA]}
  \BibitemShut {NoStop}%
\bibitem [{\citenamefont {Haiman}(2004)}]{Haiman:2004ve}%
  \BibitemOpen
  \bibfield  {author} {\bibinfo {author} {\bibfnamefont {Z.}~\bibnamefont
  {Haiman}},\ }\href {https://doi.org/10.1086/422910} {\bibfield  {journal}
  {\bibinfo  {journal} {Astrophys. J.}\ }\textbf {\bibinfo {volume} {613}},\
  \bibinfo {pages} {36} (\bibinfo {year} {2004})},\ \Eprint
  {https://arxiv.org/abs/astro-ph/0404196} {arXiv:astro-ph/0404196}
  \BibitemShut {NoStop}%
\bibitem [{\citenamefont {Graham}\ \emph {et~al.}(2020)\citenamefont {Graham},
  \citenamefont {Ford}, \citenamefont {McKernan}, \citenamefont {Ross},
  \citenamefont {Stern}, \citenamefont {Burdge}, \citenamefont {Coughlin} \emph
  {et~al.}}]{Graham:2020gwr}%
  \BibitemOpen
  \bibfield  {author} {\bibinfo {author} {\bibfnamefont {M.~J.}\ \bibnamefont
  {Graham}}, \bibinfo {author} {\bibfnamefont {K.~E.~S.}\ \bibnamefont {Ford}},
  \bibinfo {author} {\bibfnamefont {B.}~\bibnamefont {McKernan}}, \bibinfo
  {author} {\bibfnamefont {N.~P.}\ \bibnamefont {Ross}}, \bibinfo {author}
  {\bibfnamefont {D.}~\bibnamefont {Stern}}, \bibinfo {author} {\bibfnamefont
  {K.}~\bibnamefont {Burdge}}, \bibinfo {author} {\bibfnamefont
  {M.}~\bibnamefont {Coughlin}}, \emph {et~al.},\ }\href
  {https://doi.org/10.1103/PhysRevLett.124.251102} {\bibfield  {journal}
  {\bibinfo  {journal} {Phys. Rev. Lett.}\ }\textbf {\bibinfo {volume} {124}},\
  \bibinfo {pages} {251102} (\bibinfo {year} {2020})},\ \Eprint
  {https://arxiv.org/abs/2006.14122} {arXiv:2006.14122 [astro-ph.HE]}
  \BibitemShut {NoStop}%
\bibitem [{\citenamefont {Chen}\ \emph {et~al.}(2020)\citenamefont {Chen},
  \citenamefont {Haster}, \citenamefont {Vitale}, \citenamefont {Farr},\ and\
  \citenamefont {Isi}}]{Chen:2020gek}%
  \BibitemOpen
  \bibfield  {author} {\bibinfo {author} {\bibfnamefont {H.-Y.}\ \bibnamefont
  {Chen}}, \bibinfo {author} {\bibfnamefont {C.-J.}\ \bibnamefont {Haster}},
  \bibinfo {author} {\bibfnamefont {S.}~\bibnamefont {Vitale}}, \bibinfo
  {author} {\bibfnamefont {W.~M.}\ \bibnamefont {Farr}},\ and\ \bibinfo
  {author} {\bibfnamefont {M.}~\bibnamefont {Isi}},\ }\Eprint
  {https://arxiv.org/abs/2009.14057} {arXiv:2009.14057 [astro-ph.CO]}
  (\bibinfo {year} {2020})\BibitemShut {NoStop}%
\bibitem [{\citenamefont {Samsing}\ and\ \citenamefont
  {Ramirez-Ruiz}(2017)}]{Samsing:2017rat}%
  \BibitemOpen
  \bibfield  {author} {\bibinfo {author} {\bibfnamefont {J.}~\bibnamefont
  {Samsing}}\ and\ \bibinfo {author} {\bibfnamefont {E.}~\bibnamefont
  {Ramirez-Ruiz}},\ }\href {https://doi.org/10.3847/2041-8213/aa6f0b}
  {\bibfield  {journal} {\bibinfo  {journal} {Astrophys. J. Lett.}\ }\textbf
  {\bibinfo {volume} {840}},\ \bibinfo {pages} {L14} (\bibinfo {year}
  {2017})},\ \Eprint {https://arxiv.org/abs/1703.09703} {arXiv:1703.09703
  [astro-ph.HE]} \BibitemShut {NoStop}%
\bibitem [{\citenamefont {Samsing}(2018)}]{Samsing:2017xmd}%
  \BibitemOpen
  \bibfield  {author} {\bibinfo {author} {\bibfnamefont {J.}~\bibnamefont
  {Samsing}},\ }\href {https://doi.org/10.1103/PhysRevD.97.103014} {\bibfield
  {journal} {\bibinfo  {journal} {Phys. Rev. D}\ }\textbf {\bibinfo {volume}
  {97}},\ \bibinfo {pages} {103014} (\bibinfo {year} {2018})},\ \Eprint
  {https://arxiv.org/abs/1711.07452} {arXiv:1711.07452 [astro-ph.HE]}
  \BibitemShut {NoStop}%
\bibitem [{\citenamefont {Samsing}\ \emph {et~al.}(2018)\citenamefont
  {Samsing}, \citenamefont {Askar},\ and\ \citenamefont
  {Giersz}}]{Samsing:2017oij}%
  \BibitemOpen
  \bibfield  {author} {\bibinfo {author} {\bibfnamefont {J.}~\bibnamefont
  {Samsing}}, \bibinfo {author} {\bibfnamefont {A.}~\bibnamefont {Askar}},\
  and\ \bibinfo {author} {\bibfnamefont {M.}~\bibnamefont {Giersz}},\ }\href
  {https://doi.org/10.3847/1538-4357/aaab52} {\bibfield  {journal} {\bibinfo
  {journal} {Astrophys. J.}\ }\textbf {\bibinfo {volume} {855}},\ \bibinfo
  {pages} {124} (\bibinfo {year} {2018})},\ \Eprint
  {https://arxiv.org/abs/1712.06186} {arXiv:1712.06186 [astro-ph.HE]}
  \BibitemShut {NoStop}%
\bibitem [{\citenamefont {Rodriguez}\ \emph {et~al.}(2018)\citenamefont
  {Rodriguez}, \citenamefont {{Amaro-Seoane}}, \citenamefont {Chatterjee},
  \citenamefont {Kremer}, \citenamefont {Rasio}, \citenamefont {Samsing},
  \citenamefont {Ye},\ and\ \citenamefont {Zevin}}]{Rodriguez:2018pss}%
  \BibitemOpen
  \bibfield  {author} {\bibinfo {author} {\bibfnamefont {C.~L.}\ \bibnamefont
  {Rodriguez}}, \bibinfo {author} {\bibfnamefont {P.}~\bibnamefont
  {{Amaro-Seoane}}}, \bibinfo {author} {\bibfnamefont {S.}~\bibnamefont
  {Chatterjee}}, \bibinfo {author} {\bibfnamefont {K.}~\bibnamefont {Kremer}},
  \bibinfo {author} {\bibfnamefont {F.~A.}\ \bibnamefont {Rasio}}, \bibinfo
  {author} {\bibfnamefont {J.}~\bibnamefont {Samsing}}, \bibinfo {author}
  {\bibfnamefont {C.~S.}\ \bibnamefont {Ye}},\ and\ \bibinfo {author}
  {\bibfnamefont {M.}~\bibnamefont {Zevin}},\ }\bibfield  {journal} {\bibinfo
  {journal} {Phys.Rev.D}\ }\textbf {\bibinfo {volume} {98}},\ \href
  {https://doi.org/10.1103/PhysRevD.98.123005} {10.1103/PhysRevD.98.123005}
  (\bibinfo {year} {2018}),\ \Eprint {https://arxiv.org/abs/1811.04926}
  {arXiv:1811.04926} \BibitemShut {NoStop}%
\bibitem [{\citenamefont {Samsing}\ \emph {et~al.}(2020)\citenamefont
  {Samsing}, \citenamefont {Bartos}, \citenamefont {D'Orazio}, \citenamefont
  {Haiman}, \citenamefont {Kocsis}, \citenamefont {Leigh}, \citenamefont {Liu},
  \citenamefont {Pessah},\ and\ \citenamefont {Tagawa}}]{Samsing:2020tda}%
  \BibitemOpen
  \bibfield  {author} {\bibinfo {author} {\bibfnamefont {J.}~\bibnamefont
  {Samsing}}, \bibinfo {author} {\bibfnamefont {I.}~\bibnamefont {Bartos}},
  \bibinfo {author} {\bibfnamefont {D.}~\bibnamefont {D'Orazio}}, \bibinfo
  {author} {\bibfnamefont {Z.}~\bibnamefont {Haiman}}, \bibinfo {author}
  {\bibfnamefont {B.}~\bibnamefont {Kocsis}}, \bibinfo {author} {\bibfnamefont
  {N.}~\bibnamefont {Leigh}}, \bibinfo {author} {\bibfnamefont
  {B.}~\bibnamefont {Liu}}, \bibinfo {author} {\bibfnamefont {M.}~\bibnamefont
  {Pessah}},\ and\ \bibinfo {author} {\bibfnamefont {H.}~\bibnamefont
  {Tagawa}},\ }\Eprint {https://arxiv.org/abs/2010.09765} {arXiv:2010.09765
  [astro-ph.HE]}  (\bibinfo {year} {2020})\BibitemShut {NoStop}%
\bibitem [{\citenamefont {Tagawa}\ \emph {et~al.}(2021)\citenamefont {Tagawa},
  \citenamefont {Kocsis}, \citenamefont {Haiman}, \citenamefont {Bartos},
  \citenamefont {Omukai},\ and\ \citenamefont {Samsing}}]{Tagawa:2020jnc}%
  \BibitemOpen
  \bibfield  {author} {\bibinfo {author} {\bibfnamefont {H.}~\bibnamefont
  {Tagawa}}, \bibinfo {author} {\bibfnamefont {B.}~\bibnamefont {Kocsis}},
  \bibinfo {author} {\bibfnamefont {Z.}~\bibnamefont {Haiman}}, \bibinfo
  {author} {\bibfnamefont {I.}~\bibnamefont {Bartos}}, \bibinfo {author}
  {\bibfnamefont {K.}~\bibnamefont {Omukai}},\ and\ \bibinfo {author}
  {\bibfnamefont {J.}~\bibnamefont {Samsing}},\ }\href
  {https://doi.org/10.3847/2041-8213/abd4d3} {\bibfield  {journal} {\bibinfo
  {journal} {Astrophys.J.Lett.}\ }\textbf {\bibinfo {volume} {907}},\ \bibinfo
  {pages} {L20} (\bibinfo {year} {2021})}\BibitemShut {NoStop}%
\bibitem [{\citenamefont {Cardoso}\ \emph {et~al.}(2021)\citenamefont
  {Cardoso}, \citenamefont {Macedo},\ and\ \citenamefont
  {Vicente}}]{Cardoso:2020iji}%
  \BibitemOpen
  \bibfield  {author} {\bibinfo {author} {\bibfnamefont {V.}~\bibnamefont
  {Cardoso}}, \bibinfo {author} {\bibfnamefont {C.~F.~B.}\ \bibnamefont
  {Macedo}},\ and\ \bibinfo {author} {\bibfnamefont {R.}~\bibnamefont
  {Vicente}},\ }\bibfield  {journal} {\bibinfo  {journal} {Phys. Rev. D}\
  }\textbf {\bibinfo {volume} {103}},\ \href
  {https://doi.org/10.1103/PhysRevD.103.023015} {10.1103/PhysRevD.103.023015}
  (\bibinfo {year} {2021}),\ \Eprint {https://arxiv.org/abs/2010.15151}
  {arXiv:2010.15151} \BibitemShut {NoStop}%
\bibitem [{\citenamefont {Abbott}\ \emph {et~al.}(2019)\citenamefont {Abbott}
  \emph {et~al.}}]{Salemi:2019owp}%
  \BibitemOpen
  \bibfield  {author} {\bibinfo {author} {\bibfnamefont {B.}~\bibnamefont
  {Abbott}} \emph {et~al.} (\bibinfo {collaboration} {LIGO Scientific and Virgo
  Collaborations}),\ }\href {https://doi.org/10.3847/1538-4357/ab3c2d}
  {\bibfield  {journal} {\bibinfo  {journal} {Astrophys. J.}\ }\textbf
  {\bibinfo {volume} {883}},\ \bibinfo {pages} {149} (\bibinfo {year}
  {2019})},\ \Eprint {https://arxiv.org/abs/1907.09384} {arXiv:1907.09384
  [astro-ph.HE]} \BibitemShut {NoStop}%
\bibitem [{\citenamefont {Abbott}\ \emph {et~al.}(2020)\citenamefont {Abbott}
  \emph {et~al.}}]{Abbott:2020tfl}%
  \BibitemOpen
  \bibfield  {author} {\bibinfo {author} {\bibfnamefont {R.}~\bibnamefont
  {Abbott}} \emph {et~al.} (\bibinfo {collaboration} {LIGO Scientific and Virgo
  Collaborations}),\ }\href {https://doi.org/10.1103/PhysRevLett.125.101102}
  {\bibfield  {journal} {\bibinfo  {journal} {Phys. Rev. Lett.}\ }\textbf
  {\bibinfo {volume} {125}},\ \bibinfo {pages} {101102} (\bibinfo {year}
  {2020})},\ \Eprint {https://arxiv.org/abs/2009.01075} {arXiv:2009.01075
  [gr-qc]} \BibitemShut {NoStop}%
\bibitem [{\citenamefont {{Romero-Shaw}}\ \emph {et~al.}(2020)\citenamefont
  {{Romero-Shaw}}, \citenamefont {Lasky}, \citenamefont {Thrane},\ and\
  \citenamefont {Bustillo}}]{Romero-Shaw:2020thy}%
  \BibitemOpen
  \bibfield  {author} {\bibinfo {author} {\bibfnamefont {I.~M.}\ \bibnamefont
  {{Romero-Shaw}}}, \bibinfo {author} {\bibfnamefont {P.~D.}\ \bibnamefont
  {Lasky}}, \bibinfo {author} {\bibfnamefont {E.}~\bibnamefont {Thrane}},\ and\
  \bibinfo {author} {\bibfnamefont {J.~C.}\ \bibnamefont {Bustillo}},\ }\href
  {https://doi.org/10.3847/2041-8213/abbe26} {\bibfield  {journal} {\bibinfo
  {journal} {Astrophys.J.Lett.}\ }\textbf {\bibinfo {volume} {903}},\ \bibinfo
  {pages} {L5} (\bibinfo {year} {2020})},\ \Eprint
  {https://arxiv.org/abs/2009.04771} {arXiv:2009.04771} \BibitemShut {NoStop}%
\bibitem [{\citenamefont {Gayathri}\ \emph {et~al.}(2020)\citenamefont
  {Gayathri}, \citenamefont {Healy}, \citenamefont {Lange}, \citenamefont
  {O'Brien}, \citenamefont {Szczepanczyk}, \citenamefont {Bartos},
  \citenamefont {Campanelli}, \citenamefont {Klimenko}, \citenamefont
  {Lousto},\ and\ \citenamefont {O'Shaughnessy}}]{Gayathri:2020coq}%
  \BibitemOpen
  \bibfield  {author} {\bibinfo {author} {\bibfnamefont {V.}~\bibnamefont
  {Gayathri}}, \bibinfo {author} {\bibfnamefont {J.}~\bibnamefont {Healy}},
  \bibinfo {author} {\bibfnamefont {J.}~\bibnamefont {Lange}}, \bibinfo
  {author} {\bibfnamefont {B.}~\bibnamefont {O'Brien}}, \bibinfo {author}
  {\bibfnamefont {M.}~\bibnamefont {Szczepanczyk}}, \bibinfo {author}
  {\bibfnamefont {I.}~\bibnamefont {Bartos}}, \bibinfo {author} {\bibfnamefont
  {M.}~\bibnamefont {Campanelli}}, \bibinfo {author} {\bibfnamefont
  {S.}~\bibnamefont {Klimenko}}, \bibinfo {author} {\bibfnamefont
  {C.}~\bibnamefont {Lousto}},\ and\ \bibinfo {author} {\bibfnamefont
  {R.}~\bibnamefont {O'Shaughnessy}},\ }\Eprint
  {https://arxiv.org/abs/2009.05461} {arXiv:2009.05461 [astro-ph.HE]}
  (\bibinfo {year} {2020})\BibitemShut {NoStop}%
\bibitem [{\citenamefont {Nishizawa}\ \emph {et~al.}(2016)\citenamefont
  {Nishizawa}, \citenamefont {Berti}, \citenamefont {Klein},\ and\
  \citenamefont {Sesana}}]{Nishizawa:2016jji}%
  \BibitemOpen
  \bibfield  {author} {\bibinfo {author} {\bibfnamefont {A.}~\bibnamefont
  {Nishizawa}}, \bibinfo {author} {\bibfnamefont {E.}~\bibnamefont {Berti}},
  \bibinfo {author} {\bibfnamefont {A.}~\bibnamefont {Klein}},\ and\ \bibinfo
  {author} {\bibfnamefont {A.}~\bibnamefont {Sesana}},\ }\href
  {https://doi.org/10.1103/PhysRevD.94.064020} {\bibfield  {journal} {\bibinfo
  {journal} {Phys. Rev. D}\ }\textbf {\bibinfo {volume} {94}},\ \bibinfo
  {pages} {064020} (\bibinfo {year} {2016})},\ \Eprint
  {https://arxiv.org/abs/1605.01341} {arXiv:1605.01341 [gr-qc]} \BibitemShut
  {NoStop}%
\bibitem [{\citenamefont {Breivik}\ \emph {et~al.}(2016)\citenamefont
  {Breivik}, \citenamefont {Rodriguez}, \citenamefont {Larson}, \citenamefont
  {Kalogera},\ and\ \citenamefont {Rasio}}]{Breivik:2016ddj}%
  \BibitemOpen
  \bibfield  {author} {\bibinfo {author} {\bibfnamefont {K.}~\bibnamefont
  {Breivik}}, \bibinfo {author} {\bibfnamefont {C.~L.}\ \bibnamefont
  {Rodriguez}}, \bibinfo {author} {\bibfnamefont {S.~L.}\ \bibnamefont
  {Larson}}, \bibinfo {author} {\bibfnamefont {V.}~\bibnamefont {Kalogera}},\
  and\ \bibinfo {author} {\bibfnamefont {F.~A.}\ \bibnamefont {Rasio}},\ }\href
  {https://doi.org/10.3847/2041-8205/830/1/L18} {\bibfield  {journal} {\bibinfo
   {journal} {Astrophys. J. Lett.}\ }\textbf {\bibinfo {volume} {830}},\
  \bibinfo {pages} {L18} (\bibinfo {year} {2016})},\ \Eprint
  {https://arxiv.org/abs/1606.09558} {arXiv:1606.09558 [astro-ph.GA]}
  \BibitemShut {NoStop}%
\bibitem [{\citenamefont {Nishizawa}\ \emph {et~al.}(2017)\citenamefont
  {Nishizawa}, \citenamefont {Sesana}, \citenamefont {Berti},\ and\
  \citenamefont {Klein}}]{Nishizawa:2016eza}%
  \BibitemOpen
  \bibfield  {author} {\bibinfo {author} {\bibfnamefont {A.}~\bibnamefont
  {Nishizawa}}, \bibinfo {author} {\bibfnamefont {A.}~\bibnamefont {Sesana}},
  \bibinfo {author} {\bibfnamefont {E.}~\bibnamefont {Berti}},\ and\ \bibinfo
  {author} {\bibfnamefont {A.}~\bibnamefont {Klein}},\ }\href
  {https://doi.org/10.1093/mnras/stw2993} {\bibfield  {journal} {\bibinfo
  {journal} {Mon. Not. R. Astron. Soc.}\ }\textbf {\bibinfo {volume} {465}},\
  \bibinfo {pages} {4375} (\bibinfo {year} {2017})},\ \Eprint
  {https://arxiv.org/abs/1606.09295} {arXiv:1606.09295 [astro-ph.HE]}
  \BibitemShut {NoStop}%
\bibitem [{\citenamefont {Roedig}\ and\ \citenamefont
  {Sesana}(2012)}]{Roedig:2011rn}%
  \BibitemOpen
  \bibfield  {author} {\bibinfo {author} {\bibfnamefont {C.}~\bibnamefont
  {Roedig}}\ and\ \bibinfo {author} {\bibfnamefont {A.}~\bibnamefont
  {Sesana}},\ }\bibfield  {booktitle} {\emph {\bibinfo {booktitle}
  {{Gravitational waves. Numerical relativity - data analysis. Proceedings, 9th
  Edoardo Amaldi Conference, Amaldi 9, and meeting, NRDA 2011, Cardiff, UK,
  July 10-15, 2011}}},\ }\href {https://doi.org/10.1088/1742-6596/363/1/012035}
  {\bibfield  {journal} {\bibinfo  {journal} {J. Phys. Conf. Ser.}\ }\textbf
  {\bibinfo {volume} {363}},\ \bibinfo {pages} {012035} (\bibinfo {year}
  {2012})},\ \Eprint {https://arxiv.org/abs/1111.3742} {arXiv:1111.3742
  [astro-ph.CO]} \BibitemShut {NoStop}%
\bibitem [{\citenamefont {Bonetti}\ \emph {et~al.}(2019)\citenamefont
  {Bonetti}, \citenamefont {Sesana}, \citenamefont {Haardt}, \citenamefont
  {Barausse},\ and\ \citenamefont {Colpi}}]{Bonetti:2018tpf}%
  \BibitemOpen
  \bibfield  {author} {\bibinfo {author} {\bibfnamefont {M.}~\bibnamefont
  {Bonetti}}, \bibinfo {author} {\bibfnamefont {A.}~\bibnamefont {Sesana}},
  \bibinfo {author} {\bibfnamefont {F.}~\bibnamefont {Haardt}}, \bibinfo
  {author} {\bibfnamefont {E.}~\bibnamefont {Barausse}},\ and\ \bibinfo
  {author} {\bibfnamefont {M.}~\bibnamefont {Colpi}},\ }\href
  {https://doi.org/10.1093/mnras/stz903} {\bibfield  {journal} {\bibinfo
  {journal} {Mon. Not. Roy. Astron. Soc.}\ }\textbf {\bibinfo {volume} {486}},\
  \bibinfo {pages} {4044} (\bibinfo {year} {2019})},\ \Eprint
  {https://arxiv.org/abs/1812.01011} {arXiv:1812.01011 [astro-ph.GA]}
  \BibitemShut {NoStop}%
\bibitem [{\citenamefont {Huerta}\ \emph {et~al.}(2019)\citenamefont {Huerta}
  \emph {et~al.}}]{Huerta:2019oxn}%
  \BibitemOpen
  \bibfield  {author} {\bibinfo {author} {\bibfnamefont {E.}~\bibnamefont
  {Huerta}} \emph {et~al.},\ }\href
  {https://doi.org/10.1103/PhysRevD.100.064003} {\bibfield  {journal} {\bibinfo
   {journal} {Phys. Rev. D}\ }\textbf {\bibinfo {volume} {100}},\ \bibinfo
  {pages} {064003} (\bibinfo {year} {2019})},\ \Eprint
  {https://arxiv.org/abs/1901.07038} {arXiv:1901.07038 [gr-qc]} \BibitemShut
  {NoStop}%
\bibitem [{\citenamefont {Clough}\ \emph {et~al.}(2015)\citenamefont {Clough},
  \citenamefont {Figueras}, \citenamefont {Finkel}, \citenamefont {Kunesch},
  \citenamefont {Lim},\ and\ \citenamefont {Tunyasuvunakool}}]{Clough:2015sqa}%
  \BibitemOpen
  \bibfield  {author} {\bibinfo {author} {\bibfnamefont {K.}~\bibnamefont
  {Clough}}, \bibinfo {author} {\bibfnamefont {P.}~\bibnamefont {Figueras}},
  \bibinfo {author} {\bibfnamefont {H.}~\bibnamefont {Finkel}}, \bibinfo
  {author} {\bibfnamefont {M.}~\bibnamefont {Kunesch}}, \bibinfo {author}
  {\bibfnamefont {E.~A.}\ \bibnamefont {Lim}},\ and\ \bibinfo {author}
  {\bibfnamefont {S.}~\bibnamefont {Tunyasuvunakool}},\ }\href
  {https://doi.org/10.1088/0264-9381/32/24/245011} {\bibfield  {journal}
  {\bibinfo  {journal} {Classical Quantum Gravity.}\ }\textbf {\bibinfo
  {volume} {32}},\ \bibinfo {pages} {245011} (\bibinfo {year} {2015})},\
  \Eprint {https://arxiv.org/abs/1503.03436} {arXiv:1503.03436 [gr-qc]}
  \BibitemShut {NoStop}%
\bibitem [{GRC()}]{GRChomboWebsite}%
  \BibitemOpen
  \href@noop {} {}\bibinfo {note} {\url{http://www.grchombo.org/}}\BibitemShut
  {NoStop}%
\bibitem [{\citenamefont {Sperhake}(2007)}]{Sperhake:2006cy}%
  \BibitemOpen
  \bibfield  {author} {\bibinfo {author} {\bibfnamefont {U.}~\bibnamefont
  {Sperhake}},\ }\href {https://doi.org/10.1103/PhysRevD.76.104015} {\bibfield
  {journal} {\bibinfo  {journal} {Phys. Rev. D}\ }\textbf {\bibinfo {volume}
  {76}},\ \bibinfo {pages} {104015} (\bibinfo {year} {2007})},\ \Eprint
  {https://arxiv.org/abs/gr-qc/0606079} {arXiv:gr-qc/0606079} \BibitemShut
  {NoStop}%
\bibitem [{\citenamefont {Husa}\ \emph {et~al.}(2008)\citenamefont {Husa},
  \citenamefont {Gonzalez}, \citenamefont {Hannam}, \citenamefont {Bruegmann},\
  and\ \citenamefont {Sperhake}}]{Husa:2007hp}%
  \BibitemOpen
  \bibfield  {author} {\bibinfo {author} {\bibfnamefont {S.}~\bibnamefont
  {Husa}}, \bibinfo {author} {\bibfnamefont {J.~A.}\ \bibnamefont {Gonzalez}},
  \bibinfo {author} {\bibfnamefont {M.}~\bibnamefont {Hannam}}, \bibinfo
  {author} {\bibfnamefont {B.}~\bibnamefont {Bruegmann}},\ and\ \bibinfo
  {author} {\bibfnamefont {U.}~\bibnamefont {Sperhake}},\ }\href
  {https://doi.org/10.1088/0264-9381/25/10/105006} {\bibfield  {journal}
  {\bibinfo  {journal} {Classical Quantum Gravity.}\ }\textbf {\bibinfo
  {volume} {25}},\ \bibinfo {pages} {105006} (\bibinfo {year} {2008})},\
  \Eprint {https://arxiv.org/abs/0706.0740} {arXiv:0706.0740 [gr-qc]}
  \BibitemShut {NoStop}%
\bibitem [{\citenamefont {Alic}\ \emph {et~al.}(2012)\citenamefont {Alic},
  \citenamefont {Bona-Casas}, \citenamefont {Bona}, \citenamefont {Rezzolla},\
  and\ \citenamefont {Palenzuela}}]{Alic:2011gg}%
  \BibitemOpen
  \bibfield  {author} {\bibinfo {author} {\bibfnamefont {D.}~\bibnamefont
  {Alic}}, \bibinfo {author} {\bibfnamefont {C.}~\bibnamefont {Bona-Casas}},
  \bibinfo {author} {\bibfnamefont {C.}~\bibnamefont {Bona}}, \bibinfo {author}
  {\bibfnamefont {L.}~\bibnamefont {Rezzolla}},\ and\ \bibinfo {author}
  {\bibfnamefont {C.}~\bibnamefont {Palenzuela}},\ }\href
  {https://doi.org/10.1103/PhysRevD.85.064040} {\bibfield  {journal} {\bibinfo
  {journal} {Phys. Rev. D}\ }\textbf {\bibinfo {volume} {85}},\ \bibinfo
  {pages} {064040} (\bibinfo {year} {2012})},\ \Eprint
  {https://arxiv.org/abs/1106.2254} {arXiv:1106.2254 [gr-qc]} \BibitemShut
  {NoStop}%
\bibitem [{\citenamefont {Alic}\ \emph {et~al.}(2013)\citenamefont {Alic},
  \citenamefont {Kastaun},\ and\ \citenamefont {Rezzolla}}]{Alic:2013xsa}%
  \BibitemOpen
  \bibfield  {author} {\bibinfo {author} {\bibfnamefont {D.}~\bibnamefont
  {Alic}}, \bibinfo {author} {\bibfnamefont {W.}~\bibnamefont {Kastaun}},\ and\
  \bibinfo {author} {\bibfnamefont {L.}~\bibnamefont {Rezzolla}},\ }\href
  {https://doi.org/10.1103/PhysRevD.88.064049} {\bibfield  {journal} {\bibinfo
  {journal} {Phys. Rev. D}\ }\textbf {\bibinfo {volume} {88}},\ \bibinfo
  {pages} {064049} (\bibinfo {year} {2013})},\ \Eprint
  {https://arxiv.org/abs/1307.7391} {arXiv:1307.7391 [gr-qc]} \BibitemShut
  {NoStop}%
\bibitem [{\citenamefont {Adams}\ \emph {et~al.}(2019)\citenamefont {Adams}
  \emph {et~al.}}]{ChomboReport}%
  \BibitemOpen
  \bibfield  {author} {\bibinfo {author} {\bibfnamefont {M.}~\bibnamefont
  {Adams}} \emph {et~al.},\ }\href
  {https://commons.lbl.gov/download/attachments/73468344/chomboDesign.pdf?version=1&modificationDate=1554672305006&api=v2}
  {\emph {\bibinfo {title} {{Chombo Software Package for AMR Applications -
  Design Document}}}},\ \bibinfo {type} {Tech. Rep.}\ \bibinfo {number}
  {LBNL-6616E}\ (\bibinfo  {institution} {Lawrence Berkeley National
  Laboratory},\ \bibinfo {year} {2019})\BibitemShut {NoStop}%
\bibitem [{\citenamefont {{Berger}}\ and\ \citenamefont
  {{Rigoutsos}}(1991)}]{Berger1991}%
  \BibitemOpen
  \bibfield  {author} {\bibinfo {author} {\bibfnamefont {M.}~\bibnamefont
  {{Berger}}}\ and\ \bibinfo {author} {\bibfnamefont {I.}~\bibnamefont
  {{Rigoutsos}}},\ }\href {https://doi.org/10.1109/21.120081} {\bibfield
  {journal} {\bibinfo  {journal} {IEEE Trans. Sys. Man \& Cybernet.}\ }\textbf
  {\bibinfo {volume} {21}},\ \bibinfo {pages} {1278} (\bibinfo {year}
  {1991})}\BibitemShut {NoStop}%
\bibitem [{\citenamefont {Goodale}\ \emph {et~al.}(2003)\citenamefont
  {Goodale}, \citenamefont {Allen}, \citenamefont {Lanfermann}, \citenamefont
  {Mass{\'o}}, \citenamefont {Radke}, \citenamefont {Seidel},\ and\
  \citenamefont {Shalf}}]{Goodale2002a}%
  \BibitemOpen
  \bibfield  {author} {\bibinfo {author} {\bibfnamefont {T.}~\bibnamefont
  {Goodale}}, \bibinfo {author} {\bibfnamefont {G.}~\bibnamefont {Allen}},
  \bibinfo {author} {\bibfnamefont {G.}~\bibnamefont {Lanfermann}}, \bibinfo
  {author} {\bibfnamefont {J.}~\bibnamefont {Mass{\'o}}}, \bibinfo {author}
  {\bibfnamefont {T.}~\bibnamefont {Radke}}, \bibinfo {author} {\bibfnamefont
  {E.}~\bibnamefont {Seidel}},\ and\ \bibinfo {author} {\bibfnamefont
  {J.}~\bibnamefont {Shalf}},\ }in\ \href {http://edoc.mpg.de/3341} {\emph
  {\bibinfo {booktitle} {Vector and Parallel Processing -- VECPAR'2002, 5th
  International Conference, Lecture Notes in Computer Science}}}\ (\bibinfo
  {publisher} {Springer},\ \bibinfo {address} {Berlin},\ \bibinfo {year}
  {2003})\BibitemShut {NoStop}%
\bibitem [{\citenamefont {Nakamura}\ \emph {et~al.}(1987)\citenamefont
  {Nakamura}, \citenamefont {Oohara},\ and\ \citenamefont
  {Kojima}}]{Nakamura:1987zz}%
  \BibitemOpen
  \bibfield  {author} {\bibinfo {author} {\bibfnamefont {T.}~\bibnamefont
  {Nakamura}}, \bibinfo {author} {\bibfnamefont {K.}~\bibnamefont {Oohara}},\
  and\ \bibinfo {author} {\bibfnamefont {Y.}~\bibnamefont {Kojima}},\ }\href
  {https://doi.org/10.1143/PTPS.90.1} {\bibfield  {journal} {\bibinfo
  {journal} {Prog. Theor. Phys. Suppl.}\ }\textbf {\bibinfo {volume} {90}},\
  \bibinfo {pages} {1} (\bibinfo {year} {1987})}\BibitemShut {NoStop}%
\bibitem [{\citenamefont {Shibata}\ and\ \citenamefont
  {Nakamura}(1995)}]{Shibata:1995we}%
  \BibitemOpen
  \bibfield  {author} {\bibinfo {author} {\bibfnamefont {M.}~\bibnamefont
  {Shibata}}\ and\ \bibinfo {author} {\bibfnamefont {T.}~\bibnamefont
  {Nakamura}},\ }\href {https://doi.org/10.1103/PhysRevD.52.5428} {\bibfield
  {journal} {\bibinfo  {journal} {Phys. Rev. D}\ }\textbf {\bibinfo {volume}
  {52}},\ \bibinfo {pages} {5428} (\bibinfo {year} {1995})}\BibitemShut
  {NoStop}%
\bibitem [{\citenamefont {Baumgarte}\ and\ \citenamefont
  {Shapiro}(1998)}]{Baumgarte:1998te}%
  \BibitemOpen
  \bibfield  {author} {\bibinfo {author} {\bibfnamefont {T.~W.}\ \bibnamefont
  {Baumgarte}}\ and\ \bibinfo {author} {\bibfnamefont {S.~L.}\ \bibnamefont
  {Shapiro}},\ }\href {https://doi.org/10.1103/PhysRevD.59.024007} {\bibfield
  {journal} {\bibinfo  {journal} {Phys. Rev. D}\ }\textbf {\bibinfo {volume}
  {59}},\ \bibinfo {pages} {024007} (\bibinfo {year} {1998})},\ \Eprint
  {https://arxiv.org/abs/gr-qc/9810065} {arXiv:gr-qc/9810065} \BibitemShut
  {NoStop}%
\bibitem [{\citenamefont {Campanelli}\ \emph
  {et~al.}(2006{\natexlab{b}})\citenamefont {Campanelli}, \citenamefont
  {Lousto}, \citenamefont {Marronetti},\ and\ \citenamefont
  {Zlochower}}]{Campanelli:2005dd}%
  \BibitemOpen
  \bibfield  {author} {\bibinfo {author} {\bibfnamefont {M.}~\bibnamefont
  {Campanelli}}, \bibinfo {author} {\bibfnamefont {C.}~\bibnamefont {Lousto}},
  \bibinfo {author} {\bibfnamefont {P.}~\bibnamefont {Marronetti}},\ and\
  \bibinfo {author} {\bibfnamefont {Y.}~\bibnamefont {Zlochower}},\ }\href
  {https://doi.org/10.1103/PhysRevLett.96.111101} {\bibfield  {journal}
  {\bibinfo  {journal} {Phys. Rev. Lett.}\ }\textbf {\bibinfo {volume} {96}},\
  \bibinfo {pages} {111101} (\bibinfo {year} {2006}{\natexlab{b}})},\ \Eprint
  {https://arxiv.org/abs/gr-qc/0511048} {arXiv:gr-qc/0511048} \BibitemShut
  {NoStop}%
\bibitem [{\citenamefont {Baker}\ \emph
  {et~al.}(2006{\natexlab{b}})\citenamefont {Baker}, \citenamefont {Centrella},
  \citenamefont {Choi}, \citenamefont {Koppitz},\ and\ \citenamefont {van
  Meter}}]{Baker:2005vv}%
  \BibitemOpen
  \bibfield  {author} {\bibinfo {author} {\bibfnamefont {J.~G.}\ \bibnamefont
  {Baker}}, \bibinfo {author} {\bibfnamefont {J.}~\bibnamefont {Centrella}},
  \bibinfo {author} {\bibfnamefont {D.-I.}\ \bibnamefont {Choi}}, \bibinfo
  {author} {\bibfnamefont {M.}~\bibnamefont {Koppitz}},\ and\ \bibinfo {author}
  {\bibfnamefont {J.}~\bibnamefont {van Meter}},\ }\href
  {https://doi.org/10.1103/PhysRevLett.96.111102} {\bibfield  {journal}
  {\bibinfo  {journal} {Phys. Rev. Lett.}\ }\textbf {\bibinfo {volume} {96}},\
  \bibinfo {pages} {111102} (\bibinfo {year} {2006}{\natexlab{b}})},\ \Eprint
  {https://arxiv.org/abs/gr-qc/0511103} {arXiv:gr-qc/0511103} \BibitemShut
  {NoStop}%
\bibitem [{\citenamefont {Schnetter}\ \emph {et~al.}(2004)\citenamefont
  {Schnetter}, \citenamefont {Hawley},\ and\ \citenamefont
  {Hawke}}]{Schnetter:2003rb}%
  \BibitemOpen
  \bibfield  {author} {\bibinfo {author} {\bibfnamefont {E.}~\bibnamefont
  {Schnetter}}, \bibinfo {author} {\bibfnamefont {S.~H.}\ \bibnamefont
  {Hawley}},\ and\ \bibinfo {author} {\bibfnamefont {I.}~\bibnamefont
  {Hawke}},\ }\href {https://doi.org/10.1088/0264-9381/21/6/014} {\bibfield
  {journal} {\bibinfo  {journal} {Classical Quantum Gravity.}\ }\textbf
  {\bibinfo {volume} {21}},\ \bibinfo {pages} {1465} (\bibinfo {year}
  {2004})},\ \Eprint {https://arxiv.org/abs/gr-qc/0310042}
  {arXiv:gr-qc/0310042} \BibitemShut {NoStop}%
\bibitem [{\citenamefont {Thornburg}(1996)}]{Thornburg:1995cp}%
  \BibitemOpen
  \bibfield  {author} {\bibinfo {author} {\bibfnamefont {J.}~\bibnamefont
  {Thornburg}},\ }\href {https://doi.org/10.1103/PhysRevD.54.4899} {\bibfield
  {journal} {\bibinfo  {journal} {Phys. Rev. D}\ }\textbf {\bibinfo {volume}
  {54}},\ \bibinfo {pages} {4899} (\bibinfo {year} {1996})},\ \Eprint
  {https://arxiv.org/abs/gr-qc/9508014} {gr-qc/9508014} \BibitemShut {NoStop}%
\bibitem [{\citenamefont {Thornburg}(2004)}]{Thornburg:2003sf}%
  \BibitemOpen
  \bibfield  {author} {\bibinfo {author} {\bibfnamefont {J.}~\bibnamefont
  {Thornburg}},\ }\href {https://doi.org/10.1088/0264-9381/21/2/026} {\bibfield
   {journal} {\bibinfo  {journal} {Classical Quantum Gravity.}\ }\textbf
  {\bibinfo {volume} {21}},\ \bibinfo {pages} {743} (\bibinfo {year} {2004})},\
  \Eprint {https://arxiv.org/abs/gr-qc/0306056} {arXiv:gr-qc/0306056}
  \BibitemShut {NoStop}%
\bibitem [{\citenamefont {Brandt}\ and\ \citenamefont
  {Bruegmann}(1997)}]{Brandt:1997tf}%
  \BibitemOpen
  \bibfield  {author} {\bibinfo {author} {\bibfnamefont {S.}~\bibnamefont
  {Brandt}}\ and\ \bibinfo {author} {\bibfnamefont {B.}~\bibnamefont
  {Bruegmann}},\ }\href {https://doi.org/10.1103/PhysRevLett.78.3606}
  {\bibfield  {journal} {\bibinfo  {journal} {Phys. Rev. Lett.}\ }\textbf
  {\bibinfo {volume} {78}},\ \bibinfo {pages} {3606} (\bibinfo {year}
  {1997})},\ \Eprint {https://arxiv.org/abs/gr-qc/9703066}
  {arXiv:gr-qc/9703066} \BibitemShut {NoStop}%
\bibitem [{\citenamefont {Bowen}\ and\ \citenamefont
  {York}(1980)}]{Bowen:1980yu}%
  \BibitemOpen
  \bibfield  {author} {\bibinfo {author} {\bibfnamefont {J.~M.}\ \bibnamefont
  {Bowen}}\ and\ \bibinfo {author} {\bibfnamefont {J.~W.}\ \bibnamefont
  {York}},\ }\href {https://doi.org/10.1103/PhysRevD.21.2047} {\bibfield
  {journal} {\bibinfo  {journal} {Phys. Rev. D}\ }\textbf {\bibinfo {volume}
  {21}},\ \bibinfo {pages} {2047} (\bibinfo {year} {1980})}\BibitemShut
  {NoStop}%
\bibitem [{\citenamefont {Ansorg}\ \emph {et~al.}(2004)\citenamefont {Ansorg},
  \citenamefont {Bruegmann},\ and\ \citenamefont {Tichy}}]{Ansorg:2004ds}%
  \BibitemOpen
  \bibfield  {author} {\bibinfo {author} {\bibfnamefont {M.}~\bibnamefont
  {Ansorg}}, \bibinfo {author} {\bibfnamefont {B.}~\bibnamefont {Bruegmann}},\
  and\ \bibinfo {author} {\bibfnamefont {W.}~\bibnamefont {Tichy}},\ }\href
  {https://doi.org/10.1103/PhysRevD.70.064011} {\bibfield  {journal} {\bibinfo
  {journal} {Phys. Rev. D}\ }\textbf {\bibinfo {volume} {70}},\ \bibinfo
  {pages} {064011} (\bibinfo {year} {2004})},\ \Eprint
  {https://arxiv.org/abs/gr-qc/0404056} {arXiv:gr-qc/0404056} \BibitemShut
  {NoStop}%
\bibitem [{\citenamefont {Paschalidis}\ \emph {et~al.}(2013)\citenamefont
  {Paschalidis}, \citenamefont {Etienne}, \citenamefont {Gold},\ and\
  \citenamefont {Shapiro}}]{Paschalidis:2013oya}%
  \BibitemOpen
  \bibfield  {author} {\bibinfo {author} {\bibfnamefont {V.}~\bibnamefont
  {Paschalidis}}, \bibinfo {author} {\bibfnamefont {Z.~B.}\ \bibnamefont
  {Etienne}}, \bibinfo {author} {\bibfnamefont {R.}~\bibnamefont {Gold}},\ and\
  \bibinfo {author} {\bibfnamefont {S.~L.}\ \bibnamefont {Shapiro}},\ }\Eprint
  {https://arxiv.org/abs/1304.0457} {arXiv:1304.0457 [gr-qc]}  (\bibinfo {year}
  {2013})\BibitemShut {NoStop}%
\bibitem [{\citenamefont {Sperhake}\ \emph {et~al.}(2008)\citenamefont
  {Sperhake}, \citenamefont {Berti}, \citenamefont {Cardoso}, \citenamefont
  {Gonzalez}, \citenamefont {Bruegmann},\ and\ \citenamefont
  {Ansorg}}]{Sperhake:2007gu}%
  \BibitemOpen
  \bibfield  {author} {\bibinfo {author} {\bibfnamefont {U.}~\bibnamefont
  {Sperhake}}, \bibinfo {author} {\bibfnamefont {E.}~\bibnamefont {Berti}},
  \bibinfo {author} {\bibfnamefont {V.}~\bibnamefont {Cardoso}}, \bibinfo
  {author} {\bibfnamefont {J.~A.}\ \bibnamefont {Gonzalez}}, \bibinfo {author}
  {\bibfnamefont {B.}~\bibnamefont {Bruegmann}},\ and\ \bibinfo {author}
  {\bibfnamefont {M.}~\bibnamefont {Ansorg}},\ }\href
  {https://doi.org/10.1103/PhysRevD.78.064069} {\bibfield  {journal} {\bibinfo
  {journal} {Phys. Rev. D}\ }\textbf {\bibinfo {volume} {78}},\ \bibinfo
  {pages} {064069} (\bibinfo {year} {2008})},\ \Eprint
  {https://arxiv.org/abs/0710.3823} {arXiv:0710.3823 [gr-qc]} \BibitemShut
  {NoStop}%
\bibitem [{\citenamefont {Arnowitt}\ \emph {et~al.}(1962)\citenamefont
  {Arnowitt}, \citenamefont {Deser},\ and\ \citenamefont
  {Misner}}]{Arnowitt:1962hi}%
  \BibitemOpen
  \bibfield  {author} {\bibinfo {author} {\bibfnamefont {R.}~\bibnamefont
  {Arnowitt}}, \bibinfo {author} {\bibfnamefont {S.}~\bibnamefont {Deser}},\
  and\ \bibinfo {author} {\bibfnamefont {C.~W.}\ \bibnamefont {Misner}},\ }in\
  \href@noop {} {\emph {\bibinfo {booktitle} {Gravitation an Introduction to
  Current Research}}},\ \bibinfo {editor} {edited by\ \bibinfo {editor}
  {\bibfnamefont {L.}~\bibnamefont {Witten}}}\ (\bibinfo  {publisher} {John
  Wiley, New York},\ \bibinfo {year} {1962})\ pp.\ \bibinfo {pages}
  {227--265},\ \Eprint {https://arxiv.org/abs/gr-qc/0405109}
  {arXiv:gr-qc/0405109} \BibitemShut {NoStop}%
\bibitem [{\citenamefont {Bruegmann}\ \emph {et~al.}(2008)\citenamefont
  {Bruegmann}, \citenamefont {Gonzalez}, \citenamefont {Hannam}, \citenamefont
  {Husa}, \citenamefont {Sperhake},\ and\ \citenamefont
  {Tichy}}]{Brugmann:2008zz}%
  \BibitemOpen
  \bibfield  {author} {\bibinfo {author} {\bibfnamefont {B.}~\bibnamefont
  {Bruegmann}}, \bibinfo {author} {\bibfnamefont {J.~A.}\ \bibnamefont
  {Gonzalez}}, \bibinfo {author} {\bibfnamefont {M.}~\bibnamefont {Hannam}},
  \bibinfo {author} {\bibfnamefont {S.}~\bibnamefont {Husa}}, \bibinfo {author}
  {\bibfnamefont {U.}~\bibnamefont {Sperhake}},\ and\ \bibinfo {author}
  {\bibfnamefont {W.}~\bibnamefont {Tichy}},\ }\href
  {https://doi.org/10.1103/PhysRevD.77.024027} {\bibfield  {journal} {\bibinfo
  {journal} {Phys. Rev. D}\ }\textbf {\bibinfo {volume} {77}},\ \bibinfo
  {pages} {024027} (\bibinfo {year} {2008})},\ \Eprint
  {https://arxiv.org/abs/gr-qc/0610128} {arXiv:gr-qc/0610128} \BibitemShut
  {NoStop}%
\bibitem [{\citenamefont {York}(1989)}]{York:1989jn}%
  \BibitemOpen
  \bibfield  {author} {\bibinfo {author} {\bibfnamefont {J.~W.}\ \bibnamefont
  {York}, \bibfnamefont {Jr.}},\ }\bibinfo {title} {{Initial data for
  collisions of black holes and other gravitational miscellany}}\ (\bibinfo
  {publisher} {Cambridge University Press},\ \bibinfo {address} {Cambridge,
  England},\ \bibinfo {year} {1989})\ p.~\bibinfo {pages} {89}\BibitemShut
  {NoStop}%
\bibitem [{\citenamefont {Loutrel}\ \emph {et~al.}(2019)\citenamefont
  {Loutrel}, \citenamefont {Liebersbach}, \citenamefont {Yunes},\ and\
  \citenamefont {Cornish}}]{Loutrel:2018ydu}%
  \BibitemOpen
  \bibfield  {author} {\bibinfo {author} {\bibfnamefont {N.}~\bibnamefont
  {Loutrel}}, \bibinfo {author} {\bibfnamefont {S.}~\bibnamefont
  {Liebersbach}}, \bibinfo {author} {\bibfnamefont {N.}~\bibnamefont {Yunes}},\
  and\ \bibinfo {author} {\bibfnamefont {N.}~\bibnamefont {Cornish}},\ }\href
  {https://doi.org/10.1088/1361-6382/aaf2a9} {\bibfield  {journal} {\bibinfo
  {journal} {Classical Quantum Gravity.}\ }\textbf {\bibinfo {volume} {36}},\
  \bibinfo {pages} {025004} (\bibinfo {year} {2019})},\ \Eprint
  {https://arxiv.org/abs/1810.03521} {arXiv:1810.03521 [gr-qc]} \BibitemShut
  {NoStop}%
\bibitem [{\citenamefont {Memmesheimer}\ \emph {et~al.}(2004)\citenamefont
  {Memmesheimer}, \citenamefont {Gopakumar},\ and\ \citenamefont
  {Schaefer}}]{Memmesheimer:2004cv}%
  \BibitemOpen
  \bibfield  {author} {\bibinfo {author} {\bibfnamefont {R.-M.}\ \bibnamefont
  {Memmesheimer}}, \bibinfo {author} {\bibfnamefont {A.}~\bibnamefont
  {Gopakumar}},\ and\ \bibinfo {author} {\bibfnamefont {G.}~\bibnamefont
  {Schaefer}},\ }\href {https://doi.org/10.1103/PhysRevD.70.104011} {\bibfield
  {journal} {\bibinfo  {journal} {Phys. Rev. D}\ }\textbf {\bibinfo {volume}
  {70}},\ \bibinfo {pages} {104011} (\bibinfo {year} {2004})},\ \Eprint
  {https://arxiv.org/abs/gr-qc/0407049} {arXiv:gr-qc/0407049} \BibitemShut
  {NoStop}%
\bibitem [{\citenamefont {Campanelli}\ and\ \citenamefont
  {Lousto}(1999)}]{Campanelli:1998jv}%
  \BibitemOpen
  \bibfield  {author} {\bibinfo {author} {\bibfnamefont {M.}~\bibnamefont
  {Campanelli}}\ and\ \bibinfo {author} {\bibfnamefont {C.~O.}\ \bibnamefont
  {Lousto}},\ }\href {https://doi.org/10.1103/PhysRevD.59.124022} {\bibfield
  {journal} {\bibinfo  {journal} {Phys. Rev. D}\ }\textbf {\bibinfo {volume}
  {59}},\ \bibinfo {pages} {124022} (\bibinfo {year} {1999})},\ \Eprint
  {https://arxiv.org/abs/gr-qc/9811019} {arXiv:gr-qc/9811019} \BibitemShut
  {NoStop}%
\bibitem [{\citenamefont {Lousto}\ and\ \citenamefont
  {Zlochower}(2007)}]{Lousto:2007mh}%
  \BibitemOpen
  \bibfield  {author} {\bibinfo {author} {\bibfnamefont {C.~O.}\ \bibnamefont
  {Lousto}}\ and\ \bibinfo {author} {\bibfnamefont {Y.}~\bibnamefont
  {Zlochower}},\ }\href {https://doi.org/10.1103/PhysRevD.76.041502} {\bibfield
   {journal} {\bibinfo  {journal} {Phys. Rev. D}\ }\textbf {\bibinfo {volume}
  {76}},\ \bibinfo {pages} {041502(R)} (\bibinfo {year} {2007})},\ \Eprint
  {https://arxiv.org/abs/gr-qc/0703061} {arXiv:gr-qc/0703061} \BibitemShut
  {NoStop}%
\bibitem [{\citenamefont {Ruiz}\ \emph {et~al.}(2008)\citenamefont {Ruiz},
  \citenamefont {Takahashi}, \citenamefont {Alcubierre},\ and\ \citenamefont
  {Nunez}}]{Ruiz:2007yx}%
  \BibitemOpen
  \bibfield  {author} {\bibinfo {author} {\bibfnamefont {M.}~\bibnamefont
  {Ruiz}}, \bibinfo {author} {\bibfnamefont {R.}~\bibnamefont {Takahashi}},
  \bibinfo {author} {\bibfnamefont {M.}~\bibnamefont {Alcubierre}},\ and\
  \bibinfo {author} {\bibfnamefont {D.}~\bibnamefont {Nunez}},\ }\href
  {https://doi.org/10.1007/s10714-007-0570-8} {\bibfield  {journal} {\bibinfo
  {journal} {Gen. Rel. Grav.}\ }\textbf {\bibinfo {volume} {40}},\ \bibinfo
  {pages} {2467} (\bibinfo {year} {2008})},\ \Eprint
  {https://arxiv.org/abs/0707.4654} {arXiv:0707.4654 [gr-qc]} \BibitemShut
  {NoStop}%
\bibitem [{\citenamefont {Healy}\ \emph {et~al.}(2017)\citenamefont {Healy},
  \citenamefont {Lousto},\ and\ \citenamefont {Zlochower}}]{Healy:2017mvh}%
  \BibitemOpen
  \bibfield  {author} {\bibinfo {author} {\bibfnamefont {J.}~\bibnamefont
  {Healy}}, \bibinfo {author} {\bibfnamefont {C.~O.}\ \bibnamefont {Lousto}},\
  and\ \bibinfo {author} {\bibfnamefont {Y.}~\bibnamefont {Zlochower}},\ }\href
  {https://doi.org/10.1103/PhysRevD.96.024031} {\bibfield  {journal} {\bibinfo
  {journal} {Phys. Rev. D}\ }\textbf {\bibinfo {volume} {96}},\ \bibinfo
  {pages} {024031} (\bibinfo {year} {2017})},\ \Eprint
  {https://arxiv.org/abs/1705.07034} {arXiv:1705.07034 [gr-qc]} \BibitemShut
  {NoStop}%
\bibitem [{\citenamefont {Br{\"u}gmann}\ \emph {et~al.}(2008)\citenamefont
  {Br{\"u}gmann}, \citenamefont {Gonz{\'a}lez}, \citenamefont {Hannam},
  \citenamefont {Husa},\ and\ \citenamefont {Sperhake}}]{Bruegmann:2007zj}%
  \BibitemOpen
  \bibfield  {author} {\bibinfo {author} {\bibfnamefont {B.}~\bibnamefont
  {Br{\"u}gmann}}, \bibinfo {author} {\bibfnamefont {J.~A.}\ \bibnamefont
  {Gonz{\'a}lez}}, \bibinfo {author} {\bibfnamefont {M.~D.}\ \bibnamefont
  {Hannam}}, \bibinfo {author} {\bibfnamefont {S.}~\bibnamefont {Husa}},\ and\
  \bibinfo {author} {\bibfnamefont {U.}~\bibnamefont {Sperhake}},\ }\href
  {https://doi.org/10.1103/PhysRevD.77.124047} {\bibfield  {journal} {\bibinfo
  {journal} {Phys. Rev. D}\ }\textbf {\bibinfo {volume} {77}},\ \bibinfo
  {pages} {124047} (\bibinfo {year} {2008})},\ \Eprint
  {https://arxiv.org/abs/0707.0135} {arXiv:0707.0135 [gr-qc]} \BibitemShut
  {NoStop}%
\bibitem [{\citenamefont {Lousto}\ \emph {et~al.}(2010)\citenamefont {Lousto},
  \citenamefont {Nakano}, \citenamefont {Zlochower},\ and\ \citenamefont
  {Campanelli}}]{Lousto:2009ka}%
  \BibitemOpen
  \bibfield  {author} {\bibinfo {author} {\bibfnamefont {C.~O.}\ \bibnamefont
  {Lousto}}, \bibinfo {author} {\bibfnamefont {H.}~\bibnamefont {Nakano}},
  \bibinfo {author} {\bibfnamefont {Y.}~\bibnamefont {Zlochower}},\ and\
  \bibinfo {author} {\bibfnamefont {M.}~\bibnamefont {Campanelli}},\ }\href
  {https://doi.org/10.1103/PhysRevD.81.084023} {\bibfield  {journal} {\bibinfo
  {journal} {Phys. Rev. D}\ }\textbf {\bibinfo {volume} {81}},\ \bibinfo
  {pages} {084023} (\bibinfo {year} {2010})},\ \bibinfo {note} {[Erratum: Phys.
  Rev. D 82, 129902 (2010)]},\ \Eprint {https://arxiv.org/abs/0910.3197}
  {arXiv:0910.3197 [gr-qc]} \BibitemShut {NoStop}%
\bibitem [{\citenamefont {Wiseman}(1992)}]{Wiseman:1992dv}%
  \BibitemOpen
  \bibfield  {author} {\bibinfo {author} {\bibfnamefont {A.~G.}\ \bibnamefont
  {Wiseman}},\ }\href {https://doi.org/10.1103/PhysRevD.46.1517} {\bibfield
  {journal} {\bibinfo  {journal} {Phys. Rev. D}\ }\textbf {\bibinfo {volume}
  {46}},\ \bibinfo {pages} {1517} (\bibinfo {year} {1992})}\BibitemShut
  {NoStop}%
\bibitem [{\citenamefont {Kesden}\ \emph {et~al.}(2015)\citenamefont {Kesden},
  \citenamefont {Gerosa}, \citenamefont {O'Shaughnessy}, \citenamefont
  {Berti},\ and\ \citenamefont {Sperhake}}]{Kesden:2014sla}%
  \BibitemOpen
  \bibfield  {author} {\bibinfo {author} {\bibfnamefont {M.}~\bibnamefont
  {Kesden}}, \bibinfo {author} {\bibfnamefont {D.}~\bibnamefont {Gerosa}},
  \bibinfo {author} {\bibfnamefont {R.}~\bibnamefont {O'Shaughnessy}}, \bibinfo
  {author} {\bibfnamefont {E.}~\bibnamefont {Berti}},\ and\ \bibinfo {author}
  {\bibfnamefont {U.}~\bibnamefont {Sperhake}},\ }\href
  {https://doi.org/10.1103/PhysRevLett.114.081103} {\bibfield  {journal}
  {\bibinfo  {journal} {Phys. Rev. Lett.}\ }\textbf {\bibinfo {volume} {114}},\
  \bibinfo {pages} {081103} (\bibinfo {year} {2015})},\ \Eprint
  {https://arxiv.org/abs/1411.0674} {arXiv:1411.0674 [gr-qc]} \BibitemShut
  {NoStop}%
\bibitem [{\citenamefont {Gerosa}\ \emph {et~al.}(2015)\citenamefont {Gerosa},
  \citenamefont {Kesden}, \citenamefont {Sperhake}, \citenamefont {Berti},\
  and\ \citenamefont {O’Shaughnessy}}]{Gerosa:2015tea}%
  \BibitemOpen
  \bibfield  {author} {\bibinfo {author} {\bibfnamefont {D.}~\bibnamefont
  {Gerosa}}, \bibinfo {author} {\bibfnamefont {M.}~\bibnamefont {Kesden}},
  \bibinfo {author} {\bibfnamefont {U.}~\bibnamefont {Sperhake}}, \bibinfo
  {author} {\bibfnamefont {E.}~\bibnamefont {Berti}},\ and\ \bibinfo {author}
  {\bibfnamefont {R.}~\bibnamefont {O’Shaughnessy}},\ }\href
  {https://doi.org/10.1103/PhysRevD.92.064016} {\bibfield  {journal} {\bibinfo
  {journal} {Phys. Rev. D}\ }\textbf {\bibinfo {volume} {92}},\ \bibinfo
  {pages} {064016} (\bibinfo {year} {2015})},\ \Eprint
  {https://arxiv.org/abs/1506.03492} {arXiv:1506.03492 [gr-qc]} \BibitemShut
  {NoStop}%
\bibitem [{\citenamefont {Hamilton}\ and\ \citenamefont
  {Rafikov}(2019{\natexlab{a}})}]{Hamilton:2019a}%
  \BibitemOpen
  \bibfield  {author} {\bibinfo {author} {\bibfnamefont {C.}~\bibnamefont
  {Hamilton}}\ and\ \bibinfo {author} {\bibfnamefont {R.~R.}\ \bibnamefont
  {Rafikov}},\ }\href {https://doi.org/10.1093/mnras/stz1730} {\bibfield
  {journal} {\bibinfo  {journal} {Mon. Not. R. Astron. Soc.}\ }\textbf
  {\bibinfo {volume} {488}},\ \bibinfo {pages} {5489} (\bibinfo {year}
  {2019}{\natexlab{a}})},\ \Eprint {https://arxiv.org/abs/1902.01344}
  {arXiv:1902.01344} \BibitemShut {NoStop}%
\bibitem [{\citenamefont {Hamilton}\ and\ \citenamefont
  {Rafikov}(2019{\natexlab{b}})}]{Hamilton:2019b}%
  \BibitemOpen
  \bibfield  {author} {\bibinfo {author} {\bibfnamefont {C.}~\bibnamefont
  {Hamilton}}\ and\ \bibinfo {author} {\bibfnamefont {R.~R.}\ \bibnamefont
  {Rafikov}},\ }\href {https://doi.org/10.1093/mnras/stz2026} {\bibfield
  {journal} {\bibinfo  {journal} {Mon. Not. R. Astron. Soc.}\ }\textbf
  {\bibinfo {volume} {488}},\ \bibinfo {pages} {5512} (\bibinfo {year}
  {2019}{\natexlab{b}})},\ \Eprint {https://arxiv.org/abs/1902.01345}
  {arXiv:1902.01345} \BibitemShut {NoStop}%
\bibitem [{\citenamefont {Sperhake}\ \emph {et~al.}(2011)\citenamefont
  {Sperhake}, \citenamefont {Bruegmann}, \citenamefont {Muller},\ and\
  \citenamefont {Sopuerta}}]{Sperhake:2011zz}%
  \BibitemOpen
  \bibfield  {author} {\bibinfo {author} {\bibfnamefont {U.}~\bibnamefont
  {Sperhake}}, \bibinfo {author} {\bibfnamefont {B.}~\bibnamefont {Bruegmann}},
  \bibinfo {author} {\bibfnamefont {D.}~\bibnamefont {Muller}},\ and\ \bibinfo
  {author} {\bibfnamefont {C.}~\bibnamefont {Sopuerta}},\ }\href
  {https://doi.org/10.1088/0264-9381/28/13/134004} {\bibfield  {journal}
  {\bibinfo  {journal} {Classical Quantum Gravity.}\ }\textbf {\bibinfo
  {volume} {28}},\ \bibinfo {pages} {134004} (\bibinfo {year}
  {2011})}\BibitemShut {NoStop}%
\bibitem [{\citenamefont {Bishop}\ and\ \citenamefont
  {Rezzolla}(2016)}]{Bishop:2016lgv}%
  \BibitemOpen
  \bibfield  {author} {\bibinfo {author} {\bibfnamefont {N.~T.}\ \bibnamefont
  {Bishop}}\ and\ \bibinfo {author} {\bibfnamefont {L.}~\bibnamefont
  {Rezzolla}},\ }\href {https://doi.org/10.1007/s41114-016-0001-9} {\bibfield
  {journal} {\bibinfo  {journal} {Living Rev. Relativity}\ }\textbf {\bibinfo
  {volume} {19}},\ \bibinfo {pages} {2} (\bibinfo {year} {2016})},\ \Eprint
  {https://arxiv.org/abs/1606.02532} {arXiv:1606.02532 [gr-qc]} \BibitemShut
  {NoStop}%
\end{thebibliography}%


\end{document}